\documentclass[12pt,letterpaper]{article}
\pdfoutput=1
\usepackage{amsmath,amssymb,calc, latexsym, bbm, array, amsthm}
\usepackage{graphicx}
\usepackage{subfig}
\usepackage[nosort]{cite}
\usepackage{epsfig,psfrag}

\usepackage{ulem}
\usepackage{color}
\definecolor{burgundy}{rgb}{0.8,.2,.2}

\makeatletter
\renewcommand\section{\@startsection {section}{1}{\z@}%
                                   {-3.5ex \@plus -1ex \@minus -.2ex}%nn
                                   {2.3ex \@plus.2ex}%
                                   {\normalfont\large\bfseries}}
\renewcommand\subsection{\@startsection{subsection}{2}{\z@}%
                                     {-3.25ex\@plus -1ex \@minus -.2ex}%
                                     {1.5ex \@plus .2ex}%
                                     {\normalfont\bfseries}}
\makeatother

\theoremstyle{plain}

\theoremstyle{definition}

\let\non\nonumber

\let\a=\alpha
\let\b=\beta

\let\l=\lambda

\let\s=\sigma

\let\S=\Sigma
\let\Th=\Theta

\newcommand{\del}{\partial}

\def\one{^{(1)}}

\newcommand{\bea}{\begin{eqnarray}}
\newcommand{\eea}{\end{eqnarray}}
\newcommand{\be}{\begin{equation}}
\newcommand{\ee}{\end{equation}}
\newcommand{\bma}{\begin{pmatrix}}
\newcommand{\ema}{\end{pmatrix}}

\newcommand{\hlf}{\frac{1}{2}}

\newcommand{\Z}{{\mathbb Z}}

\newcommand{\F}{{\mathbb F}}
\newcommand{\PP}{{\mathbb P}}
\newcommand{\CC}{{\mathbb C}}

\newcommand{\cA}{{\cal A}}
\newcommand{\cF}{{\cal F}}

\newcommand{\cJ}{{\cal J}}

\newcommand{\I}{{\cal I}}

\newcommand{\Om}{\Omega}

\newcommand{\SO}{\operatorname{SO}}

\newcommand{\SU}{\operatorname{SU}}

\newcommand{\U}{\operatorname{U}}

\newcommand{\La}{\Lambda}
\newcommand{\la}{\lambda}

\newcommand{\G}{\Gamma}
\newcommand{\e}{\epsilon}

\newcommand{\dd}{\delta}

\newcommand{\m}{\mu}
\newcommand{\n}{\nu}
\newcommand{\p}{\partial}

\newcommand{\f}{\psi}

\newcommand{\D}[1]{\ensuremath{\mathrm{D}#1}}

\newcommand{\C}[1]{$(\ref{#1})$}

\def\IZ{\relax\ifmmode\mathchoice
{\hbox{\cmss Z\kern-.4em Z}}{\hbox{\cmss Z\kern-.4em Z}}
{\lower.9pt\hbox{\cmsss Z\kern-.4em Z}} {\lower1.2pt\hbox{\cmsss
Z\kern-.4em Z}}\else{\cmss Z\kern-.4em Z}\fi}
\def\IR{\relax{\rm I\kern-.18em R}}

\def\one{{\hbox{ 1\kern-.8mm l}}}

\def\tr{{\rm tr\,}}
\def\Tr{{\rm Tr\,}}

\newlength{\bredde}
\def\slash#1{\settowidth{\bredde}{$#1$}\ifmmode\,\raisebox{.15ex}{/}
\hspace*{-\bredde} #1\else$\,\raisebox{.15ex}{/}\hspace*{-\bredde}
#1$\fi}

\newsavebox{\zzzbar}
\sbox{\zzzbar}
  {\setlength{\unitlength}{0.9em}
  \begin{picture}(0.6,0.7)
  \thinlines
  \put(0,0){\line(1,0){0.6}}
  \put(0,0.75){\line(1,0){0.575}}
  \multiput(0,0)(0.0125,0.025){30}{\rule{0.3pt}{0.3pt}}
  \multiput(0.2,0)(0.0125,0.025){30}{\rule{0.3pt}{0.3pt}}
  \put(0,0.75){\line(0,-1){0.15}}
  \put(0.015,0.75){\line(0,-1){0.1}}
  \put(0.03,0.75){\line(0,-1){0.075}}
  \put(0.045,0.75){\line(0,-1){0.05}}
  \put(0.05,0.75){\line(0,-1){0.025}}
  \put(0.6,0){\line(0,1){0.15}}
  \put(0.585,0){\line(0,1){0.1}}
  \put(0.57,0){\line(0,1){0.075}}
  \put(0.555,0){\line(0,1){0.05}}
  \put(0.55,0){\line(0,1){0.025}}
  \end{picture}}

\def\Im{{\rm Im\, }}
\def\Re{{\rm Re\, }}

\newcommand{\ena}{\end{eqnarray}}
\newcommand{\beqa}{\begin{eqnarray}}
\newcommand{\eeqa}{\end{eqnarray}}

\def\G{\Gamma}

\def\cD{{\cal D}}

\renewcommand{\b}{\beta}
\newcommand{\g}{\gamma}

\def\d{{\rm d}}

\newcommand{\ibar}{{\bar \imath}}
\newcommand{\jbar}{{\bar \jmath}}
\newcommand{\thbar}{{\bar \theta}}
\newcommand{\Dbar}{{\bar D}}

\newcommand{\labar}{{\bar \lambda}}

\newcommand{\hG}{{\hat \G}}
\newcommand{\Jbar}{{\bar J}}

\newfont{\goth}{ygoth.tfm scaled 1200}                   % gothic font (usual)
\def\a{\alpha}
\def\b{\beta}
\def\e{\epsilon}
\def\th{\theta}
\def\f{\phi}
\def\g{\gamma}

\def\j{\psi}

\def\l{\lambda}
\def\m{\mu}
\def\n{\nu}

\def\s{\sigma}

\def\D{\Delta}
\def\F{\Phi}
\def\G{\Gamma}

\def\L{\mathcal{L}}

\def\S{\Sigma}

\def\U{\Upsilon}

\renewcommand{\SO}{\operatorname{SO}}

 \numberwithin{equation}{section}

\def\1{{(1)}}
\def\2{{(2)}}
\def\3{{(3)}}

\def\1{{\bf 1}}
\def\a{{\alpha}}

\def\M{{\mathcal M}}

\def\CC{{\mathbb C}}

\renewcommand\sl{\!\!\!\!\diagup}

\def\p{\partial}
\def\pb{\bar{\partial}}
\def\SO{\operatorname{SO{}}}
\def\GU{\operatorname{U{}}}
\def\SU{\operatorname{SU{}}}
\def\ff#1#2{{\textstyle\frac{#1}{#2}}}

\def\normd#1{{{:\!#1\!\!:}}}

\def\la{{\langle}}
\def\ra{{\rangle}}

\newcommand\wb{\overline{w}}
\newcommand\zb{\overline{z}}
\newcommand\gammab{\overline{\gamma}}
\newcommand\psib{\overline{\psi}}

\newcommand\chib{{\overline{\chi}}}
\newcommand\lambdab{{\overline{\lambda}}}
\newcommand\vphi{{\varphi}}
\newcommand\vphib{{\overline{\vphi}}}
\newcommand\vtheta{{\vartheta}}
\newcommand\ep{{\epsilon}}

\def\cA{{\cal A}}

\newcommand\Ab{\overline{A}}
\def\tcA{{\tilde\cA}}

\def\cF{{\cal F}}

\def\Jb{{{\overline{J}}}}
\def\cJ{{\cal J}}

\def\JJ{{{\mathbb J}}}
\def\JJb{{{\overline{\JJ}}}}

\def\tint{{{\text{int}}}}

\def\tYuk{{{\text{Yuk}}}}
\def\tct{{{{\text{c.t.}}}}}
\def\teff{{{{\text{eff}}}}}

\newcommand\al{{(\alpha)}}
\newcommand\bt{{(\beta)}}
\newcommand\ab{{(\alpha\beta)}}

\begin{document}
\begin{titlepage}

\begin{center}

{December 5, 2012}
%\today
\hfill         \phantom{xxx}  EFI-12-28

\vskip 2 cm {\Large \bf  Target Spaces from Chiral Gauge Theories}
%\vskip 0.3 cm {\Large \bf K\"ahler Potentials}\non\\
\vskip 1.25 cm {\bf  Ilarion Melnikov$^{a}$\footnote{ilarion.melnikov@aei.mpg.de, $^2$cquigley@uchicago.edu, $^3$sethi@uchicago.edu, $^4$stern@math.duke.edu}, Callum Quigley$^{b2}$, Savdeep Sethi$^{b3}$ and Mark Stern$^{c4}$}\non\\
%{\vskip 0.5cm  $^{a}$ {\it Dept. of Physics, University of Cincinnati,
%Cincinnati, OH 45221, USA}\non\\
\vskip 0.2 cm
{\it $^{a}$Max Planck Institute for Gravitational Physics, Am M\"uhlenberg 1, D-14476 Golm, Germany}\non\\ \vskip 0.2cm
 {\it $^{b}$Enrico Fermi Institute, University of Chicago, Chicago, IL 60637, USA}\non\\ \vskip 0.2cm
$^{c}${\it Department of Mathematics, Duke University, Durham, NC 27708, USA\non\\}
%Valckenierstraat 65, 1018 XE Amsterdam, The Netherlands} \non\\}

\end{center}
\vskip 1.5 cm

\begin{abstract}
\baselineskip=18pt

Chiral gauge theories in two dimensions with $(0,2)$ supersymmetry are central in the study of string compactifications. Remarkably little is known about generic $(0,2)$ theories.  
We consider theories with branches on which multiplets with a net gauge anomaly become massive. The simplest example is a relevant perturbation of the gauge theory that flows to the ${\CC\PP^n }$ model. 
To compute the effective action, we derive a useful set of Feynman rules for $(0,2)$ supergraphs. From the effective action, we see that the infra-red geometry reflects the gauge anomaly by the presence of a  boundary at finite distance.
In generic examples, there are  boundaries, fluxes and branes; the resulting spaces are non-K\"ahler.

\end{abstract}

\end{titlepage}

\tableofcontents
\newpage

\section{Introduction}
\label{intro}

It has been clear for a number of years that generic string geometries are quite different from the geometries that have been the main focus of study over past years. Most of the effort in understanding string compactifications has centered on Calabi-Yau spaces or closely related variants. This is for good reason: Calabi-Yau spaces solve the space-time Einstein equations which govern large volume string compactifications. These spaces form a natural set of compactifications of the type II string with ${\cal N}=2$ space-time supersymmetry. However, Calabi-Yau spaces are rather special because they do not involve any flux degrees of freedom. We expect most compactifications to involve fluxes; in the heterotic string these require non-K\"ahler geometries, typically with string scale features. Among models with ${\cal N}=1$ space-time supersymmetry, Calabi-Yau spaces should look quite distinguished.

A natural linear framework for studying non-linear sigma models is provided by two-dimensional gauge theory~\cite{Witten:1993yc}. This includes both conformal and non-conformal models. Within this framework, Calabi-Yau spaces emerge naturally as solutions to linear theories with $(2,2)$ world-sheet supersymmetry. Chiral gauge theories with $(0,2)$ supersymmetry have also been studied. These are heterotic quantum field theories. Most of the models that have been considered still involve Calabi-Yau target spaces but with a gauge bundle that differs from the tangent bundle needed for $(2,2)$ supersymmetry. Such models still have a large volume limit in which the space-time supergravity equations of motion are solved. However, these cannot be generic string compactifications. This leads to a quandary: how do we describe generic compactifications?

It has been realized that $(0,2)$ chiral gauge theories have a much richer vacuum structure than is seen in  $(2,2)$ models or their deformations.  Both $(2,2)$ and $(0,2)$ models can include field-dependent Fayet-Iliopoulos (FI) couplings. The field-dependence is via log interactions of scalar fields in a coupling of schematic form
\be\label{sketch}
S_{\rm log} =  \int d^2x  \, {\rm Im}\left(\log \s \right) F_{01},
\ee
where $\s$ is a complex scalar field and $F_{01}$ is an abelian gauge field strength.
However for $(0,2)$ models, a new twist is possible: the field $\s$ can be charged~\cite{Quigley:2011pv, Blaszczyk:2011ib}.   In this case, gauge-invariance is violated at the classical level in a way that can be compensated by a one-loop quantum anomaly. For earlier related models, see~\cite{Adams:2006kb, Adams:2009tt, Adams:2009av}. This observation led to a proposal for torsional models~\cite{Quigley:2011pv}\ that involve log couplings of the form~\C{sketch}.

However, it is unclear when such models exist as quantum field theories, as non-polynomial interactions like~\C{sketch}\ are difficult to quantize. What is really needed is a completely linear framework where the log interactions arise naturally from integrating out heavy multiplets. Such a framework was proposed in~\cite{Quigley:2012gq}, where it was realized that $(0,2)$ theories possess novel branches not found in $(2,2)$ models. There are two distinct situations to consider. The first case involves integrating out  non-anomalous combinations of heavy fields. This was the main focus of~\cite{Quigley:2012gq}. This branch corresponds to inserting NS-brane and anti-brane sources into the geometry. The resulting gauge theory vacuum equations provide a natural generalization of symplectic reduction, and the moduli spaces are complex non-K\"ahler manifolds. For example, the complex manifold $S^5\times S^1$ arises naturally from a simple example.

The other possibility involves integrating out gauge anomalous combinations of heavy fields. The physics and mathematics for this case are very different from the non-anomalous situation. Understanding how the anomaly is reflected in the infra-red geometry is the aim of this project.

%These models can appear as unconventional branches of a $(0,2)$ gauged linear sigma model (GLSM).  The goal of this project is to consider these new branches and explore their associated geometries.

\subsection{The basic idea} \label{basicidea}

A summary of our conventions and the basics of $(0,2)$ theories can be found in Appendix~\ref{superfieldconventions}. The essential physics  we wish to understand is the effective action that describes integrating out an anomalous combination of left and right-moving fermions. To explain the basic setup, let us consider a single $U(1)$ gauge field. There are two pieces of holomorphic data that must be specified in defining a $(0,2)$ theory. Each left-moving Fermi superfield $\G$ need not be chiral but can satisfy the condition~\cite{Witten:1993yc}
\be
\bar{\mathfrak{D}}_+ \G = \sqrt{2} E(\F),
\ee
where $E$ is a holomorphic function of chiral superfields $\F$ with the same charge as $\G$. This coupling produces a bosonic potential $|E|^2$, which must vanish on the moduli space. The remaining  data are the more familiar superpotential couplings that take the form
\be
S_J = -{1\over \sqrt{2}}\int\d^2x\d\th^+\, \G \cdot J(\F) + c.c.,
\ee
where $J$ is a holomorphic coupling. This coupling produces a bosonic potential $|J|^2$. In the presence of both $E$ and $J$ couplings, supersymmetry requires $E\cdot J=0$.

We are interested in the case where $E$ takes the form
\be\label{chargedE}
E = m \S P,
\ee
where both $\S$ and $P$ are charged chiral superfields. This is possible in $(0,2)$ theories but not in $(2,2)$ theories. The mass scale $m$ is needed if we assume canonical dimension $0$ for all scalar fields.

The lowest component of $\S$ is a complex scalar $\s$, while the lowest component of $P$ is $p$. In $(2,2)$ theories, $\S$ usually denotes a neutral field. Since $\S$ is charged here, there is really no reason to distinguish $\S$ from any other chiral multiplet like $\Phi$ or $P$ other than conformity to familiar notation. If $P$ has charge $Q_P$ and $\S$ has charge $Q_\S$ then
\be
Q_\G = Q_\S + Q_P.
\ee
In later discussions, it will be useful for us to note that models with just  $E$-couplings are equivalent to models with just superpotential $J$-couplings. Rather than the $E$-coupling of~\C{chargedE}, we could equally well consider the superpotential coupling
\be
S_J = -{1\over \sqrt{2}}\int\d^2x\d\th^+\, \hat\G  \S P + c.c.,
\ee
where $Q_{\hat\G} = - Q_\G$ and $\bar{\mathfrak{D}}_+ \hat\G = 0$. With this equivalence in mind, let us start by considering a model with just the $E$-coupling given in~\C{chargedE}.

To fix our conventions, note that under a gauge transformation with parameter $\a$ a field $\Phi$, with charge $Q$, and the gauge-fields transform as follows:
\be
\Phi \rightarrow e^{iQ \a} \Phi, \qquad A_\pm \rightarrow A_\pm - \p_\pm \a.
\ee
We can now consider the effect of the $E$-coupling. If $\S \neq 0$, the coupling~\C{chargedE}\ masses up the anomalous combination of left and right-moving fermions contained in $\G$ and $P$. The net anomaly from the massive $\G$ and $P$ fields,
\bea\label{expanomaly}
  {1\over 4\pi}Q_P^2 - {1\over 4\pi} Q_\G^2 = - {1\over 4\pi} Q_\S ( Q_P + Q_\G),
\eea
must be reflected in any low-energy effective action.

 Let us take the mass scale $m$ to be much larger than any other scale in the problem. % like the gauge coupling $e$.
 The other natural dimensionful parameter is the gauge coupling, $e$, with mass dimension one.
 In general, the physics depends on the dimensionless combination $em^{-1}$. We will usually work in the limit where $m\gg e$ so we can treat the gauge dynamics perturbatively.
We can then integrate out the anomalous combination of massive fields at one-loop.

It is worth noting that by scaling the charges, the anomaly can be made arbitrarily large with either a positive or negative sign. Setting the FI parameter $r\gg1$, the deep infrared theory will be in the same universality class as a non-linear sigma model. For conventional branches, the corresponding geometry is K\"ahler; for conformal models, the geometry is Calabi-Yau up to small corrections. In our case, which is really the generic situation, the sigma model geometry must reflect the UV gauge anomaly in an essential way.  This is the key issue we want to understand.

\subsection{The UV moduli space}\label{UV}

Let us examine the classical moduli space of the basic UV model of section~\ref{basicidea}, which contains a single $U(1)$ gauge multiplet and charged chiral matter $(\S,P)$ along with a charged Fermi superfield $\G$. Consistency requires an anomaly free theory and this combination of fields satisfying~\C{expanomaly}\ is anomalous. Imagine adding a collection of superfields, $\F^i$ and $\G^\al$, which supplement the basic fields $(\G, \S, P)$. The only characteristic of these additional fields is that they do not couple directly to $(\G, \S, P)$. They do, however, contribute to the gauge anomaly which must vanish:
\be\label{uvanomaly}
\left( Q_P^2 + Q_\S^2 - Q_\G^2 \right)+  \left( Q_{\F^i}^2  -  Q_{\G^\a}^2 \right)=0.
\ee
Here $Q_\F^2$ and $Q_{\G_\a}^2$ denote the contributions of potentially many fields.

The UV theory has no log interactions so the moduli space is obtained by minimizing the bosonic potential
\be
V = {1\over 2 e^2} D^2 + |E|^2.
\ee
The condition $D=0$ requires
\be
Q_P |P|^2 + Q_\S |\S|^2 + \sum Q_\F |\F|^2 = r.
\ee
After quotienting by the $U(1)$ action, this constraint gives a weighted projective space if all charges are positive. If some charges are negative, the space is a non-compact toric variety. Vanishing of the $E$-term carves out the hypersurface
\be
\S P = 0
\ee
in this projective space. This is the classical moduli space with two branches, where either $\S\neq 0$ or $P\neq 0$, and a singular locus where $\S=P=0$ and the two branches touch. We expect this classical picture to be drastically modified by quantum effects. On the branch with $\S\neq 0$, integrating out the anomalous massive  pair $(\G, P)$ at one loop generates a log interaction of the form~\C{sketch}. The two branches of the classical moduli space with either $\S\neq 0$ or $P\neq 0$ are already disconnected by this log interaction, which prevents either $\S$ or $P$ from vanishing. A study of the resulting vacuum equations with log interactions leads, however,  to a puzzle about how supersymmetry is preserved; for example, the branch with $\S\neq 0$ can be a non-complex sphere. 
The resolution of this puzzle requires a careful examination of the quantum corrections, and one of our main results is that the sphere is in fact replaced by a ball with a finite distance boundary.  The appearance of such boundaries should be a very generic feature in (0,2) target geometries.

\subsection{Summary and an outline}

The picture that emerges from our analysis is a target space constructed by a procedure that generalizes a holomorphic quotient. In conventional branches of abelian gauge theory, the moduli space is realized via a symplectic quotient: solve the $D$-term equations and quotient by the gauge group action, which is equivalent to a holomorphic quotient by the complexified gauge group. This promotion of a $U(1)$ compact gauge quotient to a $\CC^\ast$ quotient is natural in supersymmetric gauge theory. The structure of superspace automatically admits the action of the complexified gauge group as a symmetry group if we do not choose a particular gauge like Wess-Zumino gauge.

There is a tension between the supersymmetry requirement that we implement a holomorphic quotient and the inclusion of charged log couplings of the form~\C{sketch}\ in a low-energy effective action. This comes about because the solution of the $D$-term equations is no longer unique when there are charged log interactions. A unique solution is needed to complexify the gauge group action. In section~\ref{modulispace}, we begin by exploring the moduli spaces which emerge from solving the vacuum equations of gauge theories with log interactions. This is an interesting mathematical construction in its own right, with the familiar toric construction as a special case. Our analysis focuses on models with a single $U(1)$ gauge factor. We will meet a puzzle: among the target spaces is $S^4$ which does not admit any complex structure. World-sheet supersymmetry, however, requires a complex manifold. If this is the target manifold, world-sheet supersymmetry would break spontaneously, which is unexpected.  %\textbf{IVM:  why is this unexpected?  what about deformed SUSY algebra in UV theory, as in Dumitrescu and Seiberg?  That's what happens for $\PP^1$ model without left-movers.}

Before resolving this puzzle, we revisit the chiral anomaly in section~\ref{axanom}. What is of particular importance to us is the normalization of couplings in the effective action obtained by integrating out anomalous multiplets. Specifically, a subtle factor of two in~\C{expanomaly}\ when compared with the global chiral anomaly. On very general grounds, we determine the dependence of the low-energy effective action on the phase of the Higgs field which masses up the anomalous multiplet, whether or not the model has any supersymmetry. This is part of the data determining the effective action.

In section~\ref{perturbationtheory}, we start with a UV complete non-anomalous gauge theory and integrate out a single  massive $(\G, P)$ pair at one-loop. This is a fairly subtle calculation since we are dealing with a chiral supersymmetric gauge theory and there is no regulator that can preserve all the symmetries of the theory. To perform this integration and determine the Wilsonian effective action, we develop a set of Feynman rules for $(0,2)$ supergraphs. Those rules are likely to be useful in wider contexts. The effective action is computed both in a general gauge choice and in the specific case of unitary gauge.

The effective action includes a coupling like~\C{sketch}\ that reproduces the gauge anomaly of the massed up multiplet, but it also includes two additional critical contributions: the first is a correction to the metric of the $\S$-field. This modification is  similar to what one finds when integrating out non-anomalous multiplets. The second contribution comes from path-integrating over the high energy modes of the remaining light superfields, whose fermion content is anomalous. This last contribution vanishes when integrating out massive multiplets with no net gauge anomaly.

In section~\ref{NLSM}, we study the non-linear sigma model (NLSM)  obtained by classically integrating out the gauge fields of the  gauged linear sigma model (GLSM). This is a manifestly $(0,2)$ sigma model, but one defined in terms of a metric, $G$, and $B$-field which transform unconventionally under holomorphic changes of coordinate. At first sight, this is very peculiar. For example, there is a natural patch where we use the $\CC^\ast$ gauge symmetry to set the Higgs field $\S=1$. This gauge choice is always possible because the log interactions prevent $\S$ from vanishing. In this patch, we find that the metric $G$ is K\"ahler. However, in other patches the metric is not K\"ahler. This is only possible if the metric does not transform as a tensor with an invariant line element. This phenomenon does not happen for GLSMs in which anomaly-free massive multiplets are integrated out, but it does happen here.

 To find a metric with conventional transformation properties, we are forced to leave the off-shell $(0,2)$ formalism and work only with manifest $(0,1)$ supersymmetry. With some hindsight explained in section~\ref{NLSManomalies}, it is clear that this had to be the case for models with tree-level torsion. In section~\ref{invariant}, we define a natural metric $\widehat G$ invariant under holomorphic coordinate reparametrizations. It is this metric which provides a conventional notion of geometry to our target spaces. Using this metric, we see that these spaces are non-K\"ahler.

 We also see that the new couplings in the effective action, reflecting the gauge anomalous nature of the massive multiplet, lead to rather interesting behavior for the metric $\widehat G$. The scale factor for a circle in the space shrinks down to a fairly small but non-vanishing value, determined by a transcendental equation; it then begins to grow until it  diverges at a boundary located at finite distance in the target space. This large variation in the scale factor suggests that string solutions built from these spaces might exhibit hierarchies.
 %At this boundary, the scale factor for at least a circle factor of the metric diverges.
% What we find is a critical modification to the low-energy sigma model metric.  It
%collapsed boundary at finite distance in the target manifold, which is complex and even K\"ahler for the case of a single massive pair.
For the example that would have given $S^4$ without including these corrections, we find that
%restricting to the subset on which the modified metric is Riemannian,
the sphere is roughly cut in half giving a $4$-ball. Near the boundary, the form of the metric  suggests that we might want to study the theory in T-dual variables to find a weakly coupled description. That possibility will be explored elsewhere.

In addition to producing a metric on the target space, the linear model also yields an $H$-flux that should satisfy the heterotic Bianchi identity:
\be\label{bianchi}
dH  = {\alpha'\over 4}\left[ \tr R_+\wedge R_+ - \tr F\wedge F \right],
\ee
where $R_+$ is curvature of the spin connection twisted by $H$-flux.\footnote{See~\cite{Becker:2009df}\ for an explanation about why there is a preferred gravitational connection, $\Omega_+$, used to evaluate the Chern-Simons forms and curvatures.  See~\cite{Hull:1986xn}\ for the $(0,1)$ superspace counter-terms associated with these Chern-Simons corrections; a recent discussion is given in~\cite{Melnikov:2012cv}. }
The corresponding gauge transformations of the $B$-field lead to subtleties in defining the NLSM quantities, but it is clear that the non-anomalous GLSM produces a solution to the Bianchi identity.   To see this directly at the NLSM level will require a better understanding of the boundary.  It might be possible to find similar ``quantum quotient'' constructions in type II string backgrounds with orientifold planes and D-branes, which can also modify Bianchi identities.

Boundaries appear in several settings when studying string compactifications. For example, a strong coupling limit of the $E_8\times E_8$ heterotic string compactified on $\M$ develops a boundary. In that limit, the appropriate description is heterotic M-theory on $\M\times S^1/\Z_2$~\cite{Horava:1996ma}. A closer analogue for the boundary we see is found in the geometry of gauged WZW models. The simplest case is the $SU(2)/U(1)$ WZW model~\cite{Bardakci:1990ad}. The geometry of the covering space is $S^3$ so a straight geometric quotient would give
\be
{S^3 \over S^1} \sim S^2.
\ee
Because of the presence of $H$-flux threading the $S^3$, the metric on the quotient space actually degenerates at the equator of what should be $S^2$ producing a curvature singularity. This degeneration changes the topological type of the target manifold from $S^2$ to a disk. There is simply no room for $3$-form flux on $S^2$ so this topology change is the only residue of the flux present on the covering space. This has some similarities to what we see in models with a single massive anomalous pair, although our cases are typically not conformal.

The larger picture that emerges for heterotic compactifications involves three basic building blocks. The first are brane sources obtained by integrating out non-anomalous multiplets. The second are boundaries and fluxes from anomalous massive multiplets. The final ingredient is the gauge bundle specified by the choice of left-moving fermions. If we consider combinations of anomalous and non-anomalous massive multiplets, %investigated in section~\ref{moremassivemultiplets},
we will generally find target manifolds that are non-K\"ahler spaces with boundaries, branes and $H$-flux. General combinations of these ingredients should produce a large landscape of heterotic quantum field theories.

We expect compact conformal models to appear via complete intersections obtained by turning on additional superpotential or $E$-couplings. These are the models that can potentially be used as string vacua. There are many directions to pursue. A sample of questions include: what are the precise conditions for conformal invariance? This could be investigated perhaps along the lines of~\cite{Koroteev:2010ct, Cui:2011rz, Cui:2011uw, Adams:2012sh}. What are the space-time spectra and moduli spaces for these models? How many vacua exist for massive models? What is the structure of the ground ring? For a discussion of heterotic ground rings and quantum sheaf cohomology,  see~\cite{Adams:2003zy, Katz:2004nn, Adams:2005tc, Sharpe:2006qd, Tan:2006qt, Guffin:2007mp, Tan:2007bh, Donagi:2011va, Donagi:2011uz}.
Does a weakly coupled description of the high curvature boundary exist? Such a description might follow from a mirror description which generalizes~\cite{Hori:2000kt, Adams:2003zy}. For a review of $(0,2)$ mirror symmetry, see~\cite{Melnikov:2012hk}. What is the right way to describe these target manifolds? Can threshold corrections be computed in these models, perhaps along the lines of~\cite{Carlevaro:2012rz, Carlevaro:2009jx}? Can elliptic genera be computed for these generically non-K\"ahler spaces, perhaps along the lines of~\cite{ Adams:2009zg}?

\section{Moduli Spaces and A Supersymmetry Puzzle}\label{modulispace}

We would like to describe the moduli spaces that arise when we include log interactions of the form~\C{sketch}. This question is broader than models with the specific UV completion described in section~\ref{basicidea}. We will therefore allow charged log interactions and study the associated moduli space of zero energy configurations. Most of the results we derive here are for the case of a single $U(1)$ gauge factor. The general case is open and quite fascinating. It is natural to conjecture that classically gauge-invariant models will give complex moduli spaces. Some examples of this type were studied in~\cite{Quigley:2012gq}. Our interest here is primarily in models that are not classically gauge-invariant.

\subsection{The gauge group action}

Consider a  $G=U(1)^n$ abelian gauge theory.   Coupled to these gauge fields are $d$ chiral superfields $\Phi^i$ with charges $Q^a_i$. The bosonic lowest components of $\Phi^i$ are denoted $\phi^i$. Under a gauge transformation with parameters $\a^a$,
\be \label{gaugeaction}
\Phi^i \rightarrow e^{i \a^a Q^a_i} \Phi^i.
\ee
The gauge fields are arranged into gauge superfields $A^a$ and $V^a_-$ with $a=1, \ldots n$. The corresponding field strength is a fermionic superfield $\Upsilon^a$. Supersymmetric log interactions appear via superpotential couplings,
\be\label{FIparam}
S_{\rm log} =  {i\over 4} \int d^2x d\th^+ \, N_i \log (\F^i) \Upsilon + c.c.,
\ee
which modify both the $D$-term constraints and introduce $H$-flux into the resulting geometries.\footnote{This sign convention agrees with~\cite{Quigley:2012gq}. With this convention, brane-like solutions correspond to positive $N$ while anti-brane-like solutions correspond to negative $N$. } There can also be additional gauge-invariant couplings in~\C{FIparam}, but we will focus on the charged log case which is the essential new feature.

We will assume $4\pi N_i\in { \Z }$. This quantization condition is  consistent with models where the logs are obtained by integrating out massive anomalous multiplets, as we will explain in section~\ref{axanom}. The integration procedure will be described in section~\ref{perturbationtheory}; it might be possible to relax this condition for models which are not obtained from this UV completion. There is a $D$-term constraint for each gauge factor,
\be \label{dterms}
\sum_i Q^a_i |\phi^i|^2 - N_i^a \log |\phi^i| = r^a,
\ee
where the $r^a$ are the classical FI parameters.  The solution of the $D$-term constraints is a surface $W_{r,Q,N} \subset \CC^d$. The geometric moduli space $X_{r,Q,N}$ is the further quotient by the global gauge group: $X_{r,Q,N}  = W_{r,Q,N}/G$.
The basic defining data are therefore the charges $Q_i^a$, the integers $N_i^a$ and the FI parameters $r^a$.
%The charges $Q_i^a$ need not be integer. Relaxing integrality of $Q_i^a$ leads to stacks studied in~\cite{}.
We will assume integral $Q_i^a$ as usual.

This combinatorial data describes a toric variety only for the special case $N_i^a=0$. Consider the algebraic torus $\left(\CC^\ast \right)^d$ acting on  $\Phi$ by
\be
\Phi^i \rightarrow \lambda^i \Phi^i, \qquad (\lambda^1, \ldots \lambda^d) \in\left(\CC^\ast \right)^d.
\ee
The quotient by $G$ removes the compact part of a $\left(\CC^\ast \right)^n$ action determined by~\C{gaugeaction}. If all $N_i^a=0$ then we can find a unique solution to the $D$-term constraints~\C{dterms}\ in the orbit of the $\left(\CC^\ast \right)^n$ action acting on any sufficiently generic choice of $\Phi^i$. This fixes the scaling symmetry in $\left(\CC^\ast \right)^n$. We can therefore view solving the $D$-term constraints (which determine $W$) and quotienting by $G$ (which determines $X$) as gauge-fixing $\left(\CC^\ast \right)^n$. The moduli space is  a toric variety.

When some $N_i^a\neq 0$, the $D$-term constraints typically admit multiple solutions under the action of  $\left(\CC^\ast \right)^n$ and the resulting space need not be toric. Rather the moduli space of $D$-term solutions is given by the solutions of transcendental equations.
From the gauge theory, there is  a natural picture of the moduli space, $X$, in terms of a real skeleton constructed by solving the $D$-term equations. Over this skeleton are fibered compact tori from the phases of the $\Phi^i$. For models without log interactions, this is the Delzant polytope used to construct toric varieties. The way in which the compact tori degenerate over the skeleton determines the topology of the space. We will see how this works in specific examples.

\subsection{Anomaly cancelation}

The log interactions typically produce a classical violation of gauge invariance. There are many interesting cases which are gauge invariant~\cite{Quigley:2012gq}, where we expect complex moduli spaces.  Note that it is always possible to introduce a classical gauge-non-invariant superspace $AV$ coupling,
\be \label{AVcoupling}
S_{\rm anom} = - {1\over 4\pi} \int d^2x d^2\theta^+ Q_i^{[a} N_i^{b]} \, A^a V_-^b,
\ee
to cancel any antisymmetric classical gauge anomaly. So if we insist on theories which are anomaly free, we must  impose the condition
\be \label{anomcondition}
\cA^{ab} - \sum_i Q_i^{(a} N_i^{b)} =0,
\ee
where the quantum anomaly coefficient $\cA^{ab}$ is determined by the charges $Q^a_i$ of the right-moving fermions, $\j^i_+$, and the charges $Q^a_\a$ of the left-moving fermions, $\gamma^\a$,
\be\label{firstdefanomaly}
\cA^{ab} = {1\over 4\pi} \left( \sum_i Q^a_i Q^b_i - \sum_\a Q^a_\a Q^b_\a \right).
\ee
The left-moving fermions determine the choice of space-time gauge bundle.
%In these models, that choice can change the resulting target space geometry, as we will see in subsequent sections. This  does not happen for models which are classically gauge-invariant.
At least for the gauge invariant case, the way in which we choose to cancel the loop anomaly does not affect the classical geometries that emerge from solving~\C{dterms}\ and quotienting by the global gauge group.

%Our expectation is that the geometries that emerge from classically gauge invariant constructions should be complex spaces since the gauge theories possess a non-anomalous classical $(0,2)$ supersymmetry. We also expect $H$-flux to thread the spaces in an interesting way that we would like to explore. These spaces should be natural generalizations of toric varieties but generically non-K\"ahler.

\subsection{Compactness for a single $U(1)$ factor}

While there are many interesting non-compact toric spaces like the conifold and its torsional generalizations, we are primarily interested in compact spaces here. Let us focus on the case of a single $U(1)$ factor taking $n=1$.

%\begin{proposition}
If the collection of $U(1)$ charges, $Q$, has at least one positive and one negative component then $W$ and $X$ are non-compact.
%\end{proposition}
%\vskip 0.1in
%\noindent
To see this, suppose that $Q_1 > 0$ and $Q_2 < 0$. Restrict to the set where the remaining coordinates are $1$. The remaining equations become
\be
Q_1 |\phi^1|^2 - \log \left(|\phi^1|^{N_1} \right) = r + |Q_2| |\phi^2|^2 + \log \left(|\phi^2|^{N_2} \right) - \sum_{j>2} Q_j.
\ee
Both the left and right hand side are unbounded from above as $|\phi^1|$ or $|\phi^2| \rightarrow \infty$. Equality can therefore be achieved for arbitrarily large values of $|\phi^i|$. Hence the spaces are non-compact.
For compact models, we can therefore choose a convention and require $Q\geq 0$. Let us examine  various cases.

\begin{figure}[ht]
\centering
\subfloat[][$|\f|^2-\log|\f|$]{
\includegraphics[width=0.4\textwidth]{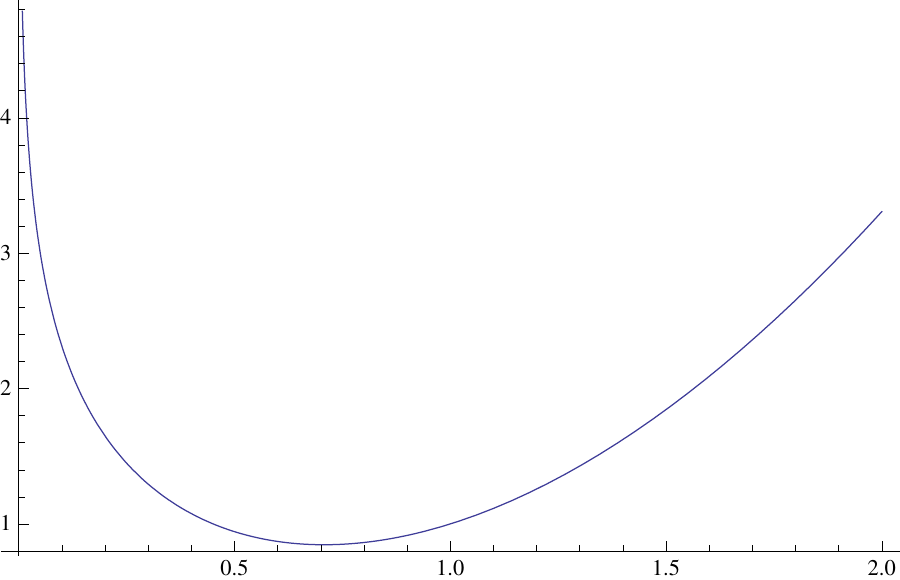}
\label{figure1}}
\qquad\qquad
\subfloat[][$|\f|^2+\log|\f|$]{
\includegraphics[width=0.4\textwidth]{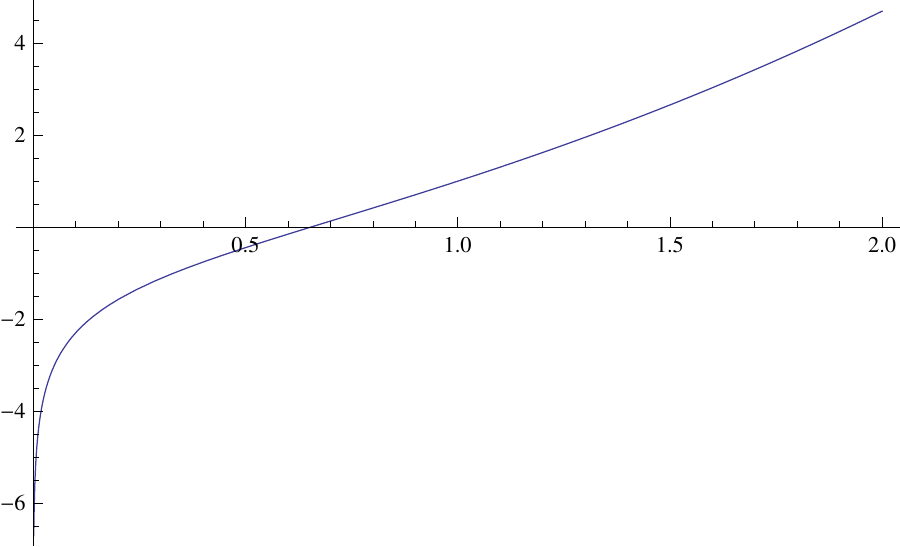}
\label{figure2}}
\caption{\textit{Plots of $|\f|^2 \mp \log |\f|$ against $|\f|$.}}
\label{figures}
\end{figure}

%\begin{figure}[ht]
%\begin{center}
%\[
%\mbox{\begin{picture}(230,180)(0,20)
%\includegraphics[scale=1]{singledterm.pdf}
%\end{picture}}
%\]
%\vskip 0.2 in \caption{\it A plot of $|\f|^2 - \log |\f|$ against $|\f|$. } \label{figure1}
%\end{center}
%\end{figure}
%
%\begin{figure}[ht]
%\begin{center}
%\[
%\mbox{\begin{picture}(230,180)(0,20)
%\includegraphics[scale=1]{singledterm2.pdf}
%\end{picture}}
%\]
%\vskip 0.2 in \caption{\it A plot of $|\f|^2 + \log |\f|$ against $|\f|$. } \label{figure2}
%\end{center}
%\end{figure}

\subsection{A single field}
\label{singlefield}
Take the case of a single field with $d=1$ and the equation
\be\label{singlelog}
Q |\phi|^2 - N \log |\phi| = r.
\ee
 If $Q=0$ then $|\phi| = e^{- r/N}.$ Assume $Q\neq 0$ and $N \neq 0$. Rescaling gives
\be
|\phi|^2 - {\hat N} \log |\phi| = {\hat r}, \qquad  {\hat N} = N/Q, \quad {\hat r} = r/Q.
\ee
This equation has a minimum at $|\phi|^2= {\hat N}/2$ if  ${\hat N} >0$.  This defines an ${\hat r}_{min}$ below which there are no solutions:
\be\label{definermin}
{\hat r}_{min} = ({{\hat N} / 2}) \left(1 - \log\left( {{\hat N}/ 2}\right)  \right).
\ee
 Note that ${\hat r}_{min}$ need not be positive! The function is drawn in figure~\ref{figure1}. The case of ${\hat N} <0$ is drawn in figure~\ref{figure2}. In this case, there are solutions for all values of ${\hat r}$.

\subsection{Two fields}

Now assume $d=2$, and for simplicity choose all charges to be $+1$. There are several possibilities.

\subsubsection{No log interaction}

\begin{figure}[ht]
\centering
\subfloat[][\textit{No log interactions}]{
\includegraphics[width=0.3\textwidth]{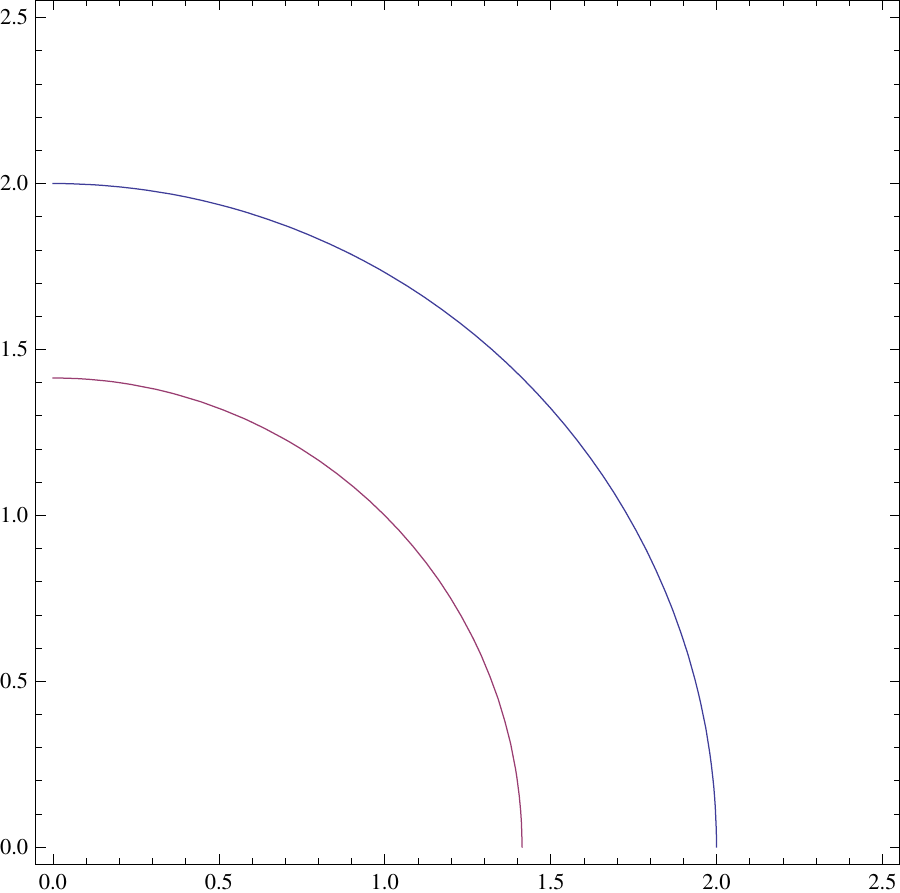}
\label{figure3}}
\
\subfloat[][\textit{A single log interactions}]{
\includegraphics[width=0.3\textwidth]{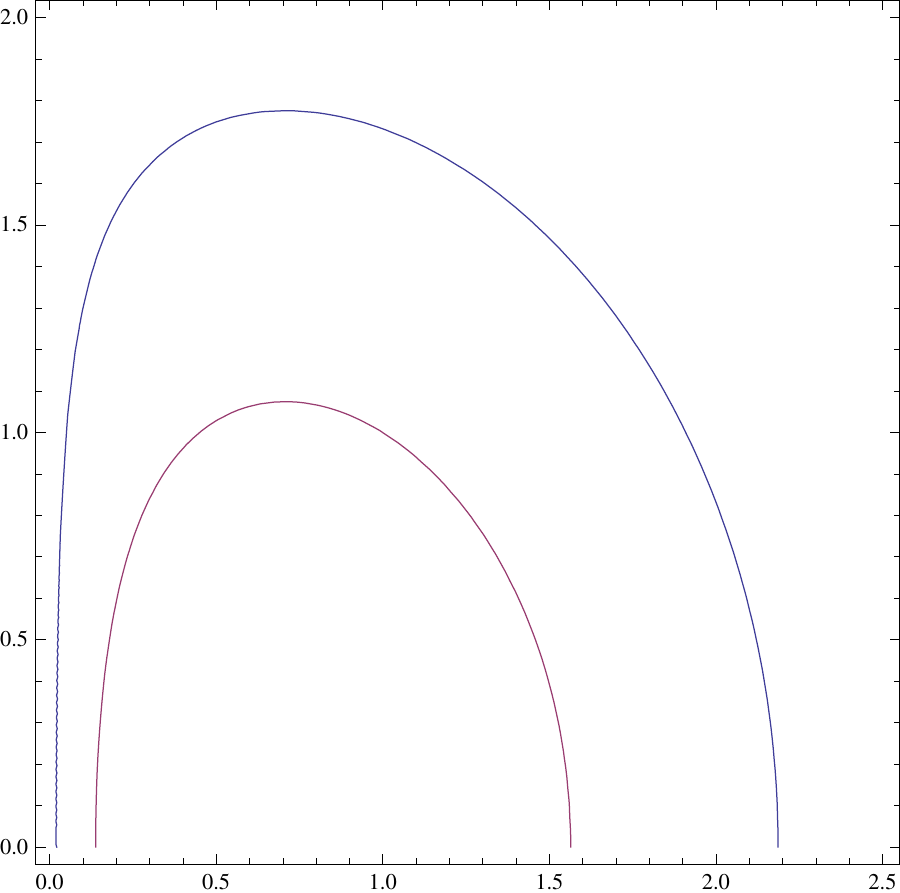}
\label{figure4}}
\
\subfloat[][\textit{Two log interactions}]{
\includegraphics[width=0.3\textwidth]{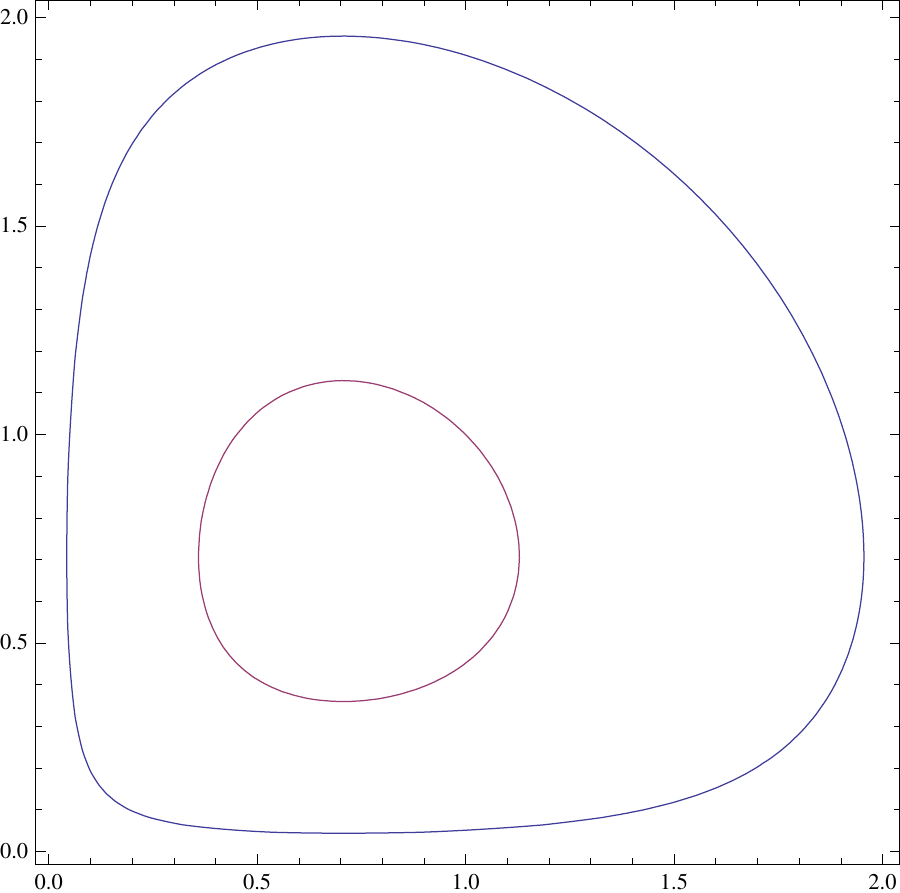}
\label{figure5}}
\caption{\textit{Contour plots of $|\f^1|$ versus $|\f^0|$ for $r=2$ and $r=4$. }}
\label{figures}
\end{figure}

 First consider the case of two fields with no log interactions. Take
\be  \label{nologdterm}
|\f^0|^2 + |\f^1|^2  = r
\ee
This describes  $S^3$ covered by two patches with either $\f^0\neq 0$ or $\f^1\neq 0$. We will define the skeleton for this space to be the contour in the $(|\f^0|, |\f^1|)$ plane solving~\C{nologdterm}. The skeleton is depicted in figure~\ref{figure3}. Over the depicted contour is fibered the phase of $\f^0$ and the phase of $\f^1$. At each axis, one of these two circles degenerates since either $\f^0=0$ or $\f^1=0$. Quotienting by the $U(1)$ gauge group amounts to removing either the phase of $\f^0$ or $\f^1$, depending on the patch. This removed circle is the topologically non-trivial circle of the Hopf fibration of $S^3$. The resulting space is $S^2$.

%\begin{figure}[ht]
%\begin{center}
%\[
%\mbox{\begin{picture}(230,240)(0,20)
%\includegraphics[scale=1]{skeleton-1.pdf}
%\end{picture}}
%\]
%\vskip 0.2 in \caption{\it A contour plot of $|\f^1|$ versus $|\f^0|$ for $r=2$ and $r=4$. } \label{figure3}
%\end{center}
%\end{figure}

\subsubsection{One log interaction}

Now let us consider a single log interaction,
\be \label{twosphere}
|\f^0|^2 + |\f^1|^2 - N_0 \log |\f^0| = r.
\ee
Note that the case $N_0 <0$ gives a non-compact solution space from the region where $\f^0$ becomes very small and $\f^1$ becomes large. We therefore restrict to $N_0>0$. The log interaction prevents $\f^0$ from vanishing. This means that the phase of $\f^0$ is a globally defined $S^1$. The skeleton for this space is depicted in figure~\ref{figure4}. Note that the skeleton begins and ends on one axis reflecting the fact that $\f^0$ can now never vanish.

%\begin{figure}[ht]
%\begin{center}
%\[
%\mbox{\begin{picture}(230,230)(0,20)
%\includegraphics[scale=1]{skeleton-2.pdf}
%\end{picture}}
%\]
%\vskip 0.2 in \caption{\it A contour plot of $|\f^1|$ versus $|\f^0|$ for $r=2$ and $r=4$, with a single log interaction. } \label{figure4}
%\end{center}
%\end{figure}

Ignoring the phase of $\f^0$ leaves a space with coordinates $(|\f^0|, \f^1)$. The coordinate $|\f^0|$ takes values in an interval. At the endpoints of the interval, the circle parametrized by the phase of $\f^1$ degenerates. This space is $S^2$. The single log interaction therefore gives $S^2\times S^1$ rather than $S^3$.

We can fix the gauge action by simply setting the phase of $\f^0$ to zero.  The resulting space is an $S^2$, albeit constructed in a way quite different from the preceding $N_0=0$ case. Since $S^2$ is complex, this result is not a priori puzzling.

% though it is strange that the $H$-flux threading $S^2 \times S^1$ appears to have had little effect on the quotient space. We will  see later that the sigma model metric is significantly affected.

\subsubsection{Two log interactions}

Now let us consider two log interactions,
\be \label{twotorus}
|\f^0|^2 + |\f^1|^2 - N_0 \log |\f^0| - N_1 \log |\f^1| = r,
\ee
with $N_0, N_1>0$.  The log interactions prevent both $\f^0$ and $\f^1$ from vanishing. The phases of $\f^0$ and $\f^1$ define a globally defined $T^2$. The skeleton for this space is depicted in figure~\ref{figure5}. In this case, the skeleton itself is a circle! This space is $T^3$. Gauge-fixing the $U(1)$ action amounts to removing one circle, leaving $T^2$.

 It is really quite surprising that we can construct a torus via a standard Lorentz invariant gauge theory without superpotential couplings.\footnote{By introducing a superpotential, it is easy to build a (2,2) GLSM describing a non-linear sigma model (NLSM) for an elliptic curve in $\PP^2$; however, this cannot be achieved via standard D-term couplings alone as $T^2$ is not toric!} Usually the introduction of circles in the moduli space of gauge theories requires either (Lorentz breaking) impurities introduced in~\cite{Sethi:1997zza, Kapustin:1998pb}, axial gauging~\cite{Gibbons:1996nt}, a special feature of three-dimensional gauge theory (the ability to dualize the photon), or compactification from higher dimensions via Wilson line moduli. Here the log interactions automatically provide globally defined circles.

%\begin{figure}[ht]
%\begin{center}
%\[
%\mbox{\begin{picture}(230,220)(0,20)
%\includegraphics[scale=1]{skeleton-3.pdf}
%\end{picture}}
%\]
%\vskip 0.2 in \caption{\it A contour plot of $|\f^1|$ versus $|\f^0|$ for $r=2$ and $r=4$, with $N_0=N_1=1$. } \label{figure5}
%\end{center}
%\end{figure}

\subsection{Many fields}\label{manyfields}

Let us generalize the preceding discussion to $d>2$. Again choose all charges to be $+1$.
Take
\be  \label{generaldterm}
\sum_{i=0}^{d-1} |\f^i|^2 - \sum_i N_i  \log |\f^i|  = r
\ee
where each $N_i \geq 0$.

First assume only $N_0 \neq 0$. The phase of $\f^0$ gives a globally defined $S^1$. We can gauge away this phase with the $U(1)$ action. The residual space is a sphere $S^{2d-2}$. This leads to an immediate worry. For $d=3$, the vacuum manifold appears to be $S^4$ which is known to possess no complex or almost complex structure. How is $(0,2)$ supersymmetry preserved? This strongly suggests that the quantum anomaly must alter the target space topology for any model with a complete UV description that preserves supersymmetry. We will return to this central issue in section~\ref{perturbationtheory}.

Now take the case where $N_i >0$ for $i=0, \ldots, m$. At most, $m=d-1$. Each log gives a global circle. One circle can be gauged away with the $U(1)$ action. The resulting space is $S^{2d-2-m}\times (S^1)^m$. Including the case with no log interactions gives the following sequence of possible spaces, ranging from $0$ to $d-1$ log interactions:
\be
\PP^{d-1}, \quad S^{2d-2}, \quad S^{2d-3} \times S^1, \quad S^{2d-4} \times (S^1)^2, \quad \cdots \quad , \, S^{d-1} \times (S^1)^{d-1}.
\ee

\section{ The Abelian Gauge Anomaly}
\label{axanom}

In this section we review the familiar  problem of the abelian gauge anomaly in two dimensions; our aim is to give a careful treatment of normalizations of terms in the effective action that are involved in the anomaly cancelation central to our work.

\subsection{Conventions} \label{ss:anconv}
We work with a flat infinite volume Euclidean world-sheet in conventions
of~\cite{Polchinski:1998rr}.\footnote{That is $z \equiv y^1+iy^2$; $\p_z \equiv \ff{1}{2}(\p_1-i\p_2)$;  $d^2z \equiv i dz\wedge d\zb = 2 d^2 y$; $\delta^2(z,\zb) \equiv \ff{1}{2} \delta^2(y)$, and $\p_z \zb^{-1} = 2\pi \delta^2(z,\zb)$.}  Our starting point is a free action for $r$ left-moving fermions $\gamma^\alpha$ and $n$ right-moving fermions $\psi^i$:
\be
S_0 = \int \frac{d^2z}{2\pi}~ \left[ \gammab^\alpha \pb_{\zb} \gamma^\alpha + \psib^i \p_z \psi^i \right].
\ee
The non-zero two-point functions are
\be
\la \gammab^\alpha_1 \gamma^\beta_2 \ra_0 = \delta^{\alpha\beta} z_{12}^{-1}, \qquad
\la \psib^i_1 \psi^j_2 \ra_0 = \delta^{ij} \zb_{12}^{-1},
\ee
where the subscript on a field indicates the insertion point, e.g. $\gamma^\alpha_1 \equiv \gamma^\alpha(z_1,\zb_1)$.  This theory has a large global symmetry group $\SO(r)\times\SO(n)$, but we will concentrate on its $\GU(1)^r \times\GU(1)^n$ subgroup with chiral currents $\cJ_L^\alpha = J^\alpha dz$ and $\cJ_R^i = \Jb^i d\zb$, where the operators $J^\alpha$ and $\Jb^i$ are defined by free-field normal ordering:
\begin{align}
J^\alpha = i~ \normd{\gammab^\alpha\gamma^\alpha}~, \qquad  \Jb^j = i~\normd{\psib^j\psi^j}~.
\end{align}
We will be interested in coupling this theory to a background $\GU(1)$ gauge field,
\be {\bf A} = A dz + \Ab d\zb. \ee
Note that until section~\ref{susyanomaly}, we use $A$ and $\Ab$ to refer to standard bosonic gauge-fields rather than superfields. In order to examine chiral currents in this background, we will introduce a slightly more general interaction term:
\begin{align}
S_\tint & = \int  \frac{d^2z}{2\pi}~\left[ A^i \Jb^i + \Ab^\alpha J^\alpha \right].
\end{align}
The $\GU(1)$ gauging sets $A^i = q_i A$ and $\Ab^\alpha = Q_\alpha \Ab$.

\subsection{The partition function and gauge invariance}\label{ss:partf}
There is no difficulty in evaluating the partition function $Z[{\bf A}] \equiv \la e^{-S_\tint}\ra_0$.  It is given by $Z[{\bf A}] = e^{W[{\bf A}]}$  with\footnote{There are no connected $n$-current correlation functions for $n>2$; with a more general non-abelian gauging, there will be a finite number of additional terms of higher order in the gauge field.  For instance, for $\SU(2)$ $W$ has just an additional $O({\bf A}^3)$ term.}
\begin{align}
W = -\frac{1}{2} \int \frac{d^2z_1 d^2z_2}{(2\pi)^2} \left[ \frac{\Ab^{\alpha}_{1} \Ab^{\alpha}_{2}}{z_{12}^2}+ \frac{A^{i}_{1} A^{i}_{2}}{\zb_{12}^2}\right]~.
\end{align}
While easily computed, $W$ is not gauge-invariant.  Under $\delta_\ep {\bf A} = -d \ep$, we find the local variation (we now set $A_i = q_i A$ and $\Ab_\alpha = Q_\alpha \Ab$)
\bea
\delta_\ep W &=& \int \frac{d^2z}{2\pi} \ep (k_L \p_z  \Ab + k_R \pb_z A), \cr &=& \frac{k_L+k_R}{4\pi} \int d^2z \delta_\ep( \Ab A) +\frac{(k_L -k_R)}{4\pi} \int d^2 z \ep (\p_z\Ab-\pb_z A),
\eea
where $k_L = \sum_\alpha Q_\alpha^2$ and $k_R = \sum_i q_i^2$.  The form of the gauge variation can be brought into a canonical topological form by a choice of counter-terms~(see, e.g., \cite{AlvarezGaume:1984dr} for a thorough discussion), and, indeed, the first term in $\delta_\ep W$ can be canceled by setting
\begin{align}
S_{\tct} = \int \frac{d^2z}{4\pi} A^i N_{i\alpha} \Ab^\alpha,\label{counterterm}
\end{align}
where $N_{i\alpha}$ satisfies $q_i N_{i\alpha} Q_\alpha = (k_L+k_R)$.  We parametrize $N$ by
\begin{align}
N_{i\alpha} = \frac{q_i Q_\alpha(k_L+k_R)}{k_L k_R} + M_{i\alpha},
\end{align}
where $M_{i\alpha}$ is annihilated by $q_i$ and $Q_\alpha$.
Including this counter-term, we obtain an improved partition function ${\widetilde Z}[{\bf A}] = e^{ {\widetilde W}[{\bf A}]}$, with
\begin{align}
\label{eq:gaugetransfW}
\delta_\ep {\widetilde W} = \frac{i(k_L-k_R)}{4\pi} \int \ep \cF,
\end{align}
where $\cF = \cF_{12} dy^1\wedge dy^2$, and $\cF_{12} = -2i (\p_z \Ab-\pb_z A)$. This is the reason for the factor of ${1\over 4\pi}$ appearing in~\C{expanomaly}.

We have reached the familiar conclusion that the partition function will be gauge-variant unless $k_L = k_R$.  Although it has been noted that the chiral Schwinger model remains unitary despite the anomaly~\cite{Jackiw:1984zi,Boyanovsky:1987ad}, for our applications we will insist that the total gauge anomaly of the GLSM cancels.

\subsection{Chiral currents}\label{ss:chicur}
Even when $k_L = k_R$, so that the gauge symmetry is non-anomalous, there will be an anomaly in the global chiral symmetries.  To study these, we define improved currents
\begin{align}
\JJ^\alpha \equiv 2\pi \left. \frac{\delta S}{\delta \Ab^\alpha}\right|_{\cA} = J^\alpha + Q_\alpha A, \qquad
\JJb^i \equiv 2\pi \left. \frac{\delta S}{\delta A^i} \right|_{\cA} = \Jb^i + q_i \Ab,
\end{align}
where we have included contributions from $S_{\tct}$.  To leading order in the gauge field,
\be
\la J^\alpha(z)\ra_{\cA}  = \int \frac{d^2 w}{2\pi} \frac{Q_\alpha \p_{w} \Ab (w)}{w-z} + O({\bf A}^2), \qquad
\la \Jb^i (z) \ra_{\cA} =  \int \frac{d^2 w}{2\pi} \frac{q_i  \pb_{\wb} A(w)}{\wb-\zb} + O({\bf A}^2),
\ee
so that
\begin{align}
\pb_z \la \JJ^\alpha(z) \ra = -\frac{i Q_\alpha}{2} \cF_{12},\qquad
\p_z \la \JJb^i (z) \ra = \frac{iq_i}{2} \cF_{12}.
\end{align}
It follows that the theory retains  $\GU(1)^{r-1} \times\GU(1)^{n-1}$ chiral currents; the non-chiral gauge current is conserved, and one chiral current is anomalous.\footnote{The reader may worry that our expressions for the currents appear to be non-gauge invariant.  The resolution is simple: the normal ordering prescription $\normd{\gammab\gamma}(w) = \lim_{z\to w} (\gammab(z)\gamma(w) - (z-w)^{-1})$ is not a priori gauge invariant, and the improvement terms compensate for that.}  We can give this result an interpretation \`a la Fujikawa~\cite{Fujikawa:1979ay}:  the properly regulated gauge-invariant fermion measure has non-trivial transformations under chiral rotations $\delta_{\zeta} \gamma^\alpha = i \zeta^\alpha\gamma^\alpha$ and $\delta_{\xi} \psi^i = i \xi^i \psi^i$, which are interpreted as shifts of the effective action by
\begin{align}
\label{eq:Fuji}
\delta S = \frac{i}{2\pi} \int (\xi_i q_i - \zeta_\alpha Q_\alpha) \cF.
\end{align}
Note that this is larger by a factor of two than the gauge variation in~(\ref{eq:gaugetransfW}).

\subsection{A background Higgs field}\label{ss:higgsb}
For our applications we are interested in coupling the chiral fermions to an additional gauge-charged bosonic scalar field $\vphi$ via a Yukawa interaction.  In this section we will make some observations on the effect of integrating out massive fermions on the Higgs branch. For simplicity, we concentrate on the case where just two fermions, a left-moving $\lambda$ and a right-moving $\chi$, have a non-trivial Yukawa coupling:
\begin{align}
S_{\tYuk} = m \int d^2 y ~ \left[ \vphi \chi \lambda + \vphib \lambdab \chib\right].
\end{align}
We take $|m \vphi|$ to be much larger than any other scale in the theory and integrate out the massive fermions.\footnote{This is the $d=2$ abelian analogue of the well-known work~\cite{DHoker:1984ph} in $d=4$ non-abelian chiral gauge theory.}  As a final simplification, we assume that $|\vphi|$ is frozen at some value, so that only its phase $\vtheta$ plays a role.
Gauge invariance of the Yukawa coupling requires $Q_\vphi = -Q_\chi-Q_\lambda$, and the phase $\vtheta$ transforms by $\delta_\ep \vtheta = Q_\vphi \ep$ under gauge transformations.

As pointed out in~\cite{DHoker:1984ph}, a naive decoupling argument as $|m\vphi| \to \infty$ fails because both the mass of the fermions and the actual Yukawa coupling of the $\theta,\lambda,\psi$ system diverge in this limit; of course the result of integrating out the massive fields must be a set of couplings for $\vtheta$ and the gauge field ${\bf A}$.  One can actually see the terms emerge explicitly by bosonizing the $\lambda,\chi$ fermions, but we will not need that level of detail.  Instead, we observe that low energy  $\vtheta$--${\bf A}$ couplings must reproduce the contribution to the anomaly from the massive fermions, given by~(\ref{eq:gaugetransfW}) as
\begin{align}
\delta_\ep W'_{\lambda,\chi} = \frac{i(Q_\lambda^2-q_\chi^2)}{4\pi} \int \ep \cF = \frac{i(Q_\lambda^2-q_\chi^2)}{4\pi} \int d^2y~\ep \cF_{12}.
\end{align}
To leading order in derivatives and ${\bf A}$, the effective action for $\vtheta$ is fixed up to two undetermined constants, $\kappa$ and $\kappa'$,
\begin{align}
S_{\teff,\vtheta} = \frac{1}{4\pi} \int d^2y~\left[ \kappa D_z\vtheta D_{\zb} \vtheta + i\kappa' \vtheta \cF_{12} \right].
\end{align}
Here $D \vtheta = d\vtheta + Q_{\vphi} {\bf A}$ is the gauge invariant $1$-form.  To match $\delta_\ep W'_{\lambda,\chi}$ we see that
\begin{align}
\kappa' =(Q_\lambda-q_\chi).
\end{align}
We can also fix $\kappa$ by matching the chiral symmetries in the UV to those in the IR. The UV theory has a non-anomalous $\GU(1)^{n-1}$ symmetry with\footnote{The argument can be repeated with $\GU(1)^{r-1}$; if $q_\chi=Q_\lambda = 0$, then $\kappa =\kappa'=0$.}
\begin{align}
\delta_\xi \psi^j = i \xi^j \psi^j,\qquad
\delta_\xi \chi =  -i q_j q_\chi^{-1} \xi^j \chi, \qquad
\delta_\xi \vtheta = q_j q_\chi^{-1} \xi^j.
\end{align}
%In general this symmetry will not be chiral, since under a variation $\delta\vtheta = \zeta(z)$ the kinetic term $\int d^2y~ D_z\vtheta D_{\zb}\vtheta$ shifts by
%\begin{align}
%\delta_\zeta S_{\tkin,\vtheta}
%= \int d^2y ~\zeta\left[-2 \p D_{\zb}\vtheta + \ff{i}{2}Q_{\vphi} \cF_{12}\right],
%%= \int d^2y ~\zeta\left[-2 \pb D_{z}\vtheta - \ff{i}{2}Q_{\vphi} \cF_{12}\right].
%\end{align}
Since we have not introduced a kinetic term for the background field $\vphi$, these are chiral symmetries of the UV theory, and we should be able to recover them in the IR theory in the presence of the quantum-generated $\vtheta$ kinetic term.  As we will now show, this is the case if and only if $\kappa =1 $.

The variation of the effective action receives two contributions.  First, there is the contribution from the light fermions; this has a term from the classical action and a term from the measure as in~(\ref{eq:Fuji}):
\begin{align}
\Delta_1 S_{\teff} = \int \frac{d^2 y}{2\pi}~\xi^j \left[ -\p_z \Jb^j + i q_j \cF_{12} \right].
\end{align}
Second, there are terms from the variation of $S_{\teff,\vtheta}$, which yield
\begin{align}
\Delta_2 S_{\teff} = \int \frac{d^2y}{4\pi} ~\xi^j \left[ \kappa q_j q_{\chi}^{-1} (-2 \p D_{\zb}\vtheta + \ff{i}{2} Q_{\vphi} \cF_{12}) + i (Q_\lambda-q_\chi) q_j q_{\chi}^{-1}\cF_{12}
\right].
\end{align}
All together, we obtain
\begin{align}
\Delta_1 S_{\teff} + \Delta_2 S_{\teff} &= -\int \frac{d^2y}{2\pi}  \xi^j \p_z \left[ \Jb^j + q_j q_{\chi}^{-1} D_{\zb}\vtheta\right] \nonumber\\
&\qquad + \frac{i}{4\pi} \left[\frac{1}{2} + \frac{\kappa Q_{\vphi} + Q_\lambda-q_\chi}{q_\chi} \right] \int d^2y ~\xi^j q_j \cF_{12}.
\end{align}
The second line vanishes if and only if $\kappa = 1$, and the remaining term corresponds to the improved conserved chiral currents for $\GU(1)^{n-1}$.

To summarize: integrating out the massive fermions $\lambda$ and $\chi$ induces a correction to the kinetic term and axial coupling of the phase of the Higgs field $\vtheta$.  As we just argued, the exact result for these terms is
\begin{align}
S_{\vtheta,\teff} = \frac{1}{4\pi} \int d^2y \left[ D_z\vtheta D_{\zb} \vtheta + i(Q_\lambda-q_\chi) \vtheta \cF_{12} \right].
\end{align}
In the non-supersymmetric setting this of course does not determine the corrections to the kinetic term or potential of the modulus $|\vphi| = \rho$, but as we will discuss in section~\ref{perturbationtheory}, they do play an important role in determining $S_{\vphi,\teff}$ in the supersymmetric theory.

\subsection{The $(0,2)$ gauge anomaly}\label{susyanomaly}

We close our discussion by turning to the supersymmetrization of the gauge anomaly.
The $(0,2)$ supersymmetric version of the gauge anomalous variation~\C{eq:gaugetransfW}\ is
\be
\dd_\La W = {\cA\over16\pi}\int d^2xd\th^+\ \La \Upsilon + c.c.,\label{(0,2)anom}
\ee
where $\La$ is a chiral superfield gauge parameter and $\cA = \sum_i Q_i^2 -\sum_\a Q_\a^2$. This variation can be produced from the non-local effective action
\be
W[A,V_-] = {\cA\over16\pi}\int d^2xd^2\th^+\ A\, {1\over\del_+}\left(D_+\U - \Dbar_+\bar\U\right).\label{GammaSusy}
\ee
In Appendix~\ref{quad} we will compute this expression directly from a loop diagram. For now, let us note that~\C{GammaSusy}\ possesses all the characteristics we desire for a representative of the two-dimensional gauge anomaly: it is expressed entirely in terms of the gauge fields, it is quadratic in the gauge fields, it is inherently non-local and so cannot be canceled by any local counter-term; most importantly, its gauge variation agrees with~\C{(0,2)anom}. Similar non-local representations of the anomaly have appeared in non-supersymmetric and $(0,1)$ gauge theories~\cite{Hwang:1985tm}.

To expand $W[A,V_-]$ in components we do not have the luxury of working in WZ gauge because the action is not gauge invariant. Instead, we must work with the full non-gauge-fixed form of the gauge fields:
\bea
A &=& C + i\th^+\chi + i\thbar^+\bar\chi +\th^+\thbar^+ A_+, \\
V_- &=& A_- - \th^+\left( 2i\bar\l -\del_-\chi\right) -\thbar^+\left(2i\l + \del_-\bar\chi \right) +\th^+\thbar^+\left(2D +\del^2C\right),
\eea
which transform as follows:
\bea
\dd_\La A &=& {1\over2i}\left(\La-\bar\La\right) = \Im\La -{i\over\sqrt{2}}\th^+\psi_\La -{i\over\sqrt{2}}\thbar^+\bar\psi_\La -\th^+\thbar^+\del_+\Re\La \label{Avar}, \\
\dd_\La V_- &=& -{1\over2}\del_-\left(\La+\bar\La\right) \cr &=& -\del_-\Re\La - {1\over\sqrt{2}}\th^+\del_- \psi_\La +{1\over\sqrt{2}}\thbar^+\del_-\bar\psi_\La +\th^+\thbar^+\del^2\Im\La.
\eea
After performing the superspace integral in~\C{GammaSusy}\ we obtain the component action
\bea\label{GammaComp}
W &=& {\cA\over4\pi} \int d^2x \left(A_+\,{1\over\del_+}F_{01} - \bar\chi\l + \chi\bar\l -CD\right).
\eea
Using~\C{Avar} and integration by parts, we find the local gauge variation
\be
\dd_\La W = {\cA\over4\pi}\int d^2x \left(\Re(\La)\,F_{01} - {1\over\sqrt{2}}\psi_\La\l + {1\over\sqrt{2}}\bar{\psi}_{\La}\bar\l - \Im(\La) D\right),
\ee
as expected. Expanding the field strength $2F_{01}=F_{+-}=\del_+A_- - \del_-A_+$ shows that the non-locality of~\C{GammaComp}\ can be confined to a single term: $\hlf A_+\,{\del_-\over\del_+} A_+ $, with the rest of the effective action comprised of purely local terms. In superspace, we can find an analogous split into local and non-local pieces by inserting $\U = \Dbar_+(\del_-A +iV_-)$ into~\C{GammaSusy}. After some straightforward manipulations, we arrive at
\be
W[A,V_-] = {\cA\over8\pi}\int d^2xd^2\th^+\left(\Dbar_+A {\del_-\over\del_+}D_+A - AV_-\right). \label{anomaly}
\ee
This is the form of the anomaly we will use throughout this work.

\section{Computing the Effective Action}\label{perturbationtheory}

We now turn to the computation of the one-loop effective action. Rather than study a mass term for an anomalous multiplet generated by an $E$-coupling, it will be more convenient to use a $J$ superpotential coupling. The two formulations are equivalent as explained in section~\ref{basicidea}.

\subsection{The setup}

Consider a theory with charged chiral superfields $P$, $\S$ and a Fermi superfield $\hG$, coupled together by the superpotential
\be \label{superpot}
\L_J = -{m\over\sqrt{2}}\int d\th^+ \hG \S P +c.c.,
\ee
where $\bar{\mathfrak{D}}_+ \hat\G = 0$ and
\be Q_\hG + Q_\S + Q_P=0\ee
to ensure gauge invariance of the superpotential.
This set of fields is generally anomalous, so we will include additional charged fields $\left(\Phi^i,\G^\a\right)$ ensuring that the net gauge anomaly vanishes:
\be
Q_P^2 + Q_\S^2 - Q_\hG^2 +\cA = 0, \qquad {\rm with} \qquad \cA = \sum_i Q_i^2 - \sum_\a Q_\a^2.
\ee
Note that  $\cA = 2Q_\S Q_P$. In this section, the fields $\Phi^i,\G^\a$ will only act as spectators ensuring the cancelation of the gauge anomaly; for clarity, we will suppress these fields.

\subsection{The form of the effective action}

When $\S$ develops an expectation value, the gauge theory is Higgsed and the $(P,\G)$ multiplets become massive. When the mass of the $(P,\G)$ fields is large compared to the scale of the gauge coupling, $e$, they should be integrated out leaving an effective theory of the Higgs field, $\S$, and the vector multiplets $A$ and $V_-$. We therefore would like to compute the effective action
\be
e^{iW[\S,\bar\S, A,V_-]} = \int\left[\cD P\, \cD\G\right] e^{iS_0[P,\G,\S,A,V_-]}.
\ee
We know that $W$ must be a local integral over both fermionic coordinates for $(0,2)$ superspace; see Appendix~\ref{superFeynman}. Furthermore, by expanding in powers of ${1\over m^2}$, $W$ must also be expressible as a local integral in position space. Dimensional analysis and Lorentz invariance imply that
\bea \label{Gammaeff}
W &=& \int d^2x d^2\th^+  \left[ f_V(A,\S,\bar\S) V_- + f_{A}(A,\S,\bar\S) \del_- A \right. \cr
&& \left. \qquad\qquad \quad+ \left(f_\S(A,\S,\bar\S)\del_-\S +c.c.\right)  +\ldots \right],
\eea
where the ellipses denote terms that are suppressed by ${1\over m^2}$. Such terms do not contribute to $W$ in the low-energy limit. Note that the $f_A$ and $f_\S$ terms are not  uniquely defined. Rather they should be identified under the equivalence relation
\be
f_A \sim f_A + \del_A f,\qquad f_\S \sim f_\S + \del_\S f,
\ee
for any function $f=f(A,\S,\bar\S)$. This identification shifts the effective action by a total derivative.
%, and so amounts to a B-field transformation in the target space geometry.

\subsection{Unitary gauge}

Integrating out the massive charged fields requires care because they contain an anomalous set of fermions. This situation has been considered in the past by D'Hoker and Farhi in the context of integrating out the top quark from the Standard Model~\cite{DHoker:1984pc}, and more generally in~\cite{DHoker:1984ph}. One approach is to combine the phase of the Higgs field with the charged fermions to give gauge invariant fermions, which can then be integrated out without worry. This is a valid procedure, as long as the Higgs field does not vanish and so its phase is well-defined.

In a supersymmetric Higgs theory, we can go one step further. Using the enlarged gauge symmetry present in superspace, we can gauge fix the full Higgs chiral superfield $\S$ to unity while simultaneously rewriting the remaining charged fields in terms of gauge neutral fields.
%can be used to combine the entire Higgs field, $\S$, with the charged chirals to make neutral ones.
We do this by effectively fixing unitary gauge. We transform all the fields by a super-gauge transformation with parameter
\be
\La = {i\over Q_\S} \log\S \label{unitary}.
\ee
Since we are transforming \textit{all} the charged fields, including $\Phi^i$ and $\G^\a$, there is no anomalous shift of the action. We end up with a set of gauge invariant fields:
\bea
\tilde P = P\, \S^{-{Q_P/ Q_\S}}, \ && \quad \tilde A = A + {1\over Q_\S}\log|\S|, \\
\tilde \G = \hG\,\S^{-{Q_\G/ Q_\S}}, \ && \quad \tilde V_- = V_- + {1\over Q_\S}\del_-\Im\log\S \non.
\eea
Note that $\S$ has been gauged away with $\tilde\S \equiv \S/\S = 1$, so only the physical degrees of freedom remain - namely, a massive vector multiplet coupled to chiral superfields. In this gauge, the effective action simplifies tremendously:
\be
W = \int d^2xd^2\th^+ \left[f_V(\tilde A)\tilde V_- + f_A(\tilde A)\del_-\tilde A +\ldots\right].
\ee
The second term is a total derivative which can be ignored in perturbation theory. In unitary gauge, the low energy effective action is therefore completely determined by $f_V(\tilde A)$. A second useful feature of unitary gauge is that the superpotential coupling~\C{superpot}\ reduces to a standard mass term that combines $(\tilde P,\tilde\G)$ into a single massive multiplet with mass $m$. There are no higher order $F$-term interactions.

Unitary gauge is often problematic for carrying out loop computations because the massive vector propagator does not decay sufficiently rapidly at large values of momentum. However, this will not be an issue for us since we will be treating the vector multiplets as background fields and only integrating out the massive chiral fields $(\tilde P,\tilde\G)$. This approach is justified since the mass of the vector multiplets is set by the gauge coupling $e$. As described in section~\ref{basicidea}, we are considering the ratio ${e\over m} \ll1$.

\subsection{Computing the effective action in unitary gauge}

\begin{figure}[h]
\begin{center}
\[
%\mbox{\begin{picture}(370,130)(0,20)
\includegraphics[scale=0.5]{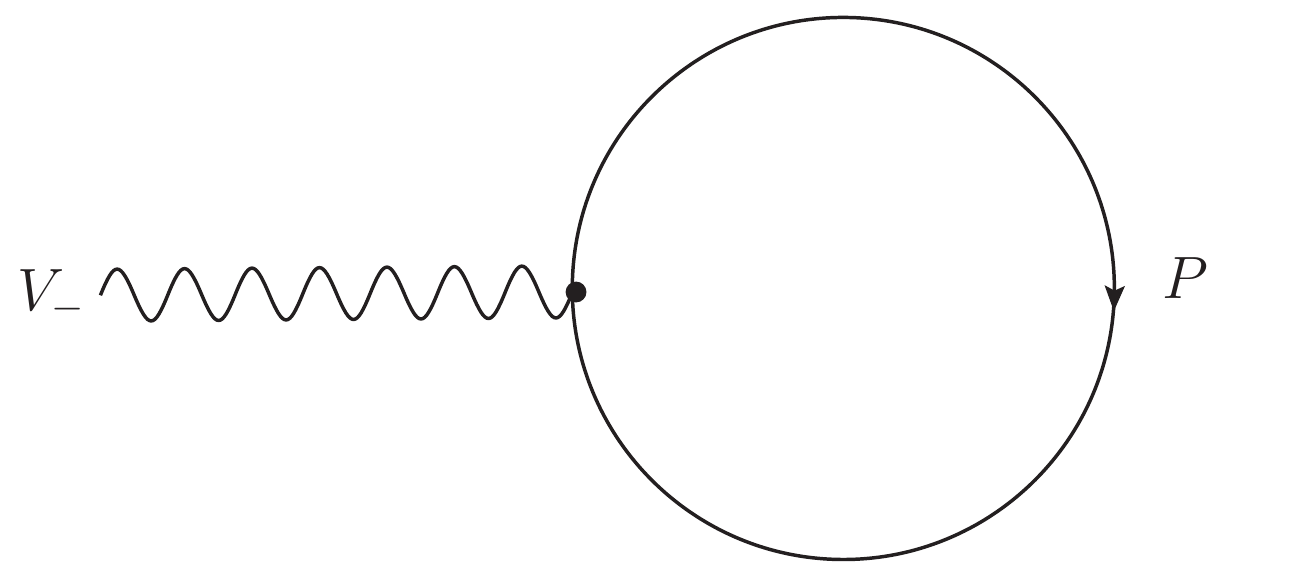}
%\end{picture}}
\]
\vskip 0.2 in \caption{\it The only contribution to the effective action in unitary gauge. } \label{V1pt}
\end{center}
\end{figure}

We can make our lives easier by noting that $f_V(\tilde A)$ is completely determined by its zero-mode dependence: if we expand $\tilde A$ about some constant value $\tilde A_0$ then
\be f_V(\tilde A_0 + \tilde A) = f_V(\tilde A_0) + \tilde A \, f'_V(\tilde A_0) +\ldots.\ee
So we really only need to determine $f_V(\tilde A_0)$, which means we only need the $\tilde A_0$-dependence of the $1$-point function $\langle \tilde V_-\rangle$. The Feynman rules for supergraphs in the presence of a constant background $\tilde A_0$ are derived in Appendix~\ref{superFeynman}. There is a single diagram to compute, shown in figure~\ref{V1pt}, which involves a loop of $P$ connected to $\tilde V_-$.
 This leads to the result,
\be
iW[\tilde A_0,\tilde V_-] = \int d^2x d^2\th^+\, \tilde V_-(x)\left(Q_P\over2\right)\I_{0,1}\left(m^2 e^{2Q_\S \tilde A_0}\right),
\ee
where the integrals
\be
\I_{p,q}(M^2) = \int_{\ell_E^2\geq\m^2} {d^2\ell\over(2\pi)^2}\, {\left(\ell^2\right)^p \over \left(\ell^2+M^2\right)^q}
\ee
are evaluated in Appendix~\ref{Loops}. The integral $\I_{0,1}$ has a logarithmic divergence. After renormalization at a scale $\mu_r$, we find
\bea
W[\tilde A_0,\tilde V_-] &=& -{Q_P\over8\pi}\int d^2x d^2\th^+\, \tilde V_-(x) \log\left(\m^2 + m^2 e^{2Q_\S \tilde A_0}\over \m_r^2\right) \\
&=& -{Q_P\over8\pi}\int d^2x d^2\th^+\, \tilde V_-(x) \left(2Q_\S \tilde A_0 +\log\left(m^2 \over \m_r^2\right) +\ldots\right), \non
\eea
where we have dropped terms that are suppressed by $\left({ \m \over m}\right)$. Restoring the full $\tilde A$-dependence and recalling that $\cA = 2Q_\S Q_P$, we find the low-energy effective action:
\be
W[\tilde A,\tilde V_-] = -{\cA\over8\pi}\int d^2x d^2\th^+\, \tilde A(x) \tilde V_-(x) \label{Gammafinal}.
\ee
We have dropped a field-independent correction to the the FI parameter. We give an alternate computation of this term by directly computing the $\langle A V_-\rangle$ correlation function in Appendix~\ref{quad}.

\subsection{Computing the effective action without unitary gauge}\label{nogauge}

While the result~\C{Gammafinal}\ is all that is needed to determine the data of the low energy sigma-model in the patch where $\S=1$, we would also like to know how this the effective action looks in other patches; for example, a patch where we set a chosen $\Phi^i=1$ instead of $\S$. Simply undoing unitary gauge, by using the inverse of the gauge transformation~\C{unitary}, turns out to be rather subtle. Instead it will prove easier to recompute $W$ without fixing unitary gauge. We will recover~\C{Gammafinal}\ by gauge fixing this more general result. As a bonus, this gauge-unfixed result will generalize straightforwardly to the case of multiple $\S$ fields giving large masses to multiple $(P,\G)$ pairs. This gives us a picture of the sigma model geometry on the cover of the $\CC^\ast$-action; i.e., a picture in terms of homogeneous rather than inhomogeneous (or gauge-fixed) coordinates.

\begin{figure}[h]
\centering
\subfloat[][]{
\includegraphics[width=0.475\textwidth]{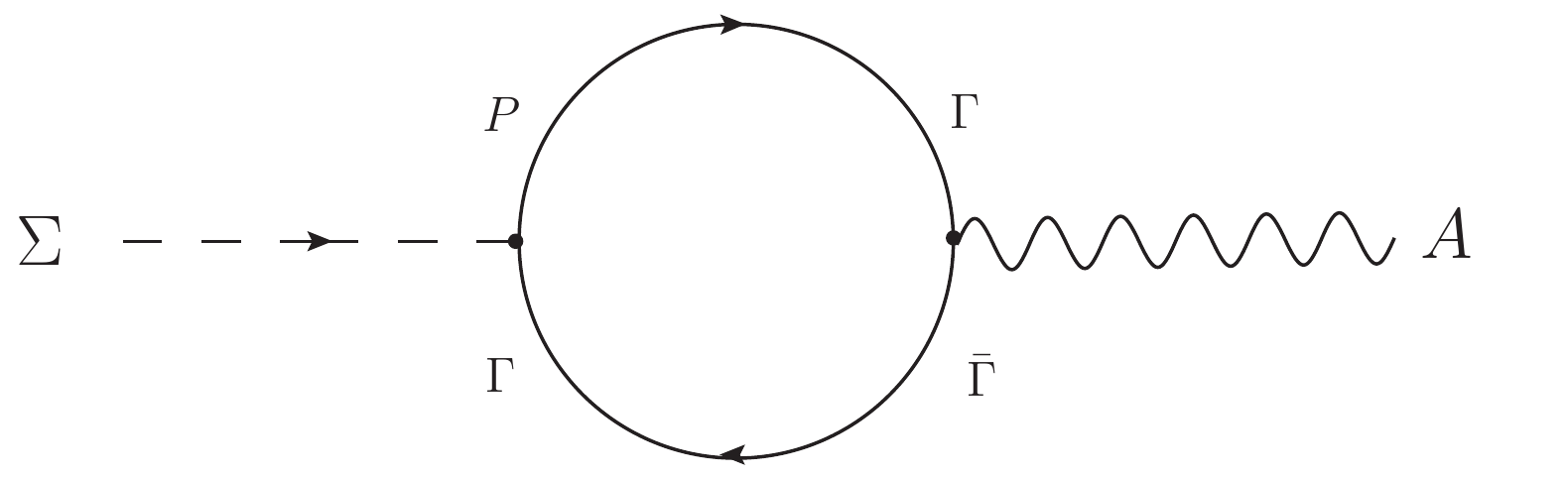}
\label{ASigma1}}
\
\subfloat[][]{
\includegraphics[width=0.475\textwidth]{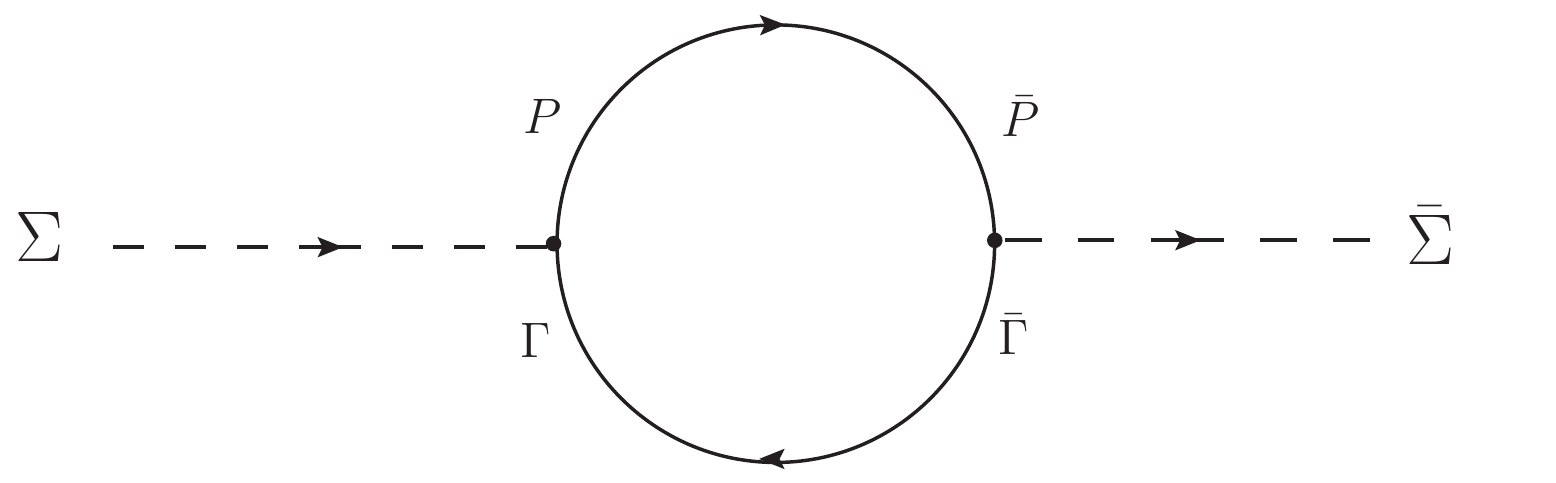}
\label{SSbar1}}
\caption{\textit{The remaining contributions to the effective action without gauge-fixing.}}
\label{otherloops}
\end{figure}

In this situation we must compute all three functions, $(f_V,f_A,f_\S)$, appearing in~\C{Gammaeff}. Once again, we will perform the computation around some constant background fields, but now we expand about a point $(A_0,\S_0)$ in moduli space, rather than just $A_0$.  The computation of $f_V$ goes through exactly as before except for the replacement $m\rightarrow m\S_0$:
\bea
f_V(A_0,\S_0,\bar\S_0) &=& -{Q_{P}\over8\pi} \log\left(\m^2+m^2|\S_0|^2e^{2Q_{\S}A_0}\over \m_r^2\right) \label{fV}\\
&=& -{1\over8\pi}\left(2Q_{P}Q_{\S}A_0 + Q_{P}\log|\S_0|^2 +\ldots \right). \non
\eea
We see that the $AV$ term is unchanged, but we now have an additional $\log|\S|$ term. This term vanishes in the gauge $\S=1$. The fact that $f_A$ and $f_\S$ appear at linear order in derivatives means we cannot get them from a $1$-point function and we have to go to $2$-point correlators. The relevant diagrams are shown in figure~\ref{otherloops}.

From figure 4\subref{ASigma1}\ we find the result
\be
Q_{\G}m^2\bar\S_0e^{2Q_\S A_0}\int {d^2q\over(2\pi)^2}d^2\th^+\ q_-A(q)\S(-q) \int dx\ (1-x) \I_{0,2}(\D),
\ee
where $\D=m^2|\S_0|^2e^{2Q_\S A_0} + x(1-x)q^2$. The same diagram with $AP\bar P$ replacing the $A\G\bar\G$ vertex happens to vanish. Expanding in $\left({\m\over m}\right)$, we can see that this correlator requires the following term in the effective action:
\be
f_A(\S_0, \bar\S_0) = -i{Q_\G\over8\pi}\log\left(\S_0\over\bar\S_0\right).
\ee
Finally, we compute the loop shown in~\ref{SSbar1}, which corrects the $\langle \S\bar\S\rangle$ propagator. The result is
\be
\hlf m^2 e^{2Q_\S A_0}\int {d^2q\over(2\pi)^2}d^2\th^+\ q_-\S(q)\bar\S(-q) \int dx\ x\I_{0,2}(\D).
\ee
Again expanding in the limit $m\gg\m$, we find that this term originates from
\be
f_\S(\S_0,\bar\S_0) = -{i\over32\pi}{\log(\bar\S_0)\over\S_0}.
\ee
These results combine very nicely into a sensible effective action. The function $f_\S$ is clearly a renormalization of the $\S$ kinetic term. To ensure gauge invariance of this term, we must extract the appropriate couplings to the vector multiplets. Writing
\bea 2Q_P &=& (Q_P-Q_\hG)-Q_\S \\
2Q_\hG &=& -(Q_P-Q_\hG)-Q_\S \non
\eea
for the remaining $\log(\S)$ terms, we can write the one-loop effective action at a scale $\mu\ll m$ in the form
\bea
W^{\rm 1-loop} &=& -{i\over16\pi}\int d^2xd^2\th^+\left[\left(\log|e^{Q_\S A}\S|\over\S\right)\nabla_-\S +c.c.\right] \non\\
&&-\left(Q_\S^2 +\cA\over8\pi\right) \int d^2xd^2\th^+ AV_- \label{Gammaunfixed}\\
&&+i{(Q_P-Q_\G)\over16\pi}\int d\th^+\ \log(\S)\Upsilon\ +c.c. ,\non
\eea
where $\nabla_-=\del_-+Q_\S \left(\del_-A+i V_- \right)$.

\subsection{The structure of the effective action}\label{structureeffective}

Before examining the structure of the effective action, we should comment on the validity of the one-loop approximation. We are integrating out a massive multiplet with mass $m |\S|$. As $|\S|$ becomes sufficiently small, there can be large corrections to a one-loop effective action. With this caveat in mind, we note that
the first line of~\C{Gammaunfixed}\ gives a gauge invariant correction to the $\S$ kinetic terms.

These kinetic terms, together with the remaining quantum corrections of~\C{Gammaunfixed}, are crucial in resolving the supersymmetry puzzle of section~\ref{modulispace}. We can sketch how this comes about: supersymmetry requires a $\CC^\ast$-action on the space of fields, but our $D$-term equations typically admit multiple solutions in the orbit of this action, as shown in figure~\ref{figure1}\ for one $D$-term. A horizontal slice of that graph typically has two solutions. This is what led to the sphere topologies of the target manifold and the apparent  supersymmetry breaking. Because of the quantum corrections of~\C{Gammaunfixed}, we find that figure~\ref{figure1}\ is basically cut in half because the low-energy metric becomes singular before one can access both solutions to the $D$-term equation. Since our one-loop effective action is reliable at large $|\S|$, the small $|\S|$ branch of solutions is not accessible via our analysis.

%For a general  non-linear $(0,2)$ model,  $K_\S$  determines the $\S$ metric via the couplings:
%\bea \label{sigmakinetic}
%S_{\S} &=& -{i\over 4} \int\d^2 xd^2\th^+ \,\left[K_\S (\S, \bar\S) \nabla_- \S - K_{\bar\S }(\S, \bar\S) \nabla_- {\bar \S} \right], \cr
%&=& \int d^2x \left[ -{1\over 2}\left( K_{\S, \bar\S} + K_{\bar\S, \S}\right) \big|\cD_\mu \S \big|^2  + { Q_\S D \over 2}\left( K_\S \S + K_{\bar\S} \bar\S \right) + \ldots \right].
%\eea
%The omitted couplings involve fermions. Gauge invariance is a very severe constraint on $K_\S$. Certainly the free case of $K_\S = \bar\S$ is consistent. There is another possibility which is important for us; namely,
%\be\label{ksigma}
%K_\S = { \log(\bar\S) \over\S}.
%\ee
%At first sight, this appears to violate gauge invariance. However, the $K_i$ couplings enjoy a generalization of K\"ahler gauge invariance.  Shifting $K_i$ by sending,
%\be
%K_i \rightarrow K_i + f_i,
%\ee
%is a symmetry of the theory if $f_i d\F^i$ is a holomorphic $1$-form. For the choice~\C{ksigma}, a gauge transformation corresponds precisely to such a transformation.

Turning to the remaining terms of~\C{Gammaunfixed}, we see that the third line is precisely the pion-like $F$-term coupling needed to reproduce the gauge anomaly of the pair $(P, \G)$. Under a gauge transformation, this $F$-term transforms anomalously like~\C{(0,2)anom}\ but with  coefficient
\be
-(Q_P-Q_{\hat\G})Q_\S = Q_P^2-Q_{\hat\G}^2,
\ee
which reproduces the anomaly of the fields we  integrated out. Finally, the term appearing in the second line has a nice interpretation as a local contribution coming from the anomalous measure of the remaining light charged fields $(\S,\Phi^i,\G^\a)$. Recall that we can represent the gauge anomaly by the non-local effective action
\be\label{repeatanomaly}
W= {\tilde\cA\over8\pi}\int d^2xd^2\th^+\left(\Dbar_+A {\del_-\over\del_+}D_+A - AV_-\right),
\ee
described in section~\ref{susyanomaly}, where $\tilde\cA=Q_\S^2+\cA$.

This 1PI effective action is non-local because we have integrated out massless degrees of freedom. We are actually studying the local Wilsonian effective action, where we only integrate down to a fixed scale $\m$. This IR cut-off has the effect of smoothing out the non-local term:
\be
\Dbar_+\del_-A {1\over\del^2}D_+\del_-A \,\, \rightarrow \,\, \int_0^1 dx\ \Dbar_+\del_-A \left({-x(1-x)\over \m^2-x(1-x)\del^2}\right)D_+\del_-A.\label{smoothing}
\ee
Indeed, in Appendix~\ref{quad} we find the local expression appearing on the right hand side of~\C{smoothing}\ when we only integrate down to a scale $\m$, rather than all the way to zero. If we go to momentum space replacing $\p^2$ by $q^2$, we see that the non-local expression~\C{smoothing}\  vanishes in the limit $\m^2\gg q^2$.

What remains is the local term of~\C{repeatanomaly}. By including the local $AV_-$ term in our effective action, we are essentially changing the anomalous variation of the measure for the light chiral fermions from the usual
\be {\tilde\cA\over16\pi}\int d\th^+\ \La\Upsilon +c.c. \ee
to
\be
\dd S = {\tilde\cA\over8\pi} \int d^2xd^2\th^+\ \dd\left(\Dbar_+A {\del_-\over\del_+}D_+A\right) = {\tilde\cA\over8\pi} \int d^2xd^2\th^+\ (\La+\bar\La)\del_-A +O(\La^2).
\ee
Generalizing the result~\C{Gammaunfixed}\ to multiple sets of $(\S,P,\hG)$ is now straightforward. Furthermore, setting $\S=1$ does indeed recover the unitary gauge result~\C{Gammafinal}\ with the correct coefficient $-\cA/8\pi$. Finally, we note that setting $Q_\S=0$ forces $\cA=0$ and $Q_\hG =-Q_P$, and then~\C{Gammaunfixed}\ reduces to the gauge invariant cases studied in~\cite{Quigley:2012gq}\ with NS-brane sources. The coefficient of the log interaction in the neutral $\S$ case is a factor of $2$ larger than the charged case considered here for reasons explained in section~\ref{axanom}.

\subsection{Effects from other fields}

Since we are computing the Wilsonian effective action at a scale $\m$, we should in principle also integrate over the high-energy modes of the light fields $\Phi^i,\G^\a,\S,A,$ and $V_-$. Fortunately, up to a field-independent shift of the FI parameter, the path integration over these light fields do not affect our results. More details can be found in Appendix~\ref{others}.

\section{The Non-Linear Sigma Model}\label{NLSM}

Now that we have evaluated the one-loop corrected effective action, including the effects from integrating out anomalous pairs of massive multiplets, we are in a position to study the non-linear sigma models that emerge at low energies. We will extract a low-energy non-linear sigma model in a semi-classical fashion by sending the gauge coupling $e^2\rightarrow\infty$. In this limit, the gauge fields are effectively non-dynamical and we can integrate them out classically. We will separately consider the case with a single $\S$ field and the case of multiple $\S$ fields.

\subsection{Effective NLSM actions, supersymmetry, and anomalies}\label{NLSManomalies}

Before we turn to explicit computations of background geometries, we should discuss a few interesting subtleties in extracting a geometric interpretation from effective actions.\footnote{This material is discussed in a number of classic papers~\cite{Hull:1986xn,Sen:1986mg,Howe:1987nw}; the last paper is particularly relevant to the (0,2) discussion.} The essential point is relatively simple:  to extract a geometric interpretation from a NLSM effective action we make a split between the local and non-local contributions, and such a split is inherently ambiguous up to choosing various local finite counter-terms.  These terms are constrained by demanding manifest (super)symmetries and other desirable properties.

To make these comments concrete, consider a (0,1) NLSM.  The defining geometric data for such a theory consist of a metric $G$ and B-field $B$, as well as a choice of connection on the left-moving gauge bundle.
When we expand the classical action in components, we find that the right-moving fermions couple to a connection with torsion given by $dB$.   In the quantum theory $B$ acquires non-trivial space-time gauge and Lorentz transformations, and $H = dB +\ff{\alpha'}{4}\textrm{CS}$ is the physical gauge-invariant field strength.  This seems to lead to a small paradox:  either the right-moving fermion kinetic term is gauge-variant, or it is not supersymmetric~\cite{Hull:1986xn,Sen:1986mg}.

The resolution follows by computing the effective action in a manifestly (0,1) supersymmetric fashion.  This is comparatively easy because of the unconstrained nature of (0,1) superfields, and the result is an explictly (0,1) SUSY form for the space-time Lorentz- and gauge-variant terms in the one-loop effective action~\cite{Hull:1986xn}.  It is then easy to see that this local anomaly can be cancelled by assigning transformation properties to \emph{both} $G$ and $B$.  The latter transformation is familiar, but the former is unusual and perhaps undesirable if one wishes to use conventional intuition from Riemannian geometry.

Fortunately, there is a simple alternative:  we can add a finite (0,1) SUSY counter-term whose variation exactly matches the $G$-variation.
Now the metric in the two-derivative action can be kept invariant under the gauge transformations; furthermore, expanding the resulting effective action in components, we find that the modification of the local terms is exactly to shift $dB \to H$ in the right-moving fermion kinetic terms.

A similar analysis has also been carried out for (0,2) NLSMs~\cite{Howe:1987nw}.  The classical (0,2) NLSM is determined in terms of a (0,2) potential $K_I$ and a Hermitian metric for the left-moving fermions.  The latter determines a holomorphic connection for the gauge bundle, while $K_I$ is the (0,2) potential that fixes the Hermitian metric and B-field via
\begin{align}
\label{eq:GBfromK}
G_{I\bar J}(x,\bar x) = \del_{(I}K_{\bar J)},\qquad B_{I\bar J}(x,\bar x) =  \del_{[I} K_{\bar J]},
\end{align}
where $X^I$ denote the complete set of (0,2) bosonic chiral fields, and $x^I$ are their scalar components.  The gauge-variant part of the effective action can be evaluated in a manifestly supersymmetric fashion (though there are complications because of the use of constrained chiral superfields), and the resulting variation can be cancelled by assigning gauge transformations to the (0,2) potential $K_I$.  However, there is no manifestly (0,2) SUSY finite local counter-term that can be used to reproduce the variation due to the shift of the Hermitian metric.  Thus, to keep (0,2) SUSY manifest, we must work with space-time Lorentz and gauge-variant (0,2) potential $K_I$; in particular, both the metric and $B$-field shift under the transformations.

This is significant:  in a manifestly (0,2)-SUSY regularization, $G_{I\Jbar}$ is in general a gauge-variant object. If we want to consider a more conventional geometry, where the metric is gauge invariant, we will need to leave the realm of manifestly off-shell $(0,2)$ supersymmetry and construct an invariant metric $\widehat G_{I\Jbar}$. Such a metric has fundamental form, $\widehat J$, that is related to $H$ via
\be
H = i \left( \p -\bar\p \right) \widehat J,
\ee
and satisfies the Bianchi identity:
\be \label{Jbianchi}
dH = 2 i \p \bar\p \widehat J = {\alpha'\over 4} \left[ \tr R\wedge R - \tr F\wedge F \right].
\ee
Alternatively, we can work with the gauge variant ``metric" $G_{I\Jbar}$, but this can lead to confusion: for instance, an apparently K\"ahler background can be gauge equivalent to a Hermitian background with torsion. We will see explicit examples of this.

Usually, we are not interested in a metric that precisely satisfies~\C{Jbianchi}, anymore than we are interested in the precise $\alpha'$-corrected metric that defines a conformal $(2,2)$ model. A metric solving~\C{Jbianchi}\ on the nose will be very complicated, since the curvatures appearing on the right hand side are evaluated with quantum corrected connections. Rather, we are usually interested in how~\C{Jbianchi}\ is solved at the level of cohomology. Renormalization group flow will take care of generating the precise set of $\alpha'$ corrections. This is a subtle question because the right hand side must be globally trivial, and yet integrate to something non-vanishing on a space with torsion. How this works for the original compact torsional solutions of~\cite{Dasgupta:1999ss}\ has been explored in detail~\cite{Fu:2006vj, Becker:2008rc, Becker:2009df, Becker:2009zx, Melnikov:2012cv, Melnikov:2010pq}.

\subsection{The $(0,2)$ metric and $B$-field}\label{singlesigma}
We begin with the complete low-energy effective action in the limit $e^2\rightarrow\infty$:
\bea\label{effectiveaction}
S &=& \hlf \int d^2xd^2\th^+\left[-{i\over2}\sum_i\left( \bar\Phi^i e^{2Q_i A}\del_-\Phi^i -c.c.\right) - \sum_\a\bar\G^\a e^{2Q_\a A}\G^\a \right. \non\\
&&\qquad -{i\over2} \left(\left(\bar\S e^{2Q_\S A} + {\log\bar\S\over8\pi\S}\right)\del_-\S -c.c.\right) +\Th(\S)\del_-A \\
&&\qquad \left. +\left(\sum_i Q_i|\Phi^i|^2 e^{2Q_i A} + Q_\S|\S|^2 e^{2Q_\S A} - {\cA\over4\pi} A -R(\S) \right)V_- \right] \non,
\eea
where we have introduced the natural field-dependent quantities
\be\label{defineR}
R(\S) = r + {Q_P\over2\pi}\log|\S|,\qquad \textrm{and}\qquad \Th(\S) = {\th\over2\pi}+{Q_{\hat\G}\over2\pi}\Im\left(\log\S\right).
\ee
These combine naturally into the complex quantity
\be\label{defineT}
T \equiv \Th+iR =  t +i\left(Q_P-Q_{\hat\G}\over4\pi\right)\log\S - i{Q_\S\over4\pi}\log\bar\S, \qquad t = ir + {\th\over 2\pi},
\ee
which is only holomorphic when $Q_\S=0$; examples with $Q_\S=0$ were studied in~\cite{Quigley:2012gq}.
Let us recall that all the effects of integrating out the massive anomalous pair $(\G, P)$ are encoded in the $\S$ couplings of~\C{effectiveaction}. The fields $\Phi^i$ and $\G^\a$ were spectators in that computation, described in section~\ref{perturbationtheory}. They appear with standard couplings in~\C{effectiveaction}.

This action is not gauge invariant. This is critical: we are studying a quantum consistent low-energy theory but not a classically consistent theory. Under an infinitesimal gauge transformation, the action changes by
\be
\Th(\S) \, \rightarrow \, \Th(\S) - {\tilde\cA\over4\pi}\, \Re(\La),\label{Thshift}
\ee
where
\be
\tilde\cA=Q_\S^2 +\cA=Q_{\hat\G}^2-Q_P^2=Q_\S(Q_P-Q_{\hat\G})
\ee
is the anomaly coefficient of the low-energy degrees of freedom. For convenience, we recall that $\cA = 2Q_\S Q_P$. As noted in section~\ref{structureeffective}, this shift in the action is compensated by an anomalous transformation of the path-integral measure.

To simplify notation, we will denote the complete set of chiral fields by
\be
X^I = (\Phi^i,\S) = X_{\bar I}.
\ee
Note that our convention for raising and lowering indices conjugates the index. $V_-$ appears as a Lagrange multiplier in~\C{effectiveaction}, enforcing the constraint
\be
\sum_I Q_I|X^I|^2 e^{2Q_I A}  - {\cA\over4\pi} A = R(\S). \label{constraint}
\ee
This constraint should be viewed as an equation that determines $A$ in terms of $X^I$ and the complex conjugate field $\bar X^{\bar I}$. Note that this equation is actually gauge invariant under the full $\CC^\ast$-action.
%It is clearly invariant under the compact $U(1)$ action but not under the full $\CC^\ast$-action \textbf{I don't understand this sentence, it is $\CC^*$ invariant}. However the change in $R(\S)$ is canceled by the term linear in $A$.

Implicitly solving the constraint~\C{constraint}\ gives a $(0,2)$ non-linear sigma model  action
\be
S = -\hlf\int d^2xd^2\th^+\left[{i\over2} \left(K_I\del_-X^I -c.c.\right) + h_{\a\bar\beta} \bar\G^{\bar\beta}\G^\a \right].
\ee
The metric on the gauge bundle over the target space is
\be
h_{\a\bar\beta} = e^{2Q_\a A}\dd_{a\bar\beta},
\ee
but we will ignore $h$ in the rest of this discussion because our primary concern is with the target space metric itself and its associated $B$-field. These objects are derived from
\bea \label{defineK}
K_I = X_I e^{2Q_I A} + 2i\Th \del_I A + \dd_{I\S}{\log\bar\S\over8\pi \S},
\eea
using~\C{eq:GBfromK}. Their evaluation is greatly facilitated by the relations,
\be
\del_I A = \D\left(\del_I R - Q_I\bar X_I e^{2Q_I A}\right), \label{delA}
\ee
where we have introduced the quantity
\be
\D = {\del A\over \del R} = \left( 2\sum_I Q_I^2 |X^I|^2 e^{2Q_I A} - {\cA\over4\pi}\right)^{-1}. \label{Delta}
\ee
These relations follow from differentiating the constraint~\C{constraint}. Away from $|\sigma|=0$,  the induced target space metric is
\be\label{inducedmetric}
G_{I\bar J} = e^{2Q_I A}\dd_{I\bar J} -2\del_I A \D^{-1} \del_{\bar J}A + i\del_I A\del_{\bar J}\bar T -i\del_I T\del_{\bar J}A +{\dd_{I\s}\dd_{\bar J\bar\s}\over8\pi|\s|^2},
\ee
%\bea
%&& G_{i\jbar} = e^{2Q_i A}\left(\dd_{i\jbar} - 2\bar\phi_i \phi_\jbar Q_i \D Q_j e^{2Q_j A} \right),\cr
%&& G_{i\bar\s} = -2\bar\phi_i \s Q_i\D Q_\S e^{2(Q_i+Q_\S) A} +i \del_i A \del_{\bar\s}(\Th+iR), \\
%&& G_{\s\bar\s} = \left(e^{2Q_\S A} +{1\over4\pi|\s|^2}\right)\left(1 - 2 |\s|^2 Q_\S\D Q_\S e^{2Q_\S A}\right) -{1\over8\pi|\s|^2}\left( 1 + {Q_P\D Q_{\hat\G}\over\pi} \right). \non
%\eea
and the induced $B$-field is
\be
B = 2i\Th\,\del\bar\del A +i\del A\bar\del T -i\bar\del A\del \bar T.\label{B}
\ee
Note that curvature of $B$ has the rather simple form,
\be\label{Bcurvature}
dB = i\left[\del T+\bar\del \bar T\right] \del\bar\del A = i\left[d\Th +i(\del-\bar\del)R\right]\del\bar\del A,
\ee
which satisfies
\be
dB = i(\bar\del-\del) J.\label{dBandJ}
\ee
This relation follows automatically because both $B$ and $J$ are derived from the same $(0,2)$ potential $K_I$.

Both~\C{inducedmetric}\ and~\C{Bcurvature}\ are gauge-variant quantities with respect to the the superspace $\CC^\ast$-action. We can see this in a very striking way: there is a very natural choice of gauge in which we set $\S=1$ using the superspace $\CC^\ast$-action. In this gauge $dB=0$ and therefore~\C{dBandJ}\ implies that the corresponding metric should be K\"ahler. This is sufficiently surprising that we will verify K\"ahlerity directly in section~\ref{patch}. In other gauge choices $dB \neq 0$, and so the metric no longer appears K\"ahler. Clearly, we are missing some important ingredient.

From the target space perspective, the chiral gauge parameter $\Lambda$ can be regarded as a holomorphic function of $X^I$, so that gauge transformations correspond to target space diffeomorphisms. What have found is that neither $G$ nor $B$ transform as tensors under this diffeomorphism. Based on the discussion above this had to be the case, because K\"ahlerity is a coordinate independent property. In hindsight, this might have been expected for reasons explained in section~\ref{NLSManomalies}. The manifestly $(0,2)$ GLSM is naturally giving us a NLSM with anomalous transformation properties for both the metric and $B$-field.

It is intriguing that this phenomenon does not appear for conventional GLSMs, where only non-anomalous multiplets mass up, but it does appear here. In the conventional case, the $G$ that results from the procedure we have followed defines a genuine metric. The redefinition of $G$ described in section~\ref{NLSManomalies}\ is still required, but the need for such a redefinition only shows up at one $\alpha'$-loop in the NLSM. In our case, the initial metric given to us by the GLSM is already unconventional. Presumably, this reflects the torsion present at tree-level in the background. To find a conventional metric with a $B$-field that transforms in the usual way under target space gauge and Lorentz transformations, we will need to leave our manifestly $(0,2)$ framework.

\subsection{A preferred patch}\label{patch}

We argued in section~\ref{singlesigma}\ that it is possible to find a local coordinate patch where the ``metric"~\C{inducedmetric},
\be
G_{I\bar J} = e^{2Q_I A}\dd_{I\bar J} -2\del_I A \D^{-1} \del_{\bar J}A + i\del_I A\del_{\bar J}\bar T -i\del_I T\del_{\bar J}A +{\dd_{I\s}\dd_{\bar J\bar\s}\over8\pi|\s|^2},
\ee
restricts to a K\"ahler metric. The argument relied on the relation
\be
dB = i(\bar\del-\del)J,\label{dBandJ3}
\ee
which is a consequence of working in a manifestly $(0,2)$ supersymmetric formalism. Here we would like to check K\"ahlerity in this patch directly without relying on~\C{dBandJ3}.

We use the $\CC^*$-action to set $\s=1$ and work with affine coordinates:
\be
z^i \equiv { \phi^i/\left(\s\right)^{Q_i/ Q_\S}}.
\ee
This choice is equivalent to fixing unitary gauge for the UV theory, as described in section~\ref{unitary}. If we ignore any potential singular behavior of the low-energy metric, this is the only patch needed to globally fix the gauge action since $\s$ is always non-vanishing. In this patch, the metric is given by
\be\label{metric1}
G_{i\jbar} = e^{2Q_i A}\left(\dd_{i\jbar} - 2\bar z_i z_\jbar Q_i \D Q_j e^{2Q_j A} \right).
\ee
Notice that with $\s=1$, $(R,\Th)$ of~\C{defineR}\ just reduce to $(r,\th)$, and  $B$ is now exact:
\be\label{kahlerB}
B= i{\th\over\pi}\del\bar\del A
\ee
since $\th$ is constant. This metric and $B$ have precisely the same form expected in a conventional $(0,2)$ sigma-model with one important exception: the function $\D$, which we recall here for convenience
\be
\D =  \left( 2\sum_I Q_I^2 |X^I|^2 e^{2Q_I A} - {\cA\over4\pi}\right)^{-1},
\ee
contains the quantum correction $-{\cA\over4\pi}$. This correction term can be seen by fixing unitary gauge in~\C{Gammafinal}, and all the anomalous behaviour of $G$ can be traced back to this term.

In this patch the $B$-field~\C{kahlerB} is exact, which suggests that the metric is K\"ahler. We can check this explicitly by computing:
\bea
\del_k G_{i\jbar} &&= 2Q_i\del_k A \dd_{i\jbar} e^{2Q_i A} - 2\bar z_i Q_iQ_j\left(\dd_{\jbar k}\D + z_\jbar\del_k\D+ 2(Q_i+Q_j)z_\jbar \D\del_k A\right)e^{2(Q_i+Q_j)A}, \non\\
&&= -2Q_j\D e^{2Q_jA}\left(Q_k\bar z_k\dd_{i\jbar} e^{2Q_k A} + Q_i \bar z_i \dd_{\jbar k} e^{2Q_i A}\right) \\
&&+ 4Q_iQ_jQ_k \bar z_i z_\jbar \bar z_k \D^2 e^{2(Q_i+Q_j+Q_k)A} \left[1-2\D\left(Q_\S^3 e^{2Q_\S A} + \sum_\ell Q_\ell^3|z^\ell|^2e^{2Q_\ell A}\right)\right], \non\cr
\eea
where where have used the fact that $Q_i\dd_{i\jbar}e^{2Q_i A} = Q_j\dd_{i\jbar}e^{2Q_j A}$ along with the relations~\C{delA}\ and~\C{Delta}. Each line is separately symmetric under the exchange $i\leftrightarrow k$ and therefore
\be
(\del J)_{ij\bar k} = i\del_{[i}G_{j]\bar k} =0.
\ee
Similarly, $\bar\del J=0$ confirming that the metric~\C{metric1}\ is in fact K\"ahler, as claimed.

Suppose we choose a different gauge-fixing, $\phi^0=1$, and work with affine local coordinates:
\be
\tilde z^i = \phi^i/\left(\phi^0\right)^{Q_i/Q_0},\qquad \tilde\s = \s/\left(\phi^0\right)^{Q_\S/Q_0}.
\ee
In this set of coordinates, $H\neq0$ and the space is non-K\"ahler. This can happen precisely because of the unusual metric transformation properties found in~\C{GLSMtransform}. We are perfectly free to work with this metric and $B$-field as long as we keep track of the unusual patching conditions. Indeed the GLSM naturally gives us this form for $G$ and $B$ in a manifestly $(0,2)$ supersymmetric way. However, if we want to assign a conventional geometry to this NLSM, we need to understand how to define a conventional metric.

\subsection{Defining an invariant metric}\label{invariant}

%Up to this point, we have worked in terms of projective coordinates on the target space. 
The projective coordinates naturally parameterize a $\CC^*$-bundle over the target, but we are really only interested in the  quotient of this total space by the $\CC^*$-action. Let us work with sections of this bundle that define local patches. We cover the target space with open neighbourhoods $U_\al = \{x^\a\neq0\}$, and within each such set define local coordinates:
\be
Z^I_{(\a)} = x^I /\left(x^\a\right)^{Q_I/Q_\a}.
\ee
On the intersections $U_\ab =U_\al\cap U_\bt$, we relate the local coordinate systems by
\be
Z^I_\al = \left(\exp{iQ_I\La_\ab}\right) Z^I_\bt,\label{gluing}
\ee
where the (holomorphic) gluing functions $\La_\ab$ are naturally identified with the $\CC^*$ gauge transformations:
\be
\La_{\ab} = {i\over Q_\a}\log Z^\a_\bt.
\ee
These transition functions define the bundle over the target space. Note that $A$, defined implicitly by~\C{constraint}, is not globally defined. On $U_\ab$, $A$ transforms according to
\be
A_\al = A_\bt + A_\ab,\qquad \textrm{with}\qquad A_\ab = {\La_\ab -\bar\La_\ab\over2i}.\label{Aab}
\ee
The one-form $\del A_\al$ acts as a (holomorphic) connection on our line bundle with $\del\bar\del A_\al=\del\bar\del A_\bt$ its invariant curvature two-form. Finally, it will be useful to note that the quantity $T$, defined in~\C{defineT}, shifts in a way similar to $A_\al$ except
\be
T_{(\a\b)} \equiv T_{(\a)} - T_{(\b)} = -{\tilde\cA\La_\ab + Q_\s^2 \bar\La_\ab \over4\pi}.\label{Tab}
\ee
Note that
\be
\del T_\ab = -i{\tilde\cA\over2\pi}\del A_{\ab} ,\qquad \textrm{and}\qquad \bar\del T_\ab = i{Q_\s^2\over2\pi}\bar\del A_\ab.\label{delT}
\ee

From our earlier discussion, we expect $G_\al$ and $B_\al$ to have anomalous transformations on the overlaps $U_\ab$. Indeed, by examining the line element
\be
ds^2 = G^\al_{I\Jbar}dZ^I_\al dZ^\Jbar_\al,
\ee
and applying the transformations~\C{gluing}-\C{Tab}, we find that the metric $G^\al$ has an anomalous transformation law:
\be \label{GLSMtransform}
G^{(\a)}_{I\Jbar} = G^{(\b)}_{I\Jbar} -{\tilde\cA\over2\pi} \left(\del_I\left( A_{(\b)} +A_{(\a\b)} \right)\del_\Jbar\left(A_{(\b)}+A_{(\a\b)}\right) -\del_I A_{(\b)}\del_\Jbar A_{(\b)}\right).
\ee
This is problematic if we wish to interpret $G$ as a metric since the line element $ds^2$ would not be an invariant. However,~\C{GLSMtransform}\ suggests a natural resolution to this puzzle because the quantity
\be\label{Ghat}
\widehat G_{I\Jbar} = G_{I\Jbar} +{\tcA\over2\pi}\del_I A \del_\Jbar A
\ee
does define an invariant line element. In particular,
\be\label{intersections}
\widehat G^\al_{I\Jbar} = G^{(\a)}_{I\Jbar} +{\tilde\cA\over2\pi} \del_I A_{(\a)}\del_\Jbar A_{(\a)} = G^{(\b)}_{I\Jbar} +{\tilde\cA\over2\pi} \del_I A_{(\b)}\del_\Jbar A_{(\b)} =\widehat G^\bt_{I\Jbar}
\ee
is a candidate metric for our target spaces.

\subsection{An alternate derivation of $\widehat{G}$}

Let us derive the result~\C{Ghat}\ for the metric from another, more systematic, approach. The idea is to consider the holomorphic vector field
\be
L = \sum_I Q_I X^I \del_I
\ee
that generates the $\CC^*$-action. Next, consider the contraction of this vector field with the fundamental form $J$ associated to $G$. If the $(0,1)$-form
\be
\bar V \equiv i_L\left(-iJ\right)
\ee
is non-zero, then the metric $G$ will not naturally descend to the quotient space. However, the improved fundamental form
\be
{\widehat J} = J - V \left(i_L V\right)^{-1}\bar V
\ee
will be invariant by construction, and we associate the metric ${\widehat G}$ with this improved fundamental form.

In order to compute $\bar V$, it will help to recall that $T=\Th+iR$, defined in~\C{defineT}, is a function only of $\S$ and $\bar\S$; in particular,
\be
\del_I T = i\left({Q_P-Q_{\hat\G}\over4\pi}\right){1\over\S} \dd_{I\S}.
\ee
Furthermore, we recall that
\be
\del_I A = \D\left(\del_I R -Q_I \bar X_I e^{2Q_I A}\right) = \D\left({Q_P\over4\pi\S}\dd_{I\S} -Q_I \bar X_I e^{2Q_I A}\right),
\ee
which leads to
\bea
L^I \del_I A &=& \sum_I Q_I X^I \D \left(\del_I R -Q_I \bar X_I e^{2Q_I A}\right) \non\\
&=& Q_\S \D \S \del_\S R - \hlf\D\left(\D^{-1} + {\cA\over4\pi} \right)\\
&=& \D{Q_\S Q_P \over4\pi} -\hlf -\D {\cA\over8\pi} \non\\
&=& -\hlf \non,
\eea
because $\cA=2Q_PQ_\S$. Now we can evaluate the components of the connection $\bar V$:
\bea
V_{\Jbar} &=& L^I G_{I\Jbar} \non\\
&=& \sum_I L^I \left(e^{2Q_I A}\dd_{I\bar J} -2\del_I A \D^{-1} \del_{\bar J}A + i\del_I A\del_{\bar J}\bar T -i\del_I T\del_{\bar J}A +{\dd_{I\S}\dd_{\bar J\bar\S}\over8\pi|\S|^2}\right)\\
&=& Q_J X_\Jbar e^{2Q_J A} +\left(-2\left(L^I \del_I A\right)\D^{-1} -iL^I\del_I T\right)\del_\Jbar A +\left(i(L^I\del_I A)\del_{\bar\S}\bar T +{Q_\S\over8\pi\bar\S}\right) \dd_{\Jbar\bar\S} \non\\
&=& Q_J X_\Jbar e^{2Q_JA} +\left(\D^{-1} +Q_\S {Q_P-Q_{\hat\G}\over4\pi}\right) \del_\Jbar A +\left(-Q_P+Q_{\hat\G} +Q_\S\over8\pi\bar\S\right)\dd_{\Jbar\bar\S} \non\\
&=& Q_J X_\Jbar e^{2Q_JA} -{Q_P\over4\pi \bar\S}\dd_{\Jbar\bar\S} +\D^{-1}\del_\Jbar A + {\tilde\cA\over4\pi}\del_\Jbar A \non\\
&=& {\tilde\cA\over4\pi}\del_\Jbar A \non,
\eea
where we used the relations $Q_P+Q_{\hat\G}+Q_\S=0$, and $\tcA=Q_\S(Q_P-Q_{\hat\G})$. Finally, we compute
\be
L^I V_I = {\tilde\cA\over4\pi}\left(L^I\del_I A\right) = -{\tilde\cA\over8\pi}.
\ee
Therefore, the correct invariant metric is
\bea
{\widehat G}_{I\Jbar} = G_{I\Jbar} - V_I \left(L^K V_K\right)^{-1} V_\Jbar = G_{I\Jbar} + {\tilde\cA\over2\pi}\del_I A\del_\Jbar A,
\eea
as claimed in~\C{Ghat}.

\subsection{The associated $H$-flux}\label{Hflux}

Now let us consider the $B$-field, and look for an invariant $H$ associated to it. Recall the relations~\C{Bcurvature}\ and~\C{dBandJ},
\be
dB_\al = i\left(\del T_\al +\bar\del\bar T_\al\right)\del\bar\del A_\al = i(\bar\del-\del)J_\al,   \label{dBandJ2}
\ee
where $J$ is the natural $(1,1)$ form associated to $G$. This relation implies
\be
2i\del\bar\del J_\al = d^2B_\al =0.
\ee
However if instead we consider $\widehat J$, the fundamental form associated to $\widehat G$, then we can define
\bea
H_\al &\equiv& i\left(\bar\del-\del\right)\widehat J_\al \non\\
&=& i\left(\bar\del -\del\right)\left(J_\al +i{\tilde\cA\over2\pi}\del A_\al \bar\del A_\al\right) \label{H} \\
&=& dB_\al +{\tilde\cA\over2\pi} \left(\bar\del-\del\right)A_\al \del\bar\del A_\al\non.
\eea
On the overlaps $U_\ab$, $dB_\al$ shifts by
\bea
dB_{\ab} &=& i\left(\del T_{(\a\b)} +\bar\del \bar T_{(\a\b)}\right)\del\bar\del A_{(\b)} \non\\
&=& -i{\tilde\cA\over4\pi}\left(\del\La_{\ab} +\bar\del\bar\La_\ab\right)\del\bar\del A_\bt \label{dBab}\\
&=& -i{\tilde\cA\over2\pi}\,d\left(\Re\La_{\ab}\right) \del\bar\del A_\bt. \non
\eea
The Chern-Simons-like form appearing in~\C{H}\ shifts in exactly the opposite way, which follows from~\C{Aab}, so that $H_\al=H_\bt$. $H$ is therefore an invariant three-form, as desired, which satisfies the Bianchi identity:
\be\label{definedH}
dH = {\tilde\cA\over\pi}\ \del\bar\del A \wedge \del\bar\del A.
\ee
It would be very interesting to understand the relation between the right hand side of~\C{definedH}\ and $\tr R\wedge R$ of the metric $\widehat G$ computed with an appropriate connection.

\subsection{An improved $B$-field}

Given the transformation~\C{dBab}\ for $dB$, it is clear that $B$ must transform by
\be
B_\ab = -i{\tilde\cA\over2\pi}\,\left(\Re\La_{\ab}\right) \del\bar\del A_\bt + \textrm{exact}.\label{Bvar}
\ee
Our intuition from the Green-Schwartz mechanism tells us that the exact term in the above equation should vanish identically. Sadly, this is not the case for the $B$-field~\C{B}\ that follows directly from the GLSM construction. However, just as we were able to construct an improved metric, $\widehat G$, so too can we construct a $\widehat B$ that transforms like~\C{Bvar}\ but without the additional exact term.

For this discussion, it will help to introduce the complex quantity
\be
M_\al \equiv T_\al - i{Q_\S^2\over2\pi}A_\al,
\ee
which shifts by
\be
M_\ab \equiv M_\al - M_\bt = T_\ab -i{Q_\S^2\over2\pi}A_\ab = -{\tilde\cA+Q_\S^2\over4\pi}\La_\ab\label{M}
\ee
on overlaps $U_\ab$. Note that this transformation property implies $\bar\del M$ is an invariant:
\be
\bar\del M_\al = \bar\del M_\bt + \bar\del M_\ab = \bar\del M_\bt,
\ee
and furthermore, since $\del\bar\del T=0$,
\be
\del\bar\del M_\al = -i{Q_\S^2\over2\pi}\del\bar\del A_\al.
\ee
With generous use of the relation~\C{delT}, we can now compute the variation of $B$:
\bea
B_\ab &=& 2i\Th_{\ab} \del\bar\del A_\bt +i\del A_\ab\bar\del M_\bt +i\del\bar M_\bt \bar\del A_\ab \non\\
&=& 2i{Q_\S Q_\G\over2\pi}(\Re\La_\ab)\del\bar\del A_\bt +{1\over2}\del \La_\ab \bar\del M_\bt +{1\over2}\bar\del\bar\La_\ab \del\bar M_\bt \\
&=&-i{\tilde \cA\over2\pi}(\Re\La_\ab)\del\bar\del A_\bt + \hlf\del\left(\La_{\ab}\bar\del M_\bt\right) +\hlf\bar\del\left(\bar\La_\ab\del \bar M_\bt\right), \non
\eea
where we have used $\tilde \cA = Q_\S(Q_P-Q_\G) = -Q_\S^2 -2Q_\S Q_\G$. As expected, $B$ shifts according to~\C{Bvar}, but thanks to~\C{M}\ we may write this in the form
\be
B_\ab = -i{\tilde \cA\over2\pi}(\Re\La_\ab)\del\bar\del A_\bt -{2\pi\over\tilde\cA +Q_\S^2}\left(\del\left(M_\ab\bar\del M_\bt\right) +\bar\del\left(\bar M_\ab\del \bar M_\bt\right)\right).
\ee
Now we see that
\be
\widehat B_\al = B_\al + {2\pi\over\tilde\cA +Q_\S^2}\left(\del\left(M_\al\bar\del M_\al\right) +\bar\del\left(\bar M_\al\del \bar M_\al\right)\right)
\ee
has the desired transformation property, namely:
\be
\widehat B_\ab = -i{\tilde \cA\over2\pi}(\Re\La_\ab)\del\bar\del A_\bt.
\ee

\subsection{A class of examples}

Rather than clutter the resulting formulae with $Q$ factors, let us consider the class of models where
\be Q_i=Q_\S=1,\ee
and $\cA>0$. An appropriate set of left-moving charges can always be found which satisfy the gauge anomaly cancelation condition~\C{uvanomaly}\ in the UV theory if $Q_P>0$. As a nice specific case, consider the model described in section~\ref{manyfields}\  that  gives $S^4$ without accounting for the metric corrections. That model is of this form with $Q_1=Q_2=Q_\S=1$ and no left-moving Fermi superfields; the anomalous set of massive fields integrated out have charges $Q_P=1$ and $Q_\hG=-2$. For that example, $\cA=2$ and $\tilde\cA=3$.

\begin{figure}[ht]
\begin{center}
\[
\mbox{\begin{picture}(230,180)(40,20)
\includegraphics[scale=1]{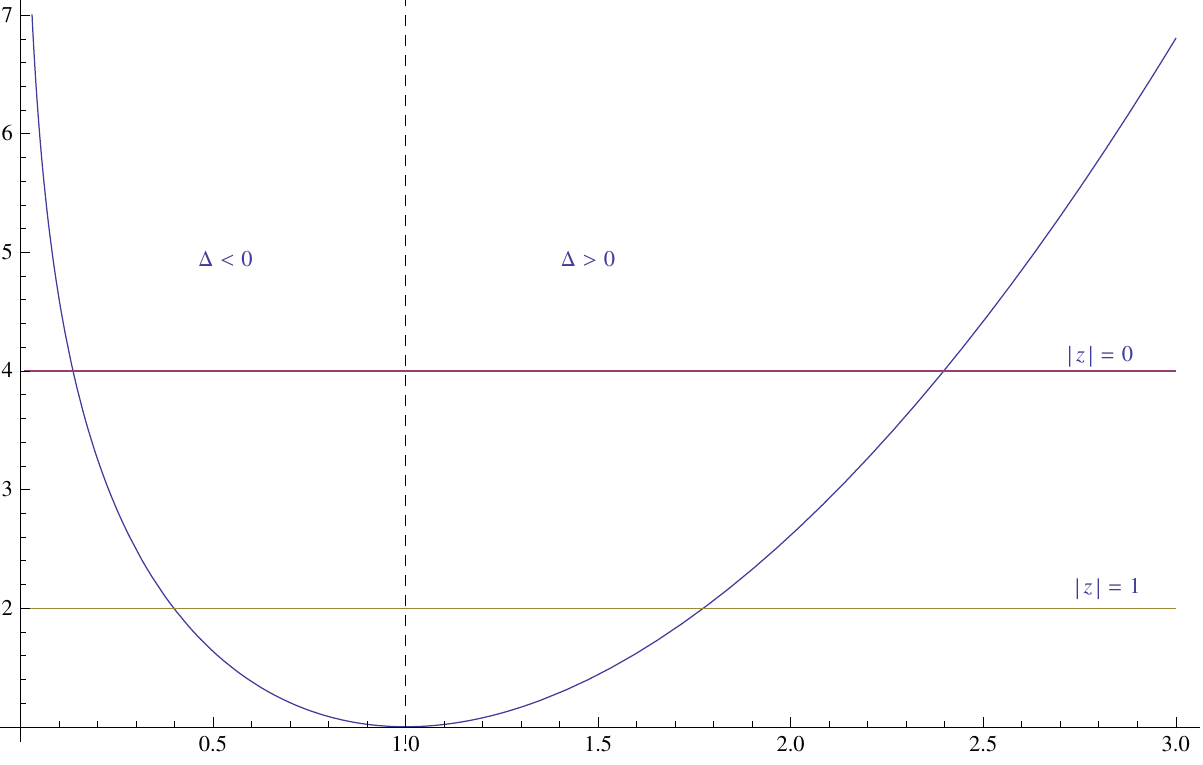}
\end{picture}}
\]
\vskip 0.2 in \caption{\it A plot depicting the solutions of~\C{defineA}\ as we increase $|z|$ from $0$ to $|z|_{max}$, with the latter value corresponding to the unique minimum. } \label{rangeA}
\end{center}
\end{figure}

It is convenient to work in the preferred patch where we set $\s=1$. The resulting metric takes the form
\be\label{samplemetric}
{\widehat G}_{i\jbar} = e^{2 A}\dd_{i\jbar} - 2 \D\left(1-{\tilde\cA\over4\pi}\D\right)  \bar z_i z_\jbar e^{4A},
\ee
where $A$ satisfies
\be\label{defineA}
e^{2A}(1+|z|^2)-{\cA\over4\pi}A =r,
\ee
with
\be
\D = \left(2e^{2A} \left(1+ |z|^2\right) -{\cA\over4\pi} \right)^{-1} = \left(2r+{\cA\over4\pi}(2A-1)\right)^{-1}.
\ee
To orient ourselves, note that~\C{defineA}\ is equivalent to~\C{singlelog}\ after setting $A=\log|\phi|$. We can view the solutions to the equation in the following way: take $|z|$ as an input. For any given $|z|$, there will generically be two solutions for $A$ as long as $r> r_{min}$. This is pictured in figure~\ref{rangeA}. The value for $r_{min}$ depends on $|z|$ via the relation~\C{definermin}. As we increase $|z|$, $r_{min}$ increases until $r=r_{min}$ at $|z|=|z|_{max}$:
\be
1+|z|_{max}^2 = {\cA \over 8\pi} e^{{8\pi r\over \cA}-1}.
\ee
We note that taking $\cA\rightarrow 0$ sends $|z|_{max} \rightarrow \infty$, which is appropriate for the projective space limit. At $|z|_{max}$, there is a unique solution for $A$ corresponding to the minimum of figure~\ref{rangeA}. At this critical point, $A$ is determined in a very nice way:
\be
A_{crit}= \left({1\over 2} -{4\pi r\over \cA}\right) .
\ee
It is very useful to note that $\Delta^{-1}$ is just the derivative of~\C{defineA}\ with respect to $A$, and so corresponds to the slope of the graph in figure~\ref{rangeA}. To the right of the minimum, $\Delta^{-1}>0$, while to the left $\Delta^{-1}<0$ with a zero at the minimum. This is very good news! The two branches of solutions for $A$ were the cause of the supersymmetry puzzles described in section~\ref{modulispace}. Whatever happens, the conformal factor of the metric~\C{samplemetric}\ will diverge at the minimum of figure~\ref{rangeA}\ disconnecting these two regions. This is what  singles out a unique solution of the $D$-term equation under the $\CC^\ast$ gauge action.

It is curious that if we approach the critical point from the left branch for $A$ with $\Delta<0$ then the metric~\C{samplemetric}\ is manifestly positive. For very large $r$, $A \sim -{ 4\pi r \over \cA}$ plus small corrections. This means the classical leading term in the metric~\C{samplemetric}\ is very small. This is the region where we do not trust our one-loop effective action, though it is intriguing that the metric is positive with a divergence when one hits the critical point at $|z|=|z|_{max}$. It would be very interesting to find an interpretation of this branch.

\begin{figure}[ht]
\begin{center}
\[
\mbox{\begin{picture}(230,180)(40,20)
\includegraphics[scale=1]{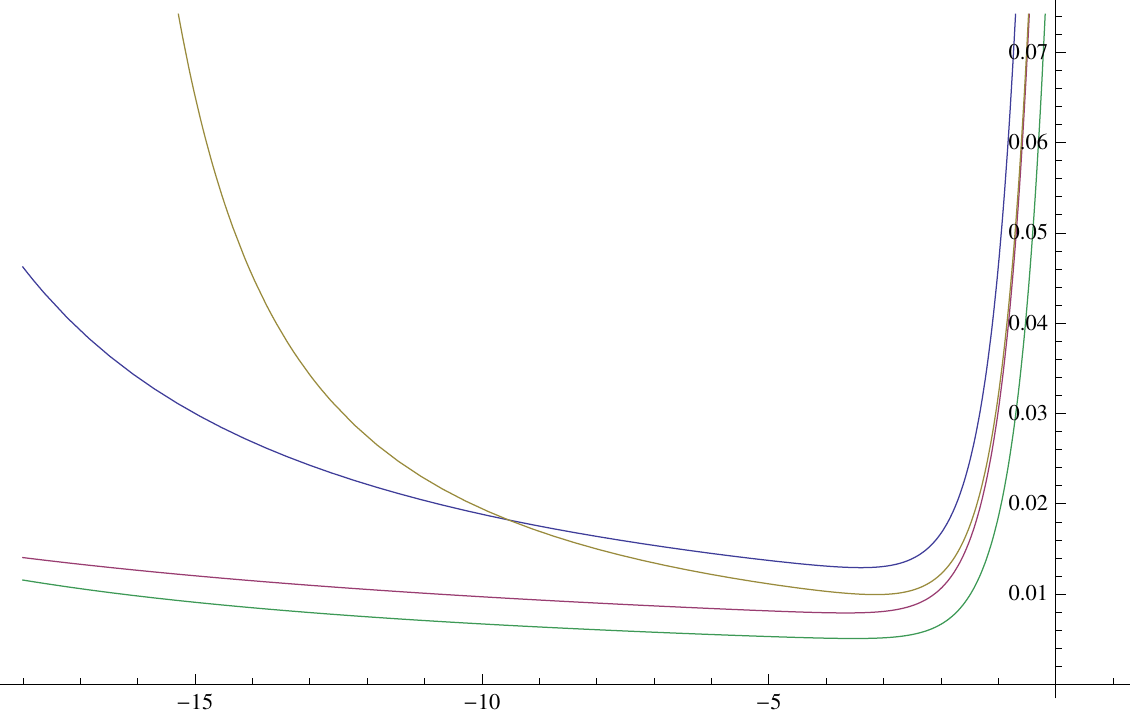}
\end{picture}}
\]
\vskip 0.2 in \caption{\it A plot of the conformal factor appearing on the left hand side of~\C{vanishingcond}\ versus $A$ for the examples $(r=4, \cA=2)$ (blue in the graph), $(r=6, \cA=2)$ (red), $(r=6, \cA=4)$ (yellow) and $(r=10, \cA=4)$ (green). The $x$-axis corresponds to vanishing conformal factor. If one extends the graph sufficiently far to the left, all the conformal factors diverge. Each conformal factor reaches some finite value to the right. } \label{conformalfactor}
\end{center}
\end{figure}

However, we want to approach from the far right where the classical metric is large and we trust our one-loop effective action. In the region to the far right, $\Delta>0$ and small if $r$ is very large. What we need to check is whether the metric~\C{samplemetric}\ encounters a singularity before we hit  $|z|_{max}$.  Because of the sign of $\Delta$, this is possible. To make our life easier, we will rotate coordinates so that
\be
(z_1, z_2, z_3, \ldots) = (z, 0, 0, \ldots).
\ee
We then need to check whether the conformal factor for the metric~\C{samplemetric}\ can vanish for some $|z|<|z|_{max}$. This requires
\be\label{vanishingcond}
1 - 2 \D\left(1-{\tilde\cA\over4\pi}\D\right) |z|^2 e^{2A} =0.
\ee
This is a transcendental equation. To see if the conformal factor vanishes, it is easiest to plot some examples. Figure~\ref{conformalfactor}\ contains a plot of the left hand side of~\C{vanishingcond}; in the examples plotted, we see that the conformal factor becomes small but does not vanish. As $|z|$ approaches $|z|_{max}$, it diverges to the far left of the plot. This behavior is quite remarkable  because generating a large variation of the conformal factor is a hint that string solutions built from these spaces might exhibit hierarchies. It is worth noting that without the correction proportional to $\tcA$ in~\C{samplemetric}, the metric would have vanished with a collapsed circle before we reach the critical point of figure~\ref{rangeA}.

We can prove positivity of the left hand side of~\C{vanishingcond}\ as follows: using the $D$-term constraint~\C{defineA}, we see that
\be
|z|^2 e^{2A} = \left( {|z|^2 \over 1+|z|^2}\right)\left( {1 \over 2 \Delta}+ {\cA \over 8\pi}\right) < \left( {1 \over 2 \Delta}+ {\cA \over 8\pi}\right) 
\ee
for $\Delta>0$. Plugging this into the left hand side of~\C{vanishingcond}\ gives
\be
1 - 2 \D\left(1-{\tilde\cA\over4\pi}\D\right) |z|^2 e^{2A} = {\tcA \cA \Delta^2 \over 16\pi^2} + {\Delta \over 8\pi} \left(2\tcA - \cA\right) >0,
\ee
demonstrating that the metric is strictly positive.

It is going to be very interesting to understand the properties of the metric~\C{samplemetric}\ in more detail, including the behavior of the conformal factor. Here, however, it will suffice to note that there is a boundary at $|z|_{max}$ at which the metric diverges. We would like to see whether this boundary is at finite distance. Near the boundary, the metric is dominated by
\be
{\widehat G}\sim {\tilde\cA\over2\pi}\D^2  |z|^2 e^{4A} d\bar z dz + \ldots,
\ee
with omitted terms non-singular at $|z|=|z|_{max}$. We note that
\be
d|z|^2 = - dA  e^{-2A} \left\{ {\cA\over  2\pi} \left(A - A_{crit} \right)\right\},
\ee
and that
\be
\Delta^{-1} = {\cA \over 2\pi} \left( A -A_{crit} \right).
\ee
These relations permit us to express the metric in terms of $A$ near the boundary at $A_{crit}$,
\bea
{\widehat G} &\sim &  {\pi \tilde\cA\over 2\cA^2 \left( A -A_{crit} \right)^2}  \left\{ {\cA\over  2\pi} \left(A - A_{crit} \right)\right\}^2 \left(dA\right)^2   + \ldots, \non\\
&\sim &  {\tcA \over 8\pi} {\left(dA\right)^2 } +\ldots.
\eea
The point $A=A_{crit}$ is therefore  at finite distance. Our target manifold has developed a finite distance boundary at which the scale factor diverges.

Specifically, the metric for the angular direction for $z$ (rather than the radial direction $|z|$) diverges at the boundary in a way highly reminiscent of the metric for the $SU(2)/U(1)$ WZW model~\cite{Bardakci:1990ad}. It seems quite possible that the metric near the boundary is regular after T-dualizing this circle direction, leading to the fascinating possibility that this space is a kind of non-geometric T-fold; that possibility will be explored further elsewhere.

\subsection{The general case}

Now that we have a basic understanding of how integrating out an anomalous multiplet affects the low energy geometry, let us take a brief look at a more general class of examples. We will let the gauge group have rank $n$, and use $a,b,\ldots$ to label the different $U(1)$ factors. We will also include multiple $\S^m$ fields, with charges $Q^a_m$. Each $\S^m$ give a large mass to a pair of $(P^m,\G^m)$ fields that we integrate out. The constraint among the charges is $Q_m^a + Q_{P^m}^a + Q_{\G^m}^a=0$. Many of the formulae from section \ref{singlesigma} generalize straightforwardly. We will still use the collective notation
\be
X^I = (\Phi^i,\S^m)
\ee
to denote the complete set of chiral fields. We define
\be
R^a(\S) = r^a + \sum_m{Q^a_{P^m}\over2\pi}\log|\S^m|,\qquad \textrm{and}\qquad \Th^a(\S) = {\th^a\over2\pi}+\sum_m{Q^a_{\G^m}\over2\pi}\Im\left(\log\S^m\right).
\ee
Under a gauge transformation, the low energy action changes by
\be
\Th^a(\S) \rightarrow \Th^a(\S) - {\tilde\cA^{ab}\over4\pi} \Re(\La^b),
\ee
where
\be
\tilde\cA^{ab}=\sum_m Q^a_m Q^b_m +\cA^{ab} = \sum_I Q^a_I Q^b_I - \sum_\a Q_\a^a Q_\a^b.
\ee
Integrating out $V^a_-$ enforces the constraints
\be
\sum_I Q_I^a|X^I|^2 e^{2Q^b_I A^b} - {\cA^{ab}\over4\pi} A^b = R^a(\S), \label{constraint2}
\ee
and we are left with a $(0,2)$ non-linear sigma model, characterized by
\bea
K_I = X_I e^{2Q_I^a A^a} + 2i\Th^a(\S)\del_I A^a + \dd_{Im}{\log\bar\S_m\over8\pi\S^m} .
\eea
The constraints~\C{constraint2}\ imply
\bea
&&\del_I A^a =  \D^{ab}\left(\del_I R^b - Q^b_I \bar X_I e^{2Q_K^c A^c}\right), \\
&&\D^{ab} = {\del A^a\over \del R^b} = \left(2\sum_I Q_I^a Q_I^b|X^I|^2 e^{2Q_I^c A^c}  - {\cA^{ab}\over4\pi}\right)^{-1},
\eea
and these allow us to compute the $(0,2)$ metric
\be
G_{I\bar J} = e^{2Q_I^a A^a}\dd_{I\bar J} -2\del_I A^a \left(\D^{-1}\right)^{ab} \del_{\bar J}A^b + i\del_I A^a\del_{\bar J}\bar T^a -i\del_I T^a\del_{\bar J}A^a +{\dd_{Im}\dd_{\bar J\bar m}\over8\pi|\s^m|^2},
\ee
and $B$-field,
\be
B = 2i\Th^a\,\del\bar\del A^a +i\del A^a\bar\del T^a -i\bar\del A^a\del \bar T^a.
\ee
These are the the objects that transform anomalously. In this case, the natural expression for an invariant metric takes the form
\be
\widehat G_{I\Jbar} = G_{I\Jbar} +{\tcA^{ab}\over2\pi} \del_I A^a \del_\Jbar A^b,
\ee
with associated $H$-flux,
\be
H = i(\bar\del-\del)\widehat J = dB + {\tcA^{ab}\over2\pi} \left(\bar\del-\del\right) A^a \del\bar\del A^b,
\ee
that satisfies the Bianchi identity:
\be
dH = {\tcA^{ab}\over\pi} \del\bar\del A^a \wedge \del\bar\del A^b.
\ee
In general, these spaces will contain boundaries and flux. As shown previously in~\cite{Quigley:2012gq}, if any $Q^a_m=0$ there will be NS-branes as well. We look forward to further studying these spaces which are simply given to us from chiral gauge theory.
%studying these more general cases in the future.
Clearly much remains to be explored.

%\vskip 0.5in
%\newpage
\subsection*{Acknowledgements}

It is our pleasure to thank Oleg Lunin, Emil Martinec, Nathan Seiberg and Stefan Theisen for helpful discussions. %I.~M. is supported by \ldots. 
I.~M. and S.~S. would like to thank the organizers and participants of the BIRS workshop on ``String Theory and Generalized Geometry'' for a stimulating atmosphere, and the BIRS center for providing a wonderful working environment. 
C.~Q. is supported in part by a Bloomenthal Fellowship and by NSF Grant No.~PHY-0758029. S.~S. is supported in part by
NSF Grant No.~PHY-0758029 and NSF Grant No.~0529954. M.~S. is funded by NSF grant DMS 1005761. All Feynman diagrams were created using Jaxodraw~\cite{Binosi:2008ig}.

\newpage
\appendix

\section{Superspace and Superfield Conventions}
\label{superfieldconventions}

\subsection{Chiral and Fermi superfields}

In this appendix, we summarize our notation and conventions. For a nice review of $(0,2)$ theories, see~\cite{McOrist:2010ae}. Throughout our discussion, we will use the language of $(0,2)$ superspace  with coordinates $(x^+,x^-,\th^+,\thbar^+)$. We define the world-sheet coordinates $x^\pm = \hlf(x^0\pm x^1)$, so that the corresponding derivatives $\del_\pm = \del_0 \pm \del_1$ satisfy $\del_\pm x^\pm =1$.
The Grassman measure is given by $\d^2\th^+ = \d\thbar^+ \d\th^+$, and  $ \int\d^2\th^+\, \th^+\thbar^+ = 1.$ The $(0,2)$ super-derivatives
\be
D_+ = \del_{\th^+} - i\thbar^+\del_+, \qquad \Dbar_+=-\del_{\thbar^+} +i\th^+\del_+,
\ee
satisfy the usual anti-commutation relations:
\be
\{D_+,D_+\} = \{\Dbar_+,\Dbar_+\} = 0, \qquad \{\Dbar_+,D_+\} = 2i\del_+ .
\ee

In the absence of gauge fields, $(0,2)$ sigma models involve two sets of superfields: chiral superfields annihilated by the $\Dbar_+$ operator,
\be
\Dbar_+\F^i=0,
\ee
and Fermi superfields $\G^\a$ that satisfy
\be
\Dbar_+\G^\a = \sqrt{2} E^\a,
\ee
where $E^\a$ is chiral: $\Dbar_+ E^\a=0$. These superfields have the following component expansions:
\bea
\F^i &=& \f^i+\sqrt{2}\th^+\j_+^i - i \th^+\thbar^+\del_+\f^i, \label{chiral}\\
\G^\a &=& \g^\a +\sqrt{2}\th^+F^\a - \sqrt{2} \thbar^+ E^\a - i\th^+\thbar^+\del_+\g^\a.
\eea

If we omit superpotential couplings, the most general  Lorentz invariant $(0,2)$ supersymmetric action involving only chiral and Fermi superfields and their complex conjugates takes the form
\be \label{(0,2) sigma}
\L =-\hlf\int\d^2\th^+\left[{i\over2}K_i \del_- \F^i - {i\over2} K_{\ibar } \del_- {\bar \F}^{\ibar} +h_{\a\bar{\b}}\bar{\G}^{\bar \b}\G^\a + h_{\a\b}\G^\a\G^\b +h_{\bar{\a}\bar{\b}}\bar{\G}^{\bar{\a}}\bar{\G}^{\bar{\b}} \right].
\ee
The one-forms $K_i$ determine the metric and $B$-field; the functions $h_{\a\b}$ and $h_{\a\bar{\b}}$ determine the bundle metric.

\subsection{Gauged linear sigma models}

We now introduce gauge fields. For a general $U(1)^n$ abelian gauge theory, we require a pair $(0,2)$  gauge superfields $A^a$ and $V_-^a$ for each abelian factor, $a=1,\ldots,n$. Let us restrict to $n=1$ for now. Under a super-gauge transformation, the vector superfields transform as follows:
\bea
\dd A &=& {i}(\bar{\La} - \La)/2, \\
\dd V_- &=& - \del_-(\La + \bar{\La})/2,
\eea
where the gauge parameter $\La$ is a chiral superfield.  In Wess-Zumino gauge, the gauge superfields take the form
\bea
A &=& \th^+\thbar^+ A_+, \\
V_- &=& A_- - 2i\th^+\labar_- -2i\thbar^+\lambda_- + 2\th^+\thbar^+ D,
\eea
where $A_\pm = A_0 \pm A_1$ are the components of the gauge field. We will denote the gauge covariant derivatives by
\be
\cD_\pm = \del_\pm + i Q A_\pm
\ee
when acting on a field of charge $Q$. This allows us to replace our usual superderivatives, $D_+$ and $\Dbar_+$ with gauge covariant ones
\be
\mathfrak{D}_+ = \del_{\th^+} - i\thbar^+\cD_+, \qquad \bar{\mathfrak{D}}_+=-\del_{\thbar^+} +i\th^+\cD_+,
\ee
which now satisfy the modified algebra
\be
\{\mathfrak{D}_+,\mathfrak{D}_+\} = \{\bar{\mathfrak{D}}_+,\bar{\mathfrak{D}}_+\} = 0, \qquad \{\bar{\mathfrak{D}}_+,\mathfrak{D}_+\} = 2i\cD_+ .
\ee
We must also introduce the supersymmetric gauge covariant derivative
\be
\nabla_- = \del_- + i Q V_-,
\ee
which contains $\cD_-$ as its lowest component. The gauge invariant Fermi multiplet containing the field strength is defined as follows:
\be
\Upsilon =[\bar{\mathfrak{D}}_+,\nabla_-] =  \Dbar_+(\del_- A + i V_-) = -2\big(\lambda_- - i\th^+(D-iF_{01}) - i\th^+\thbar^+\del_+\lambda_-\big).
\ee
Kinetic terms for the gauge field are given by
\be \label{LU}
\L = -{1\over8e^2}\int\d^2\th^+\, \bar{\Upsilon}\Upsilon = {1\over e^2}\left(\hlf F_{01}^2 + i\labar_-\del_+\lambda_- + \hlf D^2\right).
\ee
Since we are considering abelian gauge groups, we can also introduce an FI term with complex coefficient $t=ir + {\th\over2\pi}$:
\be \label{LFI}
{t\over4}\int\d\th^+ \Upsilon\Big|_{\thbar^+=0} + c.c. = -rD + {\th\over2\pi}F_{01}.
\ee

In order to charge our chiral fields under the gauge action, we should ensure that they satisfy the covariant chiral constraint $\mathfrak{\bar{D}}_+\Phi = 0$. Since $\mathfrak{\bar{D}}_+ = e^{QA}\Dbar_+e^{-QA}$ it follows that $e^{QA}\Phi_0$ is a chiral field of charge $Q$, where $\Phi_0$ is the neutral chiral field appearing in \C{chiral}. In components,
\be
\F = \f + \sqrt{2}\th^+ \j -i\th^+\thbar^+\cD_+\f.
\ee
The standard kinetic terms for charged chirals in $(0,2)$ gauged linear sigma models (GLSMs) are
\bea \label{LPhi}
\L &=& {-i\over2}\int\d^2\th^+\ \bar{\F}^i \nabla_- \F^i \\
&=& \left(-\big|\cD_\mu \f^i\big|^2 + \bar{\psi}_+i\cD_-\psi_+^i - \sqrt{2}iQ_i \bar{\f}^i\lambda_-\j^i_+ + \sqrt{2}iQ_i\f^i\bar{\j}_+^i\labar_- +  Q_i \big|\f^i\big|^2 D \right). \non
\eea
Fermi superfields are treated similarly. We promote them to charged fields by defining $\G = e^{QA}\G_{0}$ so that in components
\be
\G = \g + \sqrt{2}\th^+F - \sqrt{2}\thbar^+E -i\th^+\thbar^+\cD_+\g.
\ee
If we make the standard assumption that $E$ is a holomorphic function of the $\F^i$ then the  kinetic terms for the Fermi fields are:
\bea \label{LLa}
\L &=& -\hlf\int\d^2\th^+\,  \bar{\G}^\a \G^\a, \\
&=& \left(i\bar{\g}^\a\cD_+\g^\a + \big|F^\a\big|^2 - \big|E^\a\big|^2 - \bar{\g}^\a\del_i E^\a \j_+^i - \bar{\j}_+^i \del_\ibar \bar{E}^\a \g^\a\right). \non
\eea

\subsection{Superpotential couplings}

We can introduce superpotential couplings,
\be\label{super}
S_J = -{1\over \sqrt{2}}\int\d^2x\d\th^+\, \G \cdot J(\F) + c.c.,
\ee
which are supersymmetric if $E\cdot J=0$, and give a total bosonic potential
\be
V ={D^2 \over 2 e^2}+ |E|^2 + |J|^2.
\ee
The action consisting of the terms \C{LU},~\C{LFI},~\C{LPhi},~\C{LLa}\ and~\C{super}\ comprises the standard $(0,2)$ GLSM.

\section{Feynman Rules for the $P$ and $\G$ Superfields}\label{superFeynman}

Here we derive the Feynman rules needed to compute the one-loop effective action. Before we present the derivation, let us establish some conventions.

\subsection{Conventions}

A point in $(0,2)$ superspace will be denoted $z=(x^+,x^-;\th^+,\thbar^+)$; we will denote the difference between two points by $z_{12}\equiv z_1-z_2$. A delta-function on all of superspace is given by
\be
\dd^4(z_{12}) \equiv \dd^2(x_{12})\dd^2(\th_{12}),
\ee
where $\dd^2(x_{12})$ is the usual delta-function in two-dimensions, and the Grassmann delta-function takes the usual form:
\be \dd^2(\th_{12}) = \th_{12}^2 = \th^+_{12}\bar\th^+_{12}. \ee
Because $x^{\pm}\equiv \hlf(x^0\pm x^1)$, the non-zero components of the Minkowski metric and epsilon tensor are
\be
\eta_{+-} = \e_{-+} =-2,\qquad\qquad \eta^{+-}=\e^{+-} =-\hlf.
\ee
Finally, for performing Fourier transforms, we note that
\be
\tilde f(p)\equiv \int d^2x\ e^{-ip\cdot x}f(x),\quad f(x) \equiv \int {d^2p\over(2\pi)^2}\, e^{ip\cdot x} \tilde f(p),\quad \int d^2x\ e^{-ip\cdot x} = (2\pi)^2\dd^2(p).
\ee
This corresponds to the replacements
\be
-i\del_\pm \rightarrow p_\pm,\qquad D_+ \rightarrow \del_{\th^+} +\thbar^+p_+,\qquad \bar D_+ \rightarrow -\del_{\thbar^+} - \th^+ p_+.
\ee

\subsection{Loop integrals with IR cutoffs}\label{Loops}

Here we compile a list of the various loop integrals needed for computing the Wilsonian effective action. We follow the prescription of~\cite{Bilal:2007ne}, which requires us to impose the IR cutoff, $\mu$, on the shifted loop momenta. That is, we use Feynman parameters to combine denominators in the usual manner, and shift the integration variables to put the integrals in the form
\be
\I_{p,q}(M^2) = \int_{\ell_E^2\geq\m^2} {d^2\ell\over(2\pi)^2}\, {\left(\ell^2\right)^p \over \left(\ell^2+M^2\right)^q}. \label{integral}
\ee
After Wick-rotating ($\ell^0\rightarrow i\ell^0_E$) we integrate over the (shifted) Euclidean momenta with $\ell_E^2\geq\mu^2$. When $q\geq p+2$ the integrals are convergent; for example,
\be
\I_{0,n+2}(M^2) = {i\over4\pi} {1\over (n+1)} {1\over\left(\mu^2+M^2\right)^{n+1}},\qquad \forall n\geq0.
\ee
More generally,
\be
\I_{m,n+m+2}(M^2) = {i\over4\pi} {m!\, n! \over (n+m+1)!} \sum_{k=0}^m \left(\begin{array}{c} n+m+1 \\ k\end{array}\right) {M^{2(m-k)} \m^{2k} \over \left(\m^2+M^2\right)^{m+n+1}},\  \forall n,m\geq0.
\ee

When $q=p+1$, the integrals diverge and a UV regulator is required.\footnote{Of course, $\I_{p,q}$ diverges for $q<p+1$ as well, but those cases will not concern us.} Following~\cite{Bilal:2007ne}, we use dimensional reduction: carrying out all $D_+$-algebra in $d=2$, but continuing loop momenta to $d=2-2\e$ in order to evaluate divergent integrals. For $p=0$ we note that
\be
\I_{0,1}(M^2) = {i\over4\pi}\left(\Gamma(\e) -\log\left(\m^2+M^2\over4\pi\right) + O(\e)\right),
\ee
while for $p>0$ we have
\be
\I_{m,m+1}(M^2) ={i\over4\pi}\left(\Gamma(\e) -\log\left(\m^2+M^2\over4\pi\right) + P_m\left(M^2\over\m^2+M^2\right) + O(\e)\right),
\ee
where $P_m(x)$ is an $m$-th order polynomial given by
\be\label{P(x)}
P_m(x) = \sum_{k=1}^m \left(\begin{array}{c} m \\ k\end{array}\right) {(-x)^k\over k}.
\ee
In particular,
\be
P_1(x) = -x, \qquad P_2(x) = \hlf x^2 -2 x.
\ee

\subsection{The action}

In general, we are interested in $N$ charged triplets of chiral superfields $(\S^a,P^a,\G^a)$, where $\G^a$ are fermionic, coupled by a superpotential:
\be
S_J = -\sum_{a=1}^n \left\{ {m_a\over\sqrt{2}} \int d^2x d\th^+\, \G^a \S^a P^a +c.c. \right\}.
\ee
The charges of $(\S^a,P^a,\G^a)$ are only constrained by  gauge invariance of the superpotential: $Q_{\S^a} + Q_{P^a}+Q_{\G^a}=0$. In general there must be other charged fields $\left( \Phi^i,\G^\a \right)$ such that the total gauge anomaly vanishes. These additional fields will not concern us here.

If we restrict to loops of $(P^a,\G^a)$ there is no need to fix the gauge, though we may choose a unitary gauge by setting, say, $\S^1=1$. Next we expand the fields about a generic point in moduli space $(A_0,\S^a_0)$, which together with the expectation values $\Phi^i_0$ ensures that the $V_-$ tadpole vanishes. For simplicity, we will usually include $\S^1_0$ along with the rest of $\S^a_0$, even though $\S^1_0\equiv1$ when we fix unitary gauge. With this in mind, the terms in the action which contain $(P^a,\G^a)$ take the form,
\bea
S[P^a,\G^a] &=& -\hlf\sum_a \int d^2xd^2\th^+\left[i \bar{P}^a e^{2Q_{P^a}(A_0+A)}\nabla_-P^a + \bar{\G}^a e^{2Q_{\G^a}(A_0+A)}\G^a\right] \\
&&-{1\over\sqrt{2}}\sum_a \int d^2x d\th^+\ m_a\left( \S^a_0 +\S^a \right) \G^a P^a  +c.c., \non
\eea
where $\nabla_- = \del_- + Q_{P^a}(\del_-A + iV_-)$.

\subsection{Deriving the free field propagators}\label{deriving}

There is a well known difficulty in deriving the Feynman rules for chiral superfields because they satisfy a differential constraint:
\be \bar D_+ P^a = \bar D_+\G^a=0.\ee
This is similar to the case of electromagnetism, where the field strength satisfies $dF=0$. In this latter case, the well-known solution is to introduce a potential, $A$, such that $F=dA$, which can then be quantized easily. The penalty is, of course, that $A$ is not unique, but is instead a member of an equivalence class: $A\sim A+df$ for any real-valued function $f$. Associated with this redundancy is the fact that the kinetic operator for $A$, denote it $K$, has a kernel: $K(df)=0$. In order to find the propagator for $A$, we must invert $K$ on the orthogonal complement to this kernel. We will follow an analogous approach to derive the $(P^a,\G^a)$ propagator.

We begin by introducing (unconstrained) potential fields $(\Pi^a,G^a)$, such that\footnote{Let $A$ be a superfield with fermion number $F$; then $\overline{D_+ A} = (-)^{F} \Dbar_+ \bar A$ and $\overline{D_+\Dbar_+ A}= -\Dbar_+D_+ \bar A$.}
\be
P^a = \bar D_+ \Pi^a,\qquad \G^a = \bar D_+ G^a,\qquad \bar P^a=-D_+\bar \Pi^a,\qquad \bar\G^a = + D_+\bar G^a.
\ee
Note that the potential fields have the opposite statistics of their corresponding field strengths. The case where $\bar D_+ \G^a = E^a$ is easily adapted to this construction, though we will not pursue it here. These potential fields are not unique, since $\Pi^a\sim \Pi^a + \bar D_+ F_-^a$ for some bosonic superfields $F_-^a$, and similarly for $G^a$. The free part of the $(P^a,\G^a)$ action can be written succinctly as
\be
S_{free} = \int d^2x d^2\th^+\, \left(X^a\right)^\dagger K^{ab} X^b,
\ee
with
\be
X^a = \begin{pmatrix} \Pi^a \\ \bar G^a \end{pmatrix}, \qquad K^{ab} = \hlf\begin{pmatrix} ie^{2Q_{P^a}A_0} D_+\bar D_+\del_- & \sqrt{2}m_a \bar{\S}^a_0 D_+ \\ -\sqrt{2}m_a\S^a_0 \bar D_+ & -e^{2Q_{\G^a}A_0} \bar D_+  D_+ \end{pmatrix}\dd^{ab}.
\ee
Notice that introducing the potential fields $(\Pi^a,G^a)$ allows us to write the $F$-term mass as an integral over all of superspace.

At this point, one should expect that $K^{ab}$ has a non-trivial kernel. Indeed $\textrm{ker}(K^{ab})=\Im(L_+^{ab})$, where
\be
L_+^{ab} = \begin{pmatrix} \bar D_+ & 0 \\ 0 & D_+\end{pmatrix}\dd^{ab}.
\ee
$K^{ab}$ can only be inverted on the orthogonal complement of its kernel. To implement this restriction, consider the following dimension zero operator:
\be
\hat\Pi = {1\over2i\del_+}\begin{pmatrix} D_+\bar D_+ & 0 \\ 0 & \bar D_+ D_+ \end{pmatrix}.
\ee
It is not difficult to verify that $\hat\Pi$ defines a self-adjoint projection operator with $\hat\Pi L_+^{ab}=0$ and $\textrm{ker}(\hat\Pi) = \Im(L_+^{ab})$. Thus $\hat\Pi$ is the projection operator we need in order to invert $K^{ab}$.

The propagator for $X^a$ then satisfies the defining relation:
\be
K^{ab}(z_1) \D^{bc}(z_{12}) = \hat\Pi(z_1)\dd^{ab}\dd^4(z_{12}).
\ee
The desired solution turns out to be
\be
\D^{ab}(z_{12}) = -\begin{pmatrix} e^{-2Q_{P^a}A_0} & {\bar M_a \over\sqrt{2}i\del_+} e^{Q_{\S^a}A_0}D_+ \\ -{M_a \over\sqrt{2}i\del_+} e^{Q_{\S^a}A_0} \bar D_+ & -ie^{-2Q_{\G^a}A_0}\del_- \end{pmatrix} {\dd^4(z_{12}) \over \del_+\del_- + M_a^2}\ \dd^{ab},
\ee
where $M_a \equiv m_a \S^a_0 e^{Q_{\S^a}A_0}$. Equivalently, transforming to momentum space gives
\be
\D^{ab}(p) = -\begin{pmatrix} e^{-2Q_{P^a}A_0} & -{\bar M_a \over\sqrt{2}p_+} e^{Q_{\S^a}A_0}D_+ \\ {M_a \over\sqrt{2}p_+} e^{Q_{\S^a}A_0} \bar D_+ & e^{-2Q_{\G^a}A_0}p_- \end{pmatrix} {\dd^2(\th_{12}) \over p^2 + M_a^2 -i\e}\ \dd^{ab},
\ee
with an appropriate $i\e$ prescription. Note that by $D_+$ we mean $D_{1+}\equiv D_+(p,\th_1)$, although we can easily convert it to $D_{2+} \equiv D_+(-p,\th_2)$ by the relation
\be
D_{1+}\dd^2(\th_{12}) = -D_{2+}\dd^2(\th_{12}).
\ee

This defines the propagator for the potential fields $(\Pi^a,G^a)$. To obtain the propagator for the chiral fields $(P^a,\G^a)$, we should act on the left and right by $\bar D_+$ ($D_+$) for (anti-)chiral legs.

\subsection{Interactions}
The vertices of the theory can be read off directly from the interaction Lagrangian:
\bea
S_{int} &=& \hlf\sum_a \int d^2xd^2\th^+\left[ {i\over2} e^{2Q_{P^a}A_0} \left(e^{2Q_{P^a} A}-1\right) \left(P^a \del_-\bar P^a - \bar P^a\del_- P^a\right)\right. \non\\
&&\left.\frac{}{} + Q_{P^a} e^{2Q_{P^a}A_0} e^{2Q_{P^a} A} V_- |P^a|^2 - \bar\G^ae^{2Q_{\G^a}A_0}\left(e^{2Q_{\G^a} A}-1\right)\G^a\right] \\
&&- {1\over\sqrt{2}}\sum_{a} \int d^2xd\th^+\, m_a \S^a\G^a P^a +c.c. .\non
\eea
Each interaction vertex is accompanied by $i\int d^2\th^+$ except for $F$-term interactions, which only require $i\int d\th^+$ or $i\int d\thbar^+$. To make things more symmetric, we use one of the $\bar D_+$ or $D_+$ operators that act on an internal $(P^a,\G^a)$ propagator to convert the chiral measure into a full $\int d^2\th^+$ integral. We will follow the convention that the $\bar D_+$ or $D_+$ is pulled off from $\G^a$ or $\bar\G^a$.

\subsection{The rules}

We now summarize the rules for computing each term in the quantum effective action:
\begin{enumerate}
\item[(1)] The various propagators are given by $i\D^{ab}(p)$ with
\bea
\left(\begin{array}{cc} \langle P^a_1 \bar P^b_2 \rangle & \langle P^a_1 \G^b_2 \rangle \\ \langle\bar\G^a_1 \bar P^b_2\rangle & \langle \bar\G^a_1 \G^b_2\rangle \end{array}\right) =  -i\begin{pmatrix} e^{-2Q_{P^a}A_0} & -{\bar M_a \over\sqrt{2}p_+} e^{Q_{\S^a}A_0}D_+ \\ {M_a \over\sqrt{2}p_+} e^{Q_{\S^a}A_0} \bar D_+ & e^{-2Q_{\G^a}A_0}p_- \end{pmatrix}   {\dd^2(\th_{12}) \over p^2 + M_a^2 -i\e}\ \dd^{ab}, \non
\eea
where,
\be \langle X_1 Y_2 \rangle \equiv \langle X(p,\th_1) Y(-p,\th_2)\rangle, \qquad M_a\equiv m_a \S^a_0 e^{Q_{\S^a}A_0}, \qquad D_+\equiv D_+(p,\th_1).\ee
\item[(2)] The vertex factors are
\bea
\langle P^a\bar P^b V_- A\ldots A\rangle &=& {i\over2}Q_{P^a} \left(2Q_{P^a}\right)^n e^{2Q_{P^a}A_0}\dd^{ab}, \\
\langle P^a(p) \bar P^b(p') A \ldots A \rangle &=& {i\over4}\left(2Q_{P^a}\right)^n e^{2Q_{P^a}A_0}(p - p')_- \dd^{ab},\\
\langle \bar\G^a\G^b A\ldots A\rangle &=& -{i\over2}\left(2Q_{\G^a}\right)^n e^{2Q_{\G^a}A_0}\dd^{ab}, \\
\langle \S^a \G^b P^c\rangle = {i\over\sqrt{2}}m_a \dd^{ab} \dd^{ac},&&\quad \langle \bar\S^a \bar\G^b \bar P^c\rangle = {i\over\sqrt{2}}m_a \dd^{ab} \dd^{ac},\quad a\neq0.
\eea
Here $n$ denotes the number of $A$ legs. For each internal (anti-)chiral line, include a $\bar{D}_+$ $(D_+)$ acting on the associated propagator, \textit{except} for $\G^a$ ($\bar\G^a$) connected to a $\S\G P$ ($\bar\S\bar\G\bar P$) vertex.
\item[(3)] For each vertex, include an integral $\int d^2\th^+_{vert}$.
\item[(4)] For each loop, include an integral $\int {d^2 p\over(2\pi)^2}$.
\item[(5)] For each loop of Fermi fields $\G$, include a factor of $(-1)$.
\item[(6)] For a term in the effective action with $n$ field insertions, denoted collectively by $X(p_i)$, include an overall
\be
\prod_{i=1}^n\left( \int {d^2 p_i \over(2\pi)^2} X(p_i)\right)(2\pi)^2\dd^2\left(\sum_{i=1}^n p_i\right).
\ee
\item[(7)] Divide by the usual combinatoric factor.
\end{enumerate}
Note that we take all momenta in the vertex factors as incoming.

\subsection{Tips and tricks}\label{tricks}
In computing the effective action, it is always possible (by integration by parts) to move all the $D_+$ and $\bar D_+$ operators so that they act on either external fields or on $\dd^2(\th_{ij})$ of a single propagator. In doing so, it is helpful to convert all of the $D_+$ and $\bar D_+$ operators to be of the same ``type", by using the identity
\be
D_+(p,\th_i)\dd^2(\th_{ij}) = - D_+(-p,\th_j)\dd^2(\th_{ij}).
\ee
Note that this yields the following rule for converting products of $D_+$ and $\bar D_+$:
\be
\bar D_{i+} D_{i+}\dd^2(\th_{ij}) = -\bar D_{i+} D_{j+}\dd^2(\th_{ij}) = + D_{j+} \bar D_{i+}\dd^2(\th_{ij}) = - D_{j+} \bar D_{j+}\dd^2(\th_{ij}).
\ee
So for a product, the order is reversed and an overall sign is introduced. After these manipulations are performed, the resulting expression can be further simplified using the identities:
\bea
\dd^2(\th_{ij}) D_{i+} \bar D_{i+} \dd^2(\th_{ij}) = + \dd^2(\th_{ij}),&\qquad& \dd^2(\th_{ij}) \bar D_{i+} D_{i+} \dd^2(\th_{ij}) = - \dd^2(\th_{ij}), \\
\dd^2(\th_{ij}) D_{i+}\dd^2(\th_{ij}) = 0, &\qquad& \dd^2(\th_{ij}) \bar D_{i+} \dd^2(\th_{ij}) =0.
\eea
In the end, one is left with enough ``bare" $\dd^2(\th_{ij})$ to trivially carry all but one of the fermionic integrals. In this way, every term in the effective action can be reduced to a single $\int d^2\th^+$ and is therefore local in the $\th^+$ coordinates, even though the 1PI effective action may be non-local in $x$.

Computing a one point function requires some care since it can involve derivatives of $\dd^2(\th_{11})\equiv0$. We define these propagators, from one point to itself, as a limit of a standard propagator between two points. Thus,
\be
D_{1+}\bar D_{1+} \dd^2(\th_{11}) \equiv \lim_{2\rightarrow1} D_{1+}\bar D_{1+} \dd^2(\th_{12}) = 1.
\ee

\section{Feynman Rules for Other Superfields}\label{others}

Although the effective action, $W$, is determined solely by integrating out $(P^a,\G^a)$, for consistency we will also carry out the path integral over the high-energy modes of the other light fields. As one might expect, we will see that this leads only to a renormalization of the dimensionless FI parameter.

\subsection{Light chiral superfields}

The action for the light chiral superfields $\Phi^i$ and $\G^\a$ is identical to that of $P^a$ and $\G^a$, except there are no superpotential couplings that give rise to mass terms $(m=0)$. The derivation of the propagator is nearly identical to the discussion in section~\ref{deriving}. The result is
\be
\langle \Phi^i_1 \bar\Phi^j_2 \rangle = -i e^{-2Q_{\Phi^i} A_0}\, {  \dd^2(\th_{12}) \over p^2-i\epsilon}\,\dd^{ij},\qquad \langle \G^\a_1 \bar\G^\beta_2 \rangle = -i e^{-2Q_{\G^\a} A_0}\, {p_-\dd^2(\th_{12}) \over p^2-i\epsilon}\,\dd^{\a\beta},
\ee
and vertices
\bea
\langle \Phi^i\bar \Phi^j V_- A\ldots A\rangle &=& {i\over2}Q_{\Phi^i} \left(2Q_{\Phi^i}\right)^n e^{2Q_{\Phi^i}A_0}\dd^{ij}, \\
\langle \Phi^i(p) \bar \Phi^j(p') A \ldots A \rangle &=& {i\over4}\left(2Q_{\Phi^i}\right)^n e^{2Q_{\Phi^i}A_0}(p - p')_- \dd^{ij},\\
\langle \bar\G^\a\G^\beta A\ldots A\rangle &=& -{i\over2}\left(2Q_{\G^\a}\right)^n e^{2Q_{\G^\a}A_0}\dd^{\a\beta}.
\eea
Since there are no $F$-term interactions, we always act on the $\Phi^i$ and $\G^\a$ propagators with $-\bar D_{1+}D_{2+}\sim \bar D_{1+}D_{1+}$.

The computation of $f_V$ from a loop of $\Phi^i$ fields is exactly the same as~\C{fV}, except we set $m=0$. This leaves
\be
f_V = -{Q_{\Phi^i}\over8\pi}\log\left(\m^2\over\m_r^2\right).
\ee
This correction is field-independent, and so will not affect $W$, though it is important for understanding the beta function for the FI parameter $r$. The light chiral fields do not contribute to $f_A$ or $f_\S$ either, because they do not have any classical coupling to $\S$.

\subsection{The Higgs and vector multiplets}

Because of the spontaneous symmetry breaking that occurs, it is best to examine the Higgs and gauge sectors simultaneously. For simplicity, we will only consider a single Higgs field, $\S$. Their combined action, expanded about $(A_0,\S_0)$, is
\bea
S[\S,A,V_-] &=& \int d^2xd^2\th^+\left[-{i\over4}\left(\bar\S_0+\bar\S\right)e^{2Q_\S(A_0+A)}\nabla_-\left(\S_0+\S\right) +c.c.\right] \non\\
&&- {1\over8e^2}\int d^2xd^2\th^+\left[\bar\Upsilon\Upsilon +{1\over\xi}\bar F F\right] ,\label{higgsvector}
\eea
where we have included a gauge-fixing term $|F|^2$. $F$ must be a fermionic function; the choice $F=D_+(\del_-A+iV_-)$ leads to $-{1\over2\xi}(\del\cdot A)^2$ in the component action. However, the non-zero value of $\S_0$ gives a mixing between $\S$ and $(A,V_-)$ in the quadratic action.

What we need is a $(0,2)$ version of $R_\xi$ gauge, where the propagators are diagonal. Supersymmetric $R_\xi$ gauges for four-dimensional gauge theories were introduced in~\cite{Ovrut:1981wa}. It turns out that the correct choice for our purposes is
\be
F = D_+(\del_-A +iV_-) -i\xi {2M_A^2S\over Q_\S \S_0}, \label{gaugefix}
\ee
where
\be
M_A^2 = 2e^2 Q_\S^2 |\S_0|^2 e^{2Q_\S A_0},\qquad \textrm{and}\qquad\bar D_+S=\S_0+\S.
\ee
It will be important in the next section to notice that the potential field $S$ is defined for the total field $\S'\equiv\S_0+\S$, not just the shifted part $\S$.

With this choice of gauge-fixing, the quadratic part of the action becomes
\bea
S_{quad}[\S,A,V_-] &=& \hlf\int d^2xd^2\th^+ e^{2Q_\S A_0} \bar S\left(iD_+\bar D_+\del_- -2\xi M_A^2\right)S \\
&-&{1\over 2\xi e^2}\int d^2xd^2\th^+ \left[ A\left(\del^2-\xi M_A^2\right)V_- +\left({\xi-1\over4}\right)\bar\Upsilon\Upsilon\right].\non
\eea
For the $\S$ field, we find the propagator
\be
\langle \S_1 \bar\S_2 \rangle = ie^{-2Q_\S A_0}{D_+\bar D_+\dd^2(\th_{12}) \over 2p_+ \left(p^2+\xi M_A^2\right)} \sim -i e^{-2Q_\S A_0} {\dd^2(\th_{12})\over p^2 + \xi M_A^2 -i\e},
\ee
where in the last step we have used the fact that every internal $\S\bar\S$ propagator will be acted on by $-\bar D_{1+}D_{2+}$ to write an equivalent propagator with the pole at $p_+=0$ removed.\footnote{Note that $-\bar D_{1+} D_{2+}$ still acts on this equivalent form of the propagator.} We can recover unitary gauge by sending $\xi\rightarrow\infty$. In this limit, $\S$ does not propagate and it is effectively eliminated from the spectrum, as expected. However for the vector multiplet, $\xi=1$ is a much more natural choice since the kinetic terms simplify tremendously. This is a natural generalization of Feynman-'t Hooft gauge. We will henceforth work only in $\xi=1$ gauge, where the vector field propagator reduces to
\be
\langle A_1 V_2 \rangle = (2ie^2){\dd^2(\th_{12})\over p^2 + M_A^2 -i\e}.
\ee
The coefficient $2i=-i\eta_{+-}$ is exactly as one would expect for $\langle A_+A_-\rangle$ in Feynman gauge.

Interaction vertices can be read off directly from
\bea
S_{int} &=& \int d^2xd^2\th^+ \left[-{i\over4}e^{2Q_\S A_0}\left(e^{2Q_\S A}-1\right)\left(\bar\S\del_-\S - \S\del_-\bar\S\right) + {Q_\S\over2}e^{2Q_\S A_0}|\S|^2 e^{2Q_\S A} V_-\right.\non\\
&&\qquad\qquad+{Q_\S\over2}e^{2Q_\S A_0}\left(e^{2Q_\S A}-1\right)\left[V_-\left(\bar\S_0\S + \S_0\bar\S\right) +i\del_-A\left(\bar\S_0\S -\S_0\bar\S\right)\right] \non\\
&&\left.\qquad\qquad+{Q_\S\over2}|\S_0|^2e^{2Q_\S A_0} \left(e^{2Q_\S A}-2Q_\S A -1\right)V_- \right].
\eea
The terms in the first line are exactly the same as in the $\S_0=0$ case, while the terms in the second line are related to the first by replacing $\S$ or $\bar\S$ by its vev. Finally the terms of the third line, which come from setting $|\S|^2$ to its vev, give rise to a set of couplings between $V_-$ and $A$ only. In particular, the vertices are
\bea
\langle \S\bar \S V_- A\ldots A\rangle &=& {i\over2}Q_{\S} \left(2Q_{\S}\right)^n e^{2Q_{\S}A_0}, \label{1}\\
\langle \S(p) \bar \S(p') A \ldots A \rangle &=& {i\over4}\left(2Q_{\S}\right)^n e^{2Q_{\S}A_0}(p - p')_-,\label{2}\\
\langle \S V_- A\ldots A\rangle &=& {i\over2}Q_{\S} \left(2Q_{\S}\right)^n e^{2Q_{\S}A_0}\bar\S_0, \label{3}\\
\langle \bar \S V_- A\ldots A\rangle &=& {i\over2}Q_{\S} \left(2Q_{\S}\right)^n e^{2Q_{\S}A_0}\S_0, \label{4}\\
\langle \S(p) A \ldots A \rangle &=& {i\over4}\left(2Q_{\S}\right)^n e^{2Q_{\S}A_0}\bar\S_0 p_-,\label{5}\\
\langle \bar \S(p') A \ldots A \rangle &=& -{i\over4}\left(2Q_{\S}\right)^n e^{2Q_{\S}A_0}\S_0 p'_-,\label{7}\\
\langle V_- A\ldots A\rangle &=& {i\over2}Q_{\S} \left(2Q_{\S}\right)^n e^{2Q_{\S}A_0}|\S_0|^2.\label{self}
\eea
In~\C{1}\ $n\geq0$, while in~\C{2}-\C{4}\ we require $n\geq1$, and in~\C{5}-\C{self}\ $n\geq2$.

An important point to note in computing loops with these Feynman rules is that the $AV_-$ propagator is \textit{not} acted on by $\Dbar_+ D_+$, and so these propagators contribute ``bare" $\dd^2(\th_{12})$ to the loop integrals. To get a non-zero result, a loop with an internal vector line must also contain a line of chiral fields, otherwise it will be proportional to $\left(\dd^2(\th_{12})\right)^2=0$ in the case of two-point vertices, or $\dd^2(\th_{11})=0$ in the case of a single vertex.

\begin{figure}[h]
\centering
\subfloat[][]{
\includegraphics[width=0.475\textwidth]{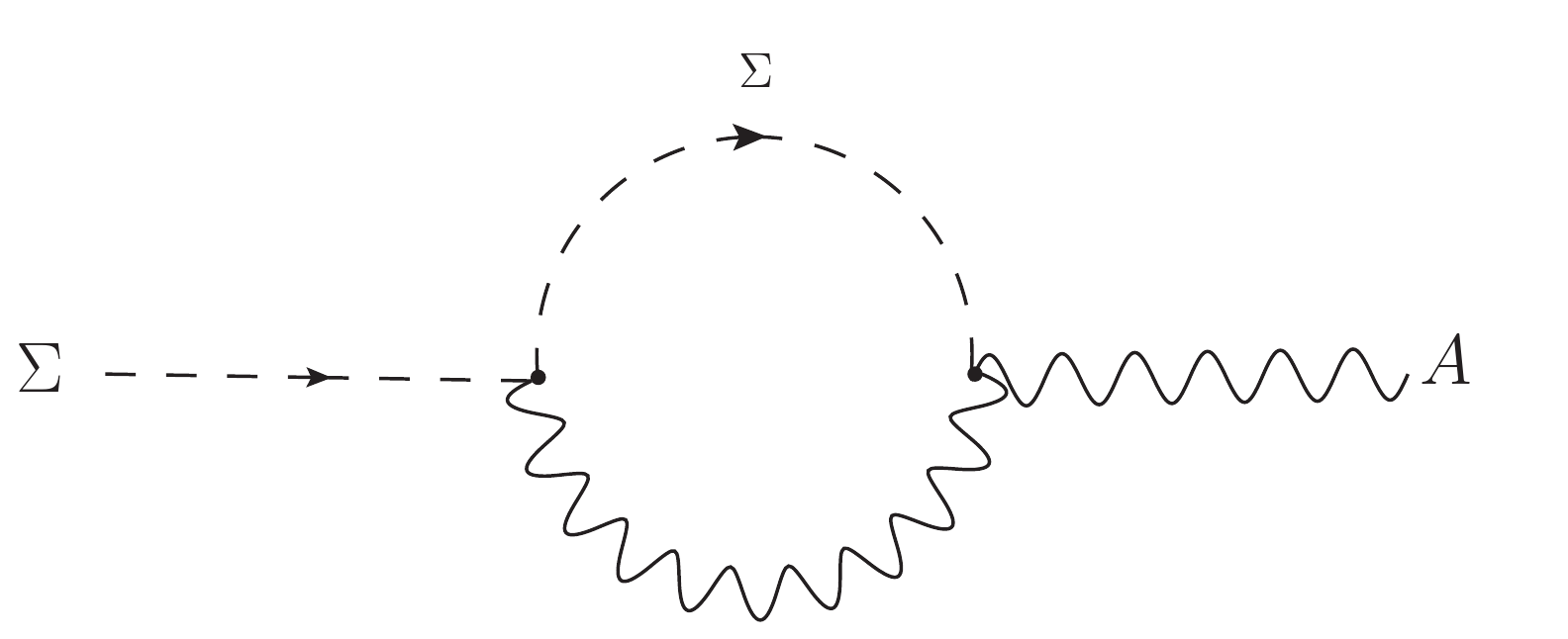}
\label{ASigma2}}
\
\subfloat[][]{
\includegraphics[width=0.475\textwidth]{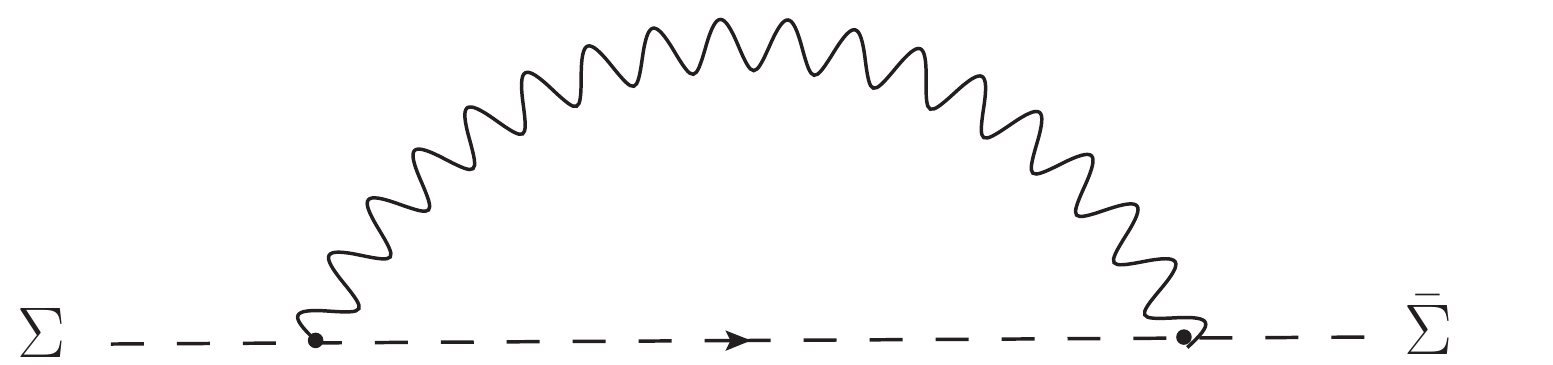}
\label{SSbar2}}
\caption{\textit{The remaining diagrams that contribute to the effective action in a general gauge.}}
\label{sigmaloops}
\end{figure}

Integrating down to a scale $\m\gg M_A$ where the gauge theory is still perturbative, the contribution to $f_V$ coming from a $\S$ loop is
\be
f_V =-{Q_{\S}\over8\pi} \log\left(\m^2+M_A^2 \over \m_r^2\right) =-{Q_{\S}\over8\pi} \log\left(\m^2\over \m_r^2\right) +\ldots.
\ee
This is a field-independent renormalization of $t$, which we can ignore.  The only diagrams which could contribute to $f_A$ and $f_\S$, and do not vanish identically, are shown in figure~\ref{sigmaloops}.
However, it is easy to check that these diagrams are suppressed by $(M_A^2/\m^2)$, and can therefore be neglected.

\subsection{Ghosts}

Even though we are dealing with an abelian gauge theory, fixing $R_\xi$ gauge leads to $F$-term interactions between both the chiral and anti-chiral ghosts, and the Higgs field $\S$. However, the ghost and matter sectors still turn out to decouple from one another; the ghosts could only possibly renormalize  $F$-terms, but we know that cannot happen. So we will find that ghosts cannot modify the effective action in this theory, despite coupling to the Higgs.

To demonstrate this claim we begin from the gauge fixed action, which naturally splits into three pieces:
\be
S = S_0[X] + S_F[\Om,F(X)] + S_{gh}[B,C,F(X)],\label{Stot}
\ee
where $S_0$ is the GLSM action, and we denote all the gauge and matter fields collectively by $X$. The second piece,
\be
S_F = \hlf\int d^2xd^2\th^+\left(\Om F(X) + \bar F(X)\bar\Om -4e^2\xi\,\bar\Om\Om\right),
\ee
is the gauge-fixing term with $\Om$ an auxiliary chiral Fermi field.\footnote{That $\Om$ is chiral follows from the fact that $F(X)$ is (essentially) anti-chiral. This is certainly true when $\xi=0$, and by exploiting the ``gauge symmetry" $S\sim S +\bar D_+T$ we can force $F$ to be anti-chiral for finite $\xi$ as well.} When $\xi=0$, $\Om$ acts as a Lagrange multiplier enforcing our gauge condition: $F(X)=0$. For non-zero $\xi$ we can solve for $\Om$ to recover the standard Gaussian average over gauge choices, as in~\C{higgsvector}. Finally, the third piece of~\C{Stot} gives the ghost action
\be
S_{gh} = -\hlf\int d^2xd^2\th^+\left[B\left(\dd_\La F\right)C - \bar B\left(\dd_{\bar\La}\bar F\right)\right],
\ee
where $B$ is a chiral commuting left-moving Fermi supermultiplet\footnote{Not to be confused with the $B$-field of a target space sigma model. Since we only discuss ghosts in this section, we hope the reader will forgive out abuse of notation.}, and $C$ is a chiral anti-commuting scalar supermultiplet:
\be
B = \beta + \th^+ b -i\th^+\thbar^+\del_+\beta,\qquad C = c + \th^+\gamma -i\th^+\thbar^+\del_+ c.
\ee
If we denote the gauge transformation of $X$ by $X^\La = X +\La\dd_\La X + \ldots,$  then
\be
\dd_\La F\equiv \left.{\dd F(X^\La) \over \dd\La}\right|_{\La=0}.
\ee
Rather than derive~\C{Stot} directly by a  $(0,2)$ version of the standard Faddeev-Poppov procedure, which can be done but has its own subtleties stemming from the fermionic nature of the gauge-fixing condition~\C{gaugefix}, we will instead offer the evidence that~\C{Stot}\ is invariant under the super-BRST symmentry:
\bea
\dd X = \e C\dd_\La X + \bar\e\bar C \dd_{\bar\La} X, &\qquad& \dd B = \e\Om,\qquad \dd \bar B = \bar\e\bar\Om, \\
\dd \Om = 0,\qquad \dd \bar \Om = 0,&\qquad& \dd C = 0,\qquad \dd \bar C = 0.
\eea
Verifying this symmetry is particularly straightforward, since $\dd C$ vanishes for an abelian gauge group.

For the gauge-fixing function~\C{gaugefix}, the ghost action is given by
\bea
S_{gh}[B,C,\S] &=& {i\over2}\int d^2xd^2\th^+\ B D_+\del_-C - \xi M_A^2 \int d^2xd\th^+ \left(1+{\S\over\S_0}\right) BC \label{ghost} \\
&&+  {i\over2}\int d^2xd^2\th^+\ \bar B \bar D_+\del_-\bar C - \xi M_A^2 \int d^2xd\bar\th^+ \left(1+{\bar\S\over\bar\S_0}\right) \bar B\bar C, \non
\eea
where we have used part of the Grassmann measure to convert the interaction with $S$ into an $F$-term interaction with $\S'=\S_0+\S$. We should stress that it is the field $\S'$ which transforms linearly under the gauge symmetry: $\S'^\La=e^{iQ_\S\La}\S'$. Since the $(B,C)$ and $(\bar B,\bar C)$ sectors decouple, it is clear that they cannot renormalize the effective action which must be a $D$-term. It should be pointed out that if $F$ were chosen so that $\dd_{\bar\La} F\neq0$ then the two sectors would be coupled and could  combine into a $D$-term.

To see this non-renormalization in greater detail, we write $C=\bar D_+\gamma$, but leave $B$ alone, giving the Feynman rules:
\bea
\langle C_1 B_2 \rangle = \langle \bar C_1 \bar B_2 \rangle = {i\dd^2(\th_{12}) \over p^2+\xi M_A^2 -i\e}, \quad \langle \S BC \rangle = -i{\xi M_A^2\over\S_0},\quad \langle \bar\S\bar B\bar C\rangle = -i {\xi M_A^2\over\bar\S_0}.
\eea
Note that internal $CB$ propagators are not acted on by  $D_+$ or $\bar D_+$. The reason is that the $\bar D_{+}$, which converts $\gamma$ to $C$ and would usually act on a $CB$ propagator, gets absorbed by the vertex factor in order to write the interaction as a $D$-term. It is then easy to see that there are no possible diagrams with these interactions that renormalize the effective action.

\section{The Full Quadratic Effective Action} \label{quad}

In this appendix, we will compute the full momentum-dependence of the effective action to quadratic order in gauge fields. Aside from general interest, there are several reasons we consider this a useful exercise. First, we wish to confirm the coefficient $-\cA/8\pi$ of~\C{Gammafinal}\ by directly computing $\langle A V_-\rangle$ without relying on the background field trick. Second, we want to demonstrate that when computing the Wilsonian effective action at a scale $\m$, the non-local term in the anomaly is smoothed out, as we claim in~\C{smoothing}. Finally, along the way we will deepen our understanding of how the local counter-term~\C{counterterm}\ arises in perturbation theory.

Let us begin by considering the light chiral fields $(\Phi^i,\G^\a)$. There are three diagrams
\begin{figure}[h]
\centering
\subfloat[][]{
\includegraphics[width=0.475\textwidth]{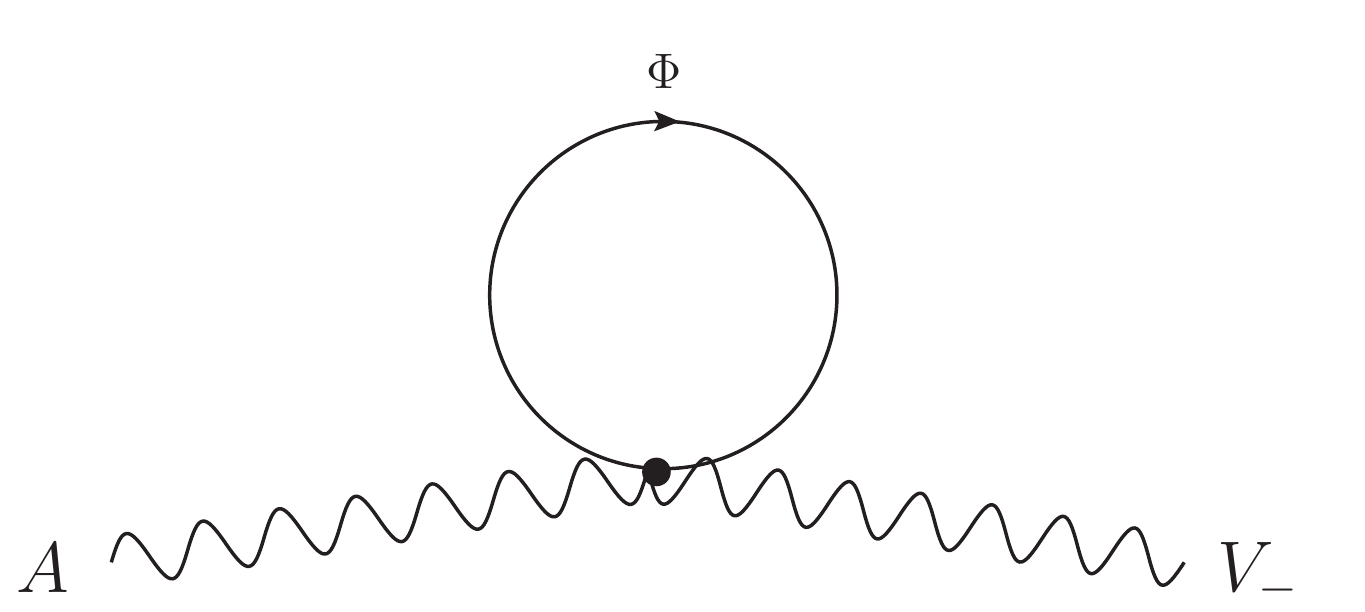}
\label{AV1}}
\
\subfloat[][]{
\includegraphics[width=0.475\textwidth]{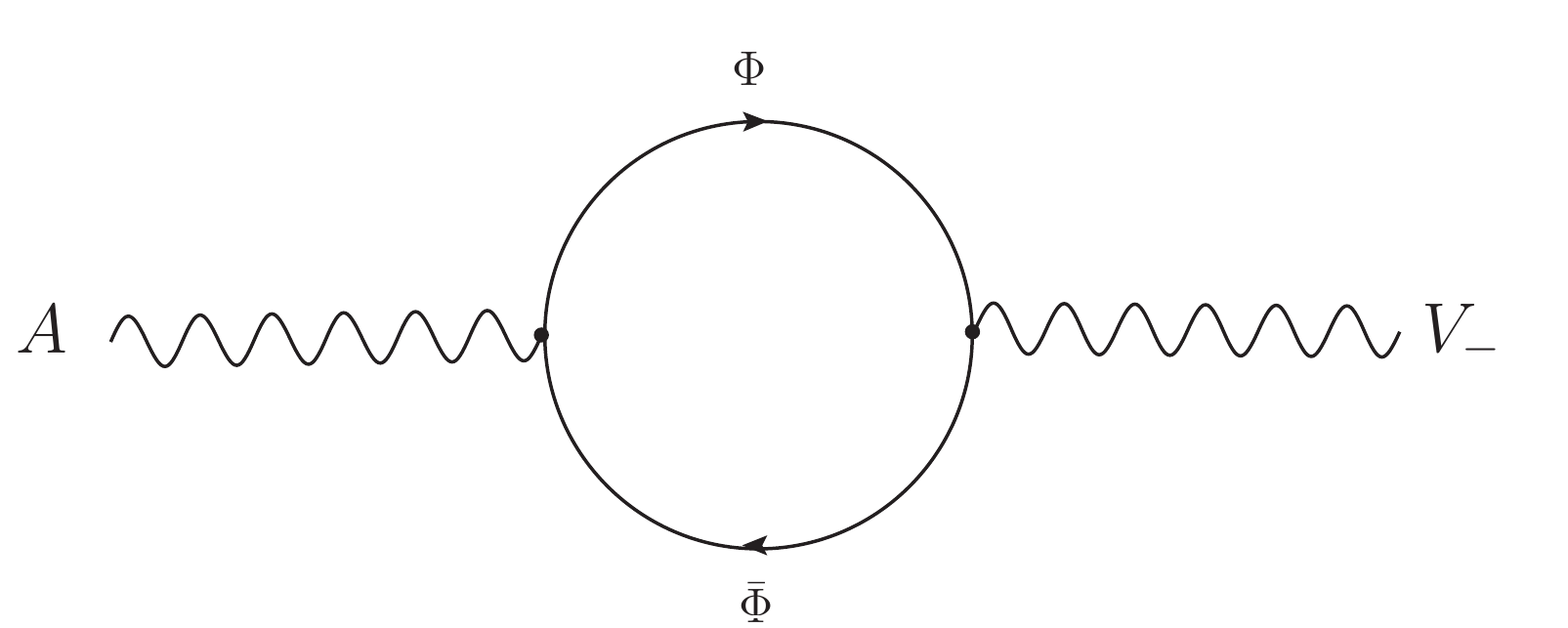}
\label{AV2}}
\\
\subfloat[][]{
\includegraphics[width=0.475\textwidth]{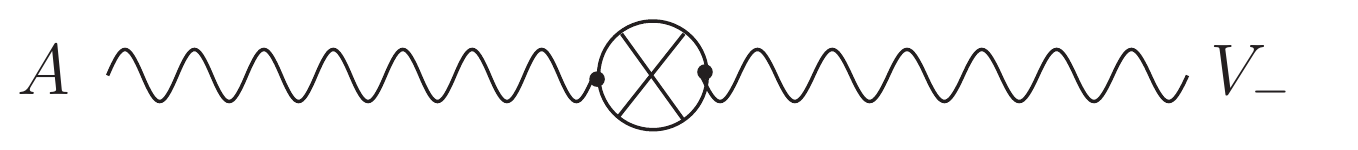}
\label{CT}}
\caption{\textit{The three loops contributions to $\langle AV_-\rangle$ coming loops of massless chiral fields.}}
\label{AVloops}
\end{figure}
that contribute to $\langle A V_- \rangle$, shown in figure~\ref{AVloops}, where figure~\ref{CT}\ corresponds to the counter-term~\C{counterterm}:
\be
W_{ct} = {1\over8\pi}\left(\sum_i Q_i^2 + \sum_\a Q_\a^2\right) \int d^2xd^2\th\ A V_- .\label{counterterm2}
\ee
To understand the origin of this counter-term, consider the computation of $\langle A_+ A_-\rangle$ in components. The integrand for the fermionic loop contains terms proportional to
\be
\sum_i Q_i^2\, \Tr\left[P_+\gamma^+ p\sl P_+ \gamma^- p\sl\right] + \sum_\a Q_\a^2\, \Tr\left[P_-\gamma^+ p\sl P_- \gamma^- p\sl\right] \equiv0,
\ee
where $P_\pm =\hlf(1\pm\g^5)$ and $\g^\pm =\hlf(\g^0\pm\g^1)$. The reason these terms vanish identically is that $P_\pm\g^\mp=0$. However if we consider the more general amplitude $\langle A_\m A_\n\rangle$ and work in $d=2-2\e$,  we find
\bea
&& \sum_i Q_i^2\, \Tr\left[P_+\gamma^\m p\sl P_+ \gamma^\n p\sl\right] + \sum_\a Q_\a^2\, \Tr\left[P_-\gamma^\m p\sl P_- \gamma^\n p\sl\right] \\
&&= \left(\sum_i Q_i^2 + \sum_\a Q_\a^2\right)  \left(d-2\over d\right)\eta^{\m\n} p^2 .\non
\eea
These terms appear inside divergent integrals so we end up with a net finite result for $\langle A_+ A_-\rangle$. This discrepancy arises when we carry out the gamma-matrix algebra in $d=2$ as opposed to $d=2-2\e$. This is the basic distinction between dimensional reduction and dimensional regularization.

It is well known that neither regularization scheme preserves supersymmetry, though in dimensional reduction the breakdown is only believed to occur at high loop order, at least for four-dimensional theories. Here we find a discrepancy already at one-loop that we can trace back to the inherently chiral structure of $(0,2)$ superspace, which cannot be continued away from $d=2$. We have already motivated the necessity of this counter-term in section~\ref{ss:partf}, and now we have pinpointed its origin.

The loops appearing in figures~\ref{AV1}\ and~\ref{AV2}\ are separately divergent, but together they yield the finite result
\be
\int {d^2q\over(2\pi)^2}d^2\th^+\, A(q)q^2V_-(-q) \left[\sum_i{Q_i^2\over4}\int_0^1 dx\, \I_{0,2}\left(x(1-x)q^2\right) \right],\label{remainder}
\ee
which requires use of the identities
\be
\int_0^1  {(2x-1)\,dx \over M^2+x(1-x)q^2} =0, \label{identity2}
\ee
and
\be\label{identity1}
\int_0^1 dx\, \log\left(M^2 +x(1-x)q^2 \over M^2\right) = \int_0^1 dx\, {x(2x-1)q^2\over M^2+x(1-x)q^2}.
\ee

\begin{figure}[h]
\centering
\subfloat[][]{
\includegraphics[width=0.475\textwidth]{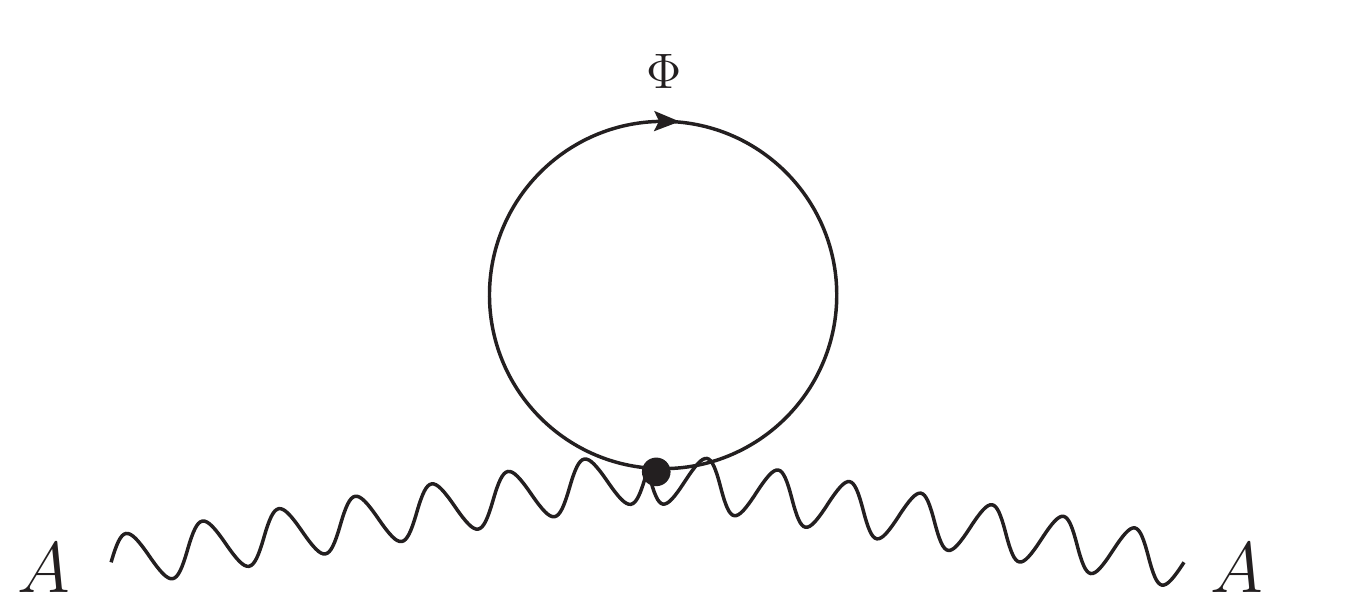}
\label{AA1}}
\
\subfloat[][]{
\includegraphics[width=0.475\textwidth]{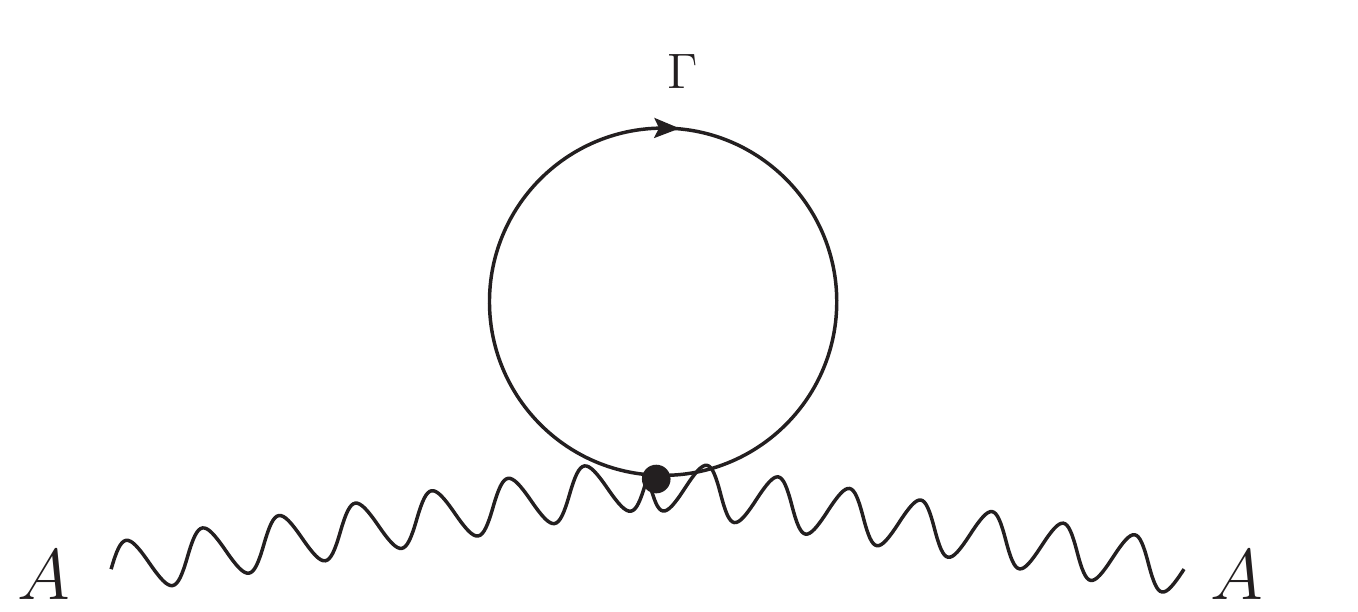}
\label{AA2}}
\\
\subfloat[][]{
\includegraphics[width=0.475\textwidth]{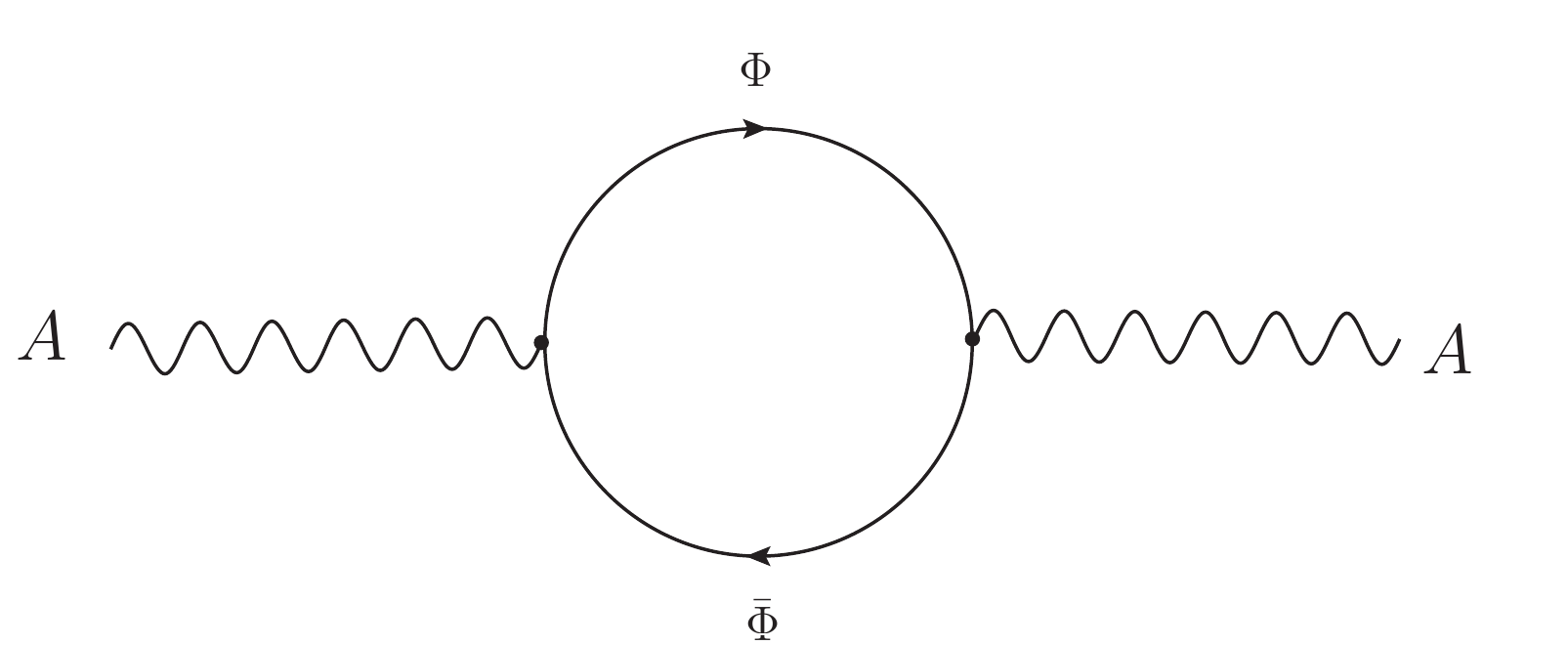}
\label{AA3}}
\
\subfloat[][]{
\includegraphics[width=0.475\textwidth]{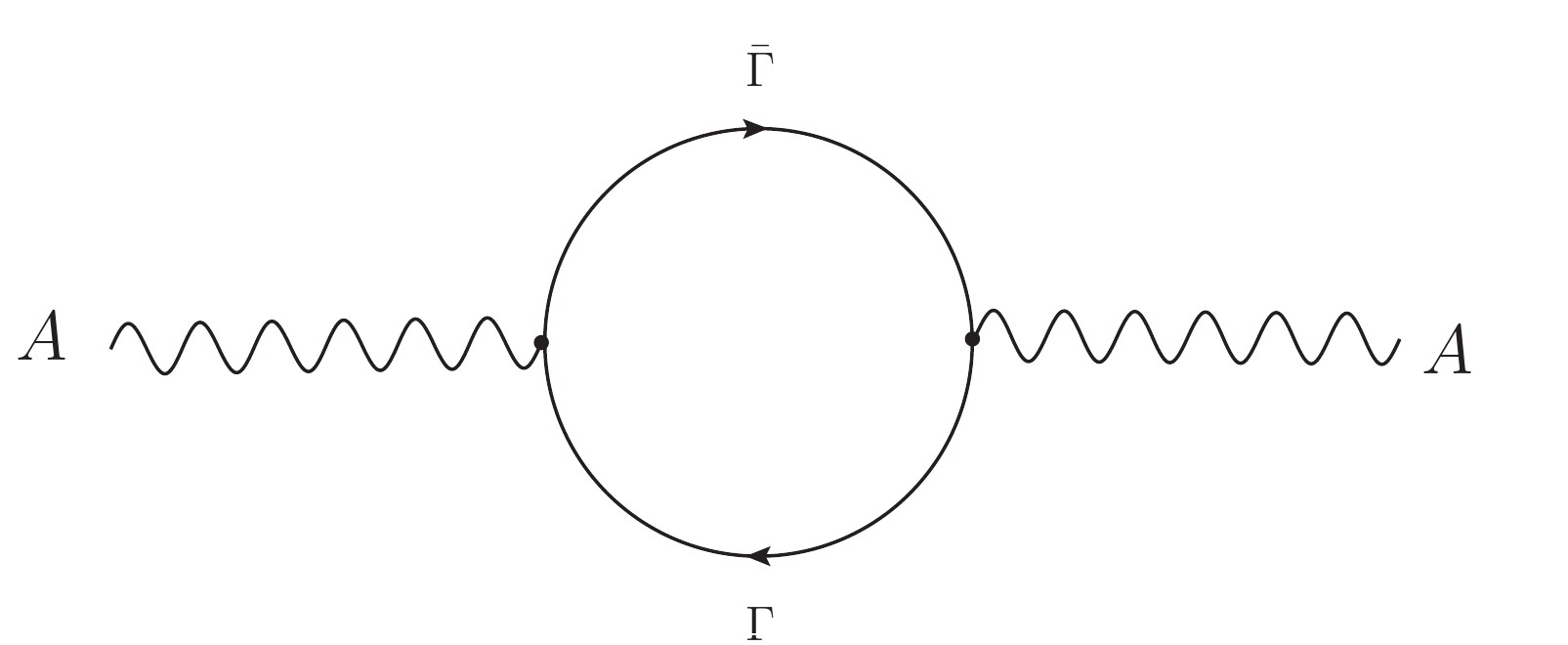}
\label{AA4}}
\caption{\textit{The diagrams contributing to $\langle AA\rangle$.}}
\label{AAloops}
\end{figure}

\begin{figure}[h]
\centering
\includegraphics[width=0.5\textwidth]{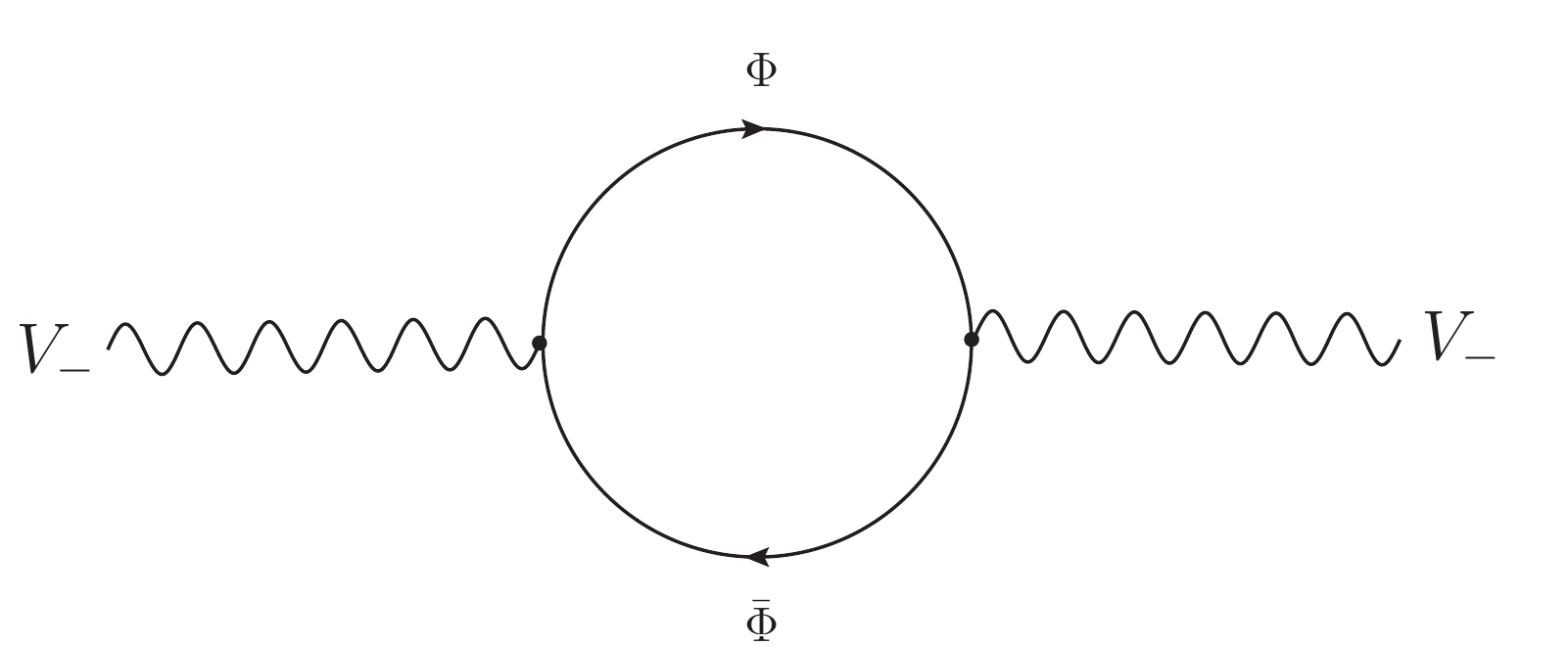}
\caption{\textit{The lone contribution to $\langle V_-V_-\rangle$.}}
\label{VV}
\end{figure}

The $\langle A A \rangle$ correlator receives contributions from the four diagrams of figure~\ref{AAloops},
though diagrams~\subref{AA1}\ and~\subref{AA2}\ are easily shown to vanish. Only the single diagram of figure~\ref{VV}\
contributes to $\langle V_- V_- \rangle$.  Together, these eight diagrams yield the following quadratic effective action:
\bea
W_{quad} &=& {1\over8\pi}\int d^2xd^2\th^+ \int_0^1 dx \left\{\sum_i{Q_i^2\over4}\bar\Upsilon\left(1\over\m^2-x(1-x)\del^2\right)\Upsilon \right. \\
&&\qquad \qquad\qquad \qquad \left. + \cA\left[\bar D_+\del_-A\left(x(1-x) \over \m^2 -x(1-x)\del^2 \right) D_+\del_-A - A V_-\right]\right\} \non,
\eea
where $\cA = \sum_i Q_i^2 - \sum_\a Q_\a^2$. Notice that when $\m=0$, $W_{quad}$ has a gauge-invariant term and a non-invariant term that produces the correct $(0,2)$ gauge anomaly. This would not have worked had we not included the counter-term~\C{counterterm2}. The non-locality of $W_{quad}$ that emerges at $\m=0$ signals that we have integrated out massless degrees of freedom. For $\m>0$ we see that the effective action has a perfectly local expansion in ${\del^2\over \m^2}$, and the non-local term in the anomaly gets smoothed out, as claimed in~\C{smoothing}.

\begin{figure}[t]
\centering
\subfloat[][]{
\includegraphics[width=0.475\textwidth]{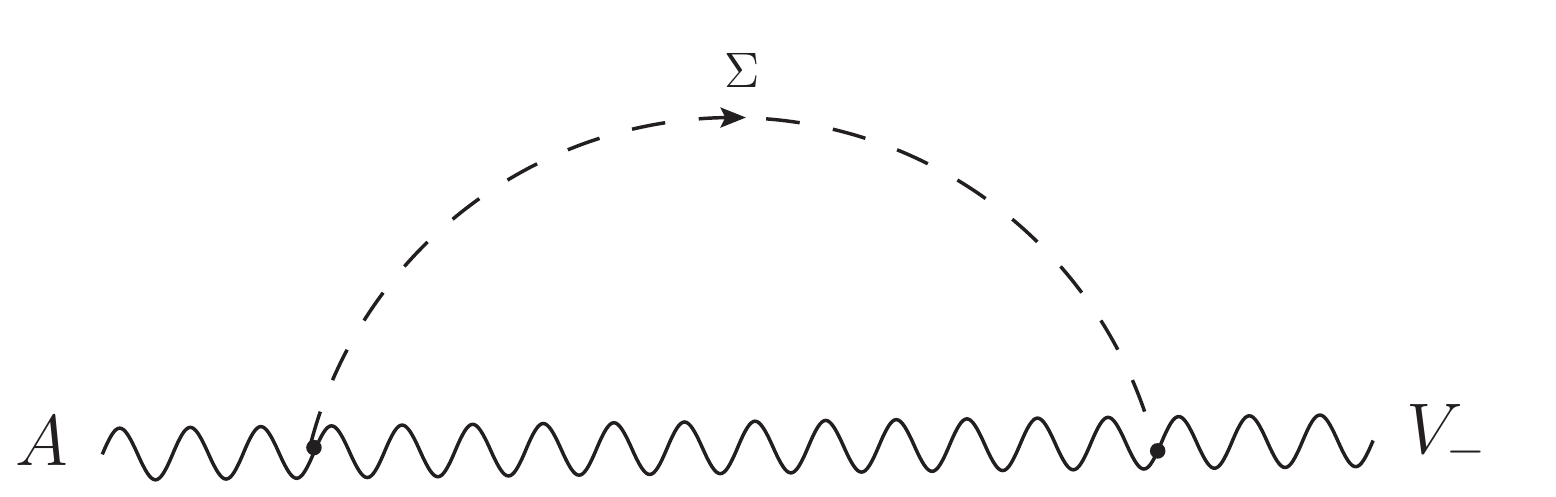}
\label{AV4}}
\
\subfloat[][]{
\includegraphics[width=0.475\textwidth]{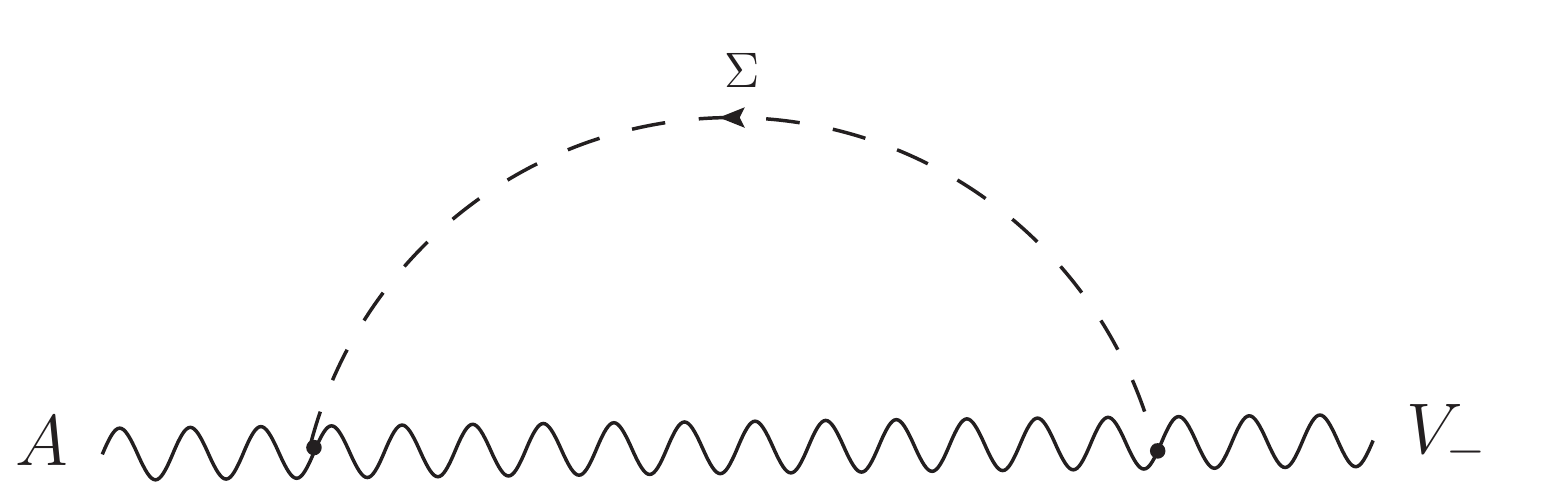}
\label{AV5}}
\caption{\textit{Novel contributions to $\langle AV_-\rangle$ from $\S$ and gauge multiplet loops.}}
\label{AVsigmaloops}
\end{figure}

Despite the self-interactions of the gauge multiplets, listed in~\C{self}, these couplings  do not give rise to any corrections to $W_{quad}$. Loops of $\S$ proceed exactly as in the case of the massless $\Phi$ fields discussed above, with two exceptions. The non-zero mass of $\S$ means we should replace $\m^2$ everywhere above with $\m^2+M_A^2$, and there are two additional diagrams to consider, shown in figure~\ref{AVsigmaloops}.
These are easy enough to work out; they give,
\be
\int {d^2q\over(2\pi)^2}d^2\th^+\, A(q)V_-(-q) \left[2M_A^2 Q_\S^2\int_0^1 dx\, \I_{0,2}\left(\D_A\right) \right],
\ee
where $\D_A = M_A^2 + x(1-x)q^2$. In the limit $M_A^2\ll \m^2$ we can neglect the mass of $\S$, so these new contributions vanish and $\S$ behaves like any of the massless chiral fields discussed above. Thus we can just replace $\sum_i Q_i^2$ with $Q_\S^2 + \sum_i Q_i^2$ in the expressions above.

\begin{figure}[h]
\centering
\subfloat[][]{
\includegraphics[width=0.475\textwidth]{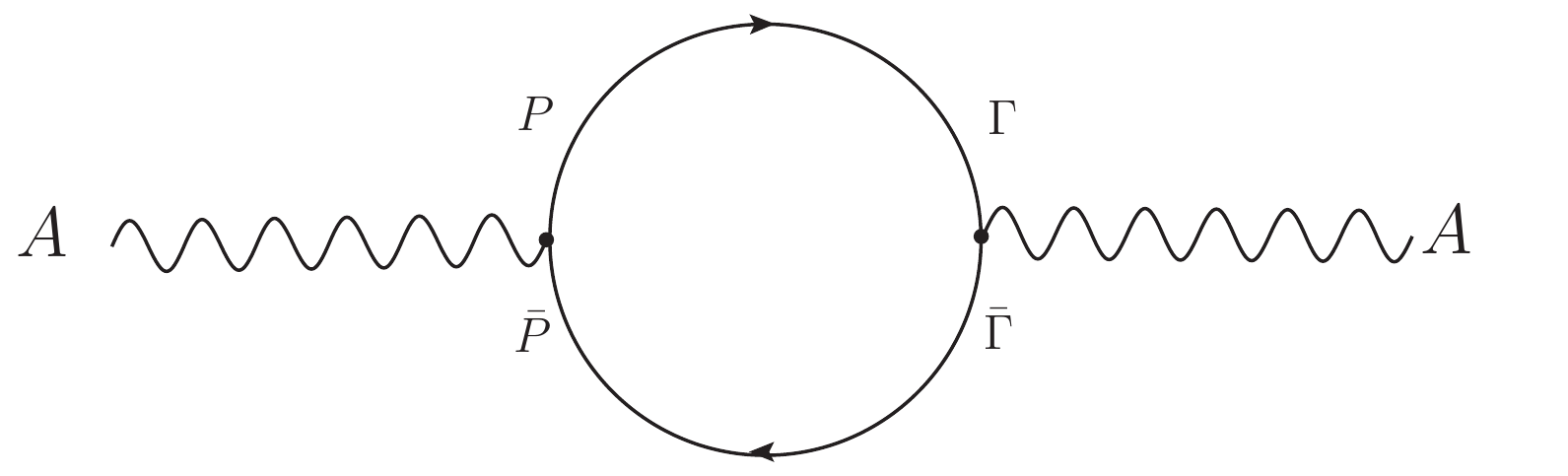}
\label{AA5}}
\
\subfloat[][]{
\includegraphics[width=0.475\textwidth]{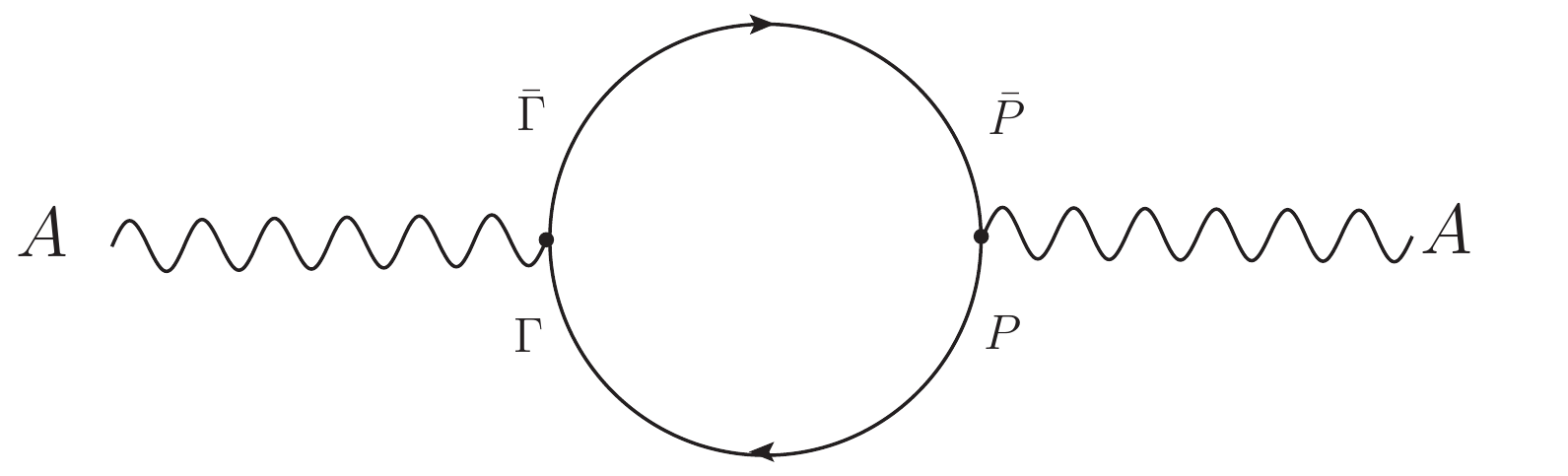}
\label{AA6}}
\
\subfloat[][]{
\includegraphics[width=0.475\textwidth]{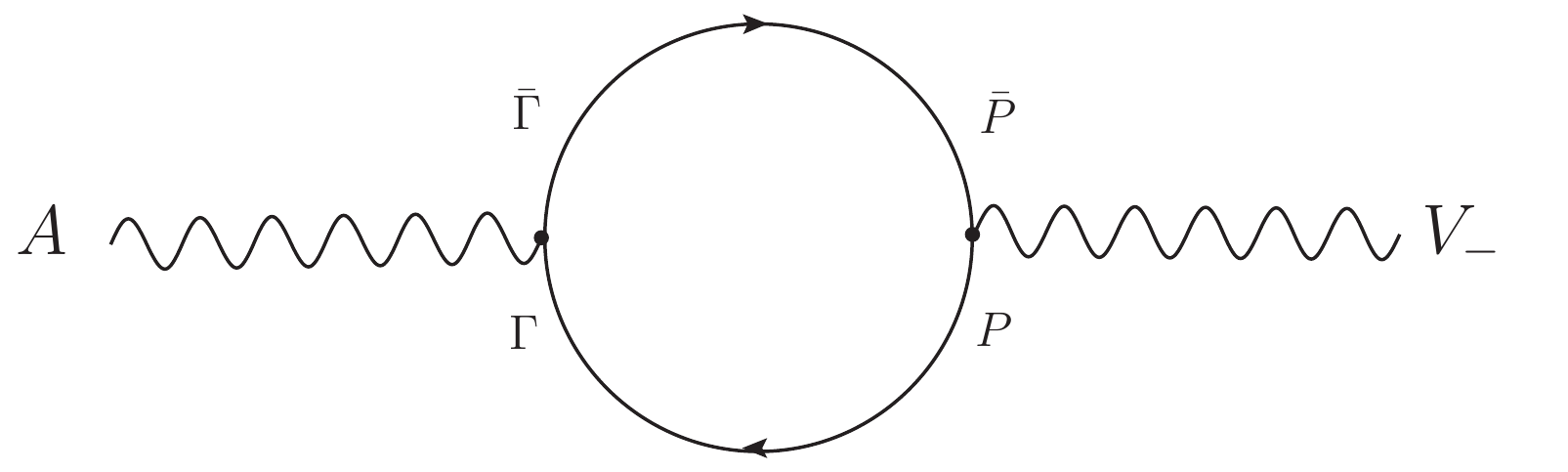}
\label{AV6}}
\caption{\textit{Novel contributions to $\langle AA\rangle$ and $\langle AV_-\rangle$ from $P$ and $\Gamma$ loops.}}
\label{sigmaloopssecond}
\end{figure}

Finally, we consider the massive fields $(P,\G)$. For simplicity we consider only one such pair with mass $M^2 = m^2|\S_0|^2 e^{2Q_\S A_0}$. Again, the main change from the massless case is the substitution $\m^2\rightarrow \m^2+ M^2$. There are also the three additional diagrams of figure~\ref{sigmaloopssecond}\ to evaluate,
which can only appear when the $P\G$ propagator is non-vanishing. In fact,~\ref{AA5}\ and~\ref{AA6}\ sum to zero. Therefore, the novel contributions coming from the $(P,\G)$ sector are
\be
\int {d^2q\over(2\pi)^2}d^2\th^+\, A(q)V_-(-q) \left[2 M^2 Q_{P}\left(Q_{P}+Q_{\G}\right)\int_0^1 dx\, \I_{0,2}\left(\D\right) \right],
\ee
where $\D= M^2 + x(1-x)q^2$. The $Q_{P} Q_{\G}$ term clearly originates from figure~\ref{AV6}, while the $Q_{P}^2$ term comes from the divergent graphs~\ref{AV1}\ and~\ref{AV2}. The latter contribution did not show up earlier in~\C{remainder}\ because it is proportional to $M^2$. In the limit $M^2\gg\m^2$,  only
\be
{Q_{\S}^2\over8\pi}  \int d^2x d^2\th^+\, A V_-,
\ee
survives from this sector, where we have used the fact that $Q_{\S} + Q_{P} + Q_{\G} =0$. This is the $AV_-$ term that arises in the one-loop correction to the $\S$ metric.

Putting everything together, we find that in the limit $M_A^2\ll \m^2 \ll M^2$ the coefficient of $AV_-$ in the effective action is
\be {1\over 8\pi} \left(Q_\S^2 -Q_\S^2 -\cA\right) = - {\cA\over 8\pi}, \ee
exactly as we found in~\C{Gammafinal}.

\newpage

%\bibliographystyle{amsunsrt-ensp}
%\bibliography{generalized}

\ifx\undefined\bysame
\newcommand{\bysame}{\leavevmode\hbox to3em{\hrulefill}\,}
\fi

\end{document}